\documentclass[twocolumn]{emulateapj}

\usepackage[usenames,dvips]{color}
\usepackage{epsfig}

\newif\ifAMStwofonts

\makeatother

\shorttitle{Correlation of structure and stellar properties of galaxies}
\shortauthors{Sachdeva et al.}

\begin{document}

\title{Correlation of structure and stellar properties of galaxies in Stripe 82}
\author{Sonali Sachdeva}
\affil{Kavli Institute for Astronomy and Astrophysics, Peking University, Beijing 100871, China}
\author{Luis C. Ho and Yang Li}
\affil{Kavli Institute for Astronomy and Astrophysics, Peking University, Beijing 100871, China}
\affil{Department of Astronomy, School of Physics, Peking University, Beijing 100871, China}

\and

\author{Francesco Shankar}
\affil{Department of Physics and Astronomy, University of Southampton, Southampton SO171BJ, UK}

\begin{abstract}
Establishing a correlation (or lack thereof) between the bimodal colour distribution of galaxies and their structural parameters is crucial to understand the origin of bimodality. To achieve that, we have performed 2D mass-based structural decomposition (bulge+disc) of all disc galaxies (total$=$1263) in the Herschel imaging area of the Stripe 82 region using $K_s$ band images from the VICS82 survey. The scaling relations thus derived are found to reflect the internal kinematics and are employed in combination to select an indubitable set of classical and pseudo bulge hosting disc galaxies. The rest of the galaxies ($<20\%$) are marked as discs with ``ambiguous" bulges. Pseudo and classical bulge disc galaxies exhibit clear bimodality in terms of all stellar parameters ($M_*$, sSFR, $r-K_s$). All pseudo bulge disc galaxies are blue and star-forming and all classical bulge disc galaxies are red and quiescent with less than $5\%$ digressions. Ambiguous bulge disc galaxies are intermittent to pseudo and classical bulge disc galaxies in the distribution of all structural and stellar parameters. $\Delta$$\langle\mu_{eb}\rangle$ - based on the placement of bulges on the Kormendy relation - is found to be the most efficient single structural indicator of both bulge type and stellar activity. The placement of ambiguous bulge disc galaxies on scaling relations and fundamental plane, in addition to their peculiar stellar properties suggest that they are dominantly a part of the green valley.
\end{abstract}

\keywords{galaxies: bulges --- galaxies: structure --- galaxies: star formation --- galaxies: stellar content}

%%%%%%%%%%%%%%%%%%%%%%%%%%%%%%%%%%%%%%%%%%%%%%%%%%%%%%%%%%%%%%%%%%%%%%%%%%%%%%%%%%%%%%%%%%%%%%%%%%%%%%%%%%%%%%%%%%%%%%%%%%%%%%%%%%%%%%%%%%%%%%%

\section{Introduction}
\label{sec:intro}

Over the past two decades, it has been emerging with increasing clarity that the structure of galaxies is correlated with their star formation history and on-going activity \citep{Kauffmannetal2003,Kauffmannetal2006,Baldryetal2006,Franxetal2008}. This conjecture is supported by the observation of statistically large samples that reveal that while redder galaxies are more early-type and bulge (or spheroid) dominated, bluer galaxies are more late-type and disc dominated \citep{Stratevaetal2001,Brinchmannetal2004}. 

Quantitatively, \citet{Driveretal2006} demonstrated that S\'ersic index of a galaxy - parameter defining the shape of its intensity profile - is most efficient in separating galaxies based on their colour. In addition to the S\'ersic index, concentration and bulge-to-total flux ratio have been demonstrated to be competitively effective for larger samples and at higher redshifts \citep{Cameronetal2009,Wuytsetal2011,Belletal2012,Mendeletal2013,Langetal2014,Blucketal2014}. In some studies, central mass density of galaxies within 1 kpc has been argued to be a more efficient separator of star forming and quiescent galaxies than other morphology indicators \citep{Cheungetal2012,Fangetal2013,Luoetal2020}. Recent studies have also argued that stellar kinematics, mainly in terms of central velocity dispersion, are a better differentiator of galaxy colour than stellar mass, surface mass density and morphology \citep{Wakeetal2012,Blucketal2016,vandeSandeetal2018}. Note that a requisite for accurate morphological decomposition is that it reflects the internal kinematics of the galaxy. Recently, \citet{Grahametal2018} reported that kinematics of 2300 galaxies obtained from the latest IFU (MANGA) survey demonstrate a tight correlation with their structure. Thus, all proposed colour differentiators - S\'ersic index ($n_b$), bulge-to-total flux ratio ($B/T$), concentration ($C$), central mass density ($\Sigma_1$) and velocity dispersion ($\sigma_o$) - are structural type indicators, firmly suggesting that the inherent structure of star forming galaxies differs from their quiescent counterparts.

Interestingly, all these structural indicators ($n_b$, $B/T$, $C$, $\Sigma_1$, $\sigma_o$) have also been found to be efficient differentiators of the bulge-type hosted by the disc galaxy \citep[reviewed in][]{FisherandDrory2016,Kormendy2016}. The higher values of these indicators for a disc galaxy suggest that the central bulge is elliptical like, i.e., ``classical", and lower values suggest that the central bulge is disc like, i.e, ``pseudo" \citep{FisherandDrory2008,Gadotti2009,SachdevaandSaha2016,Neumannetal2017,Gaoetal2020}. Involvement of the same set of indicators in the adjudication of both the bulge-type and stellar activity in disc galaxies implies that disc galaxies hosting different bulge types should exhibit distinct stellar activity. \citet{DroryandFisher2007} have argued that the underlying correlation of the colour of the galaxy is with its bulge type and correlation with all other morphological indicators is a consequence of that. 

Thus, separation of disc galaxies on the basis of their bulge-type should also result in their separation on the basis of colour, and vice-versa. However, two sets of studies have produced contradictory findings. One set of studies state that while pseudo bulges are often red, classical bulges are rarely found to be blue \citep{FisherandDrory2016,Kormendy2016}. Counter to that, other set of studies state that while pseudo bulges are rarely found to be red, classical bulges are often blue \citep{Gadotti2009,Fangetal2013,Luoetal2020}. \citet{Luoetal2020} highlighting these differences claim that even if they apply the same colour separation criteria as \citet{FisherandDrory2016}, 42\% of their classicals will be marked blue and all pseudo bulges will still be contained in the blue zone. 

The origin of contradictory results lies in the inconsistency of criteria applied in these studies for bulge classification. For example, while \citet{FisherandDrory2008} employ bulge S\'ersic index ($n_b$) to separate bulge-types, other studies have elaborated on the inefficiency of $n_b$ to separate bulge-types without severe contamination \citep{Gadotti2009,Gaoetal2020}. The reason being that while there are bulges which will satisfy multiple kinds of criteria to be unambiguously labelled as either pseudo or classical, there are also a considerable fraction of bulges which do not meet all the chosen criteria \citep[reviewed in][]{Kormendy2016}. Classification of such bulges is then subjective and both pseudo and classical samples become prone to contamination.

In this work, we attempt to overcome this issue by studying and applying multiple stringent complimentary criteria to select an indubitable set of pseudo and classical bulges. In addition to that, we prevent the contamination of these bulge sets by separating out those bulges which do not satisfy all the applied stringent criteria. We make a number of other improvisations to extract a clearer picture. Firstly, our sample consists of all local ($z<0.3$) disc dominated galaxies (total$=$$1263$) in the Herschel imaging area of the Stripe 82 region. Since this area has coverage from a large number of deep multi-wavelength surveys, high quality and large quantity of data in terms of images, spectra and derived parameters is available. This is advantageous in analysing the effect of structural transformation on the whole range of stellar, gas and dust properties. Crucially, structural parameters for all the galaxies in our sample are available in the optical bands (ugriz) from the latest decomposition performed by \citet{Bottrelletal2019} using deep co-added Sloan images. This has enabled us to perform consistency checks for our derived decomposition parameters in the $K_s$ band. 

Secondly, we have performed the decomposition in $K_s$ band. This band is the best tracer of stellar mass in galaxies since it is not biased by the dominating optical emission from young stars which account for a small fraction of galaxy's mass. It is least affected by dust obscuration and accounts for all the light emitted by middle-age and old stars which form the bulk of galaxy's baryonic mass \citep{Cowieetal1996,Bundyetal2006,Blucketal2019}. Our decomposition also benefits from the fact that the $K_s$ band images \citep[VICS82 survey,][]{Geachetal2017} boast of a high resolution (0.3"/pixel) and depth (21.4 mag). Thirdly, we have extracted all the possible structural measures of the galaxies using both parametric and non-parametric techniques. This allows us to assess the performance of all potential indicators, along with the kinematic information, to classify bulges in a robust manner. Fourthly, the stellar parameters ($M_*$, SFR, sSFR) are from a recent work \citep{Salimetal2016,Salimetal2018} that includes far-IR flux from Herschel along with the mid-IR flux from WISE to account for the dust attenuation affect which is a critical factor in the accurate estimation of stellar activity.

This paper is organized in the following manner. In Section~\ref{sec:data}, we define the sample and elaborate our computation of all defining structural parameters using both parametric and non-parametric techniques. We also detail the source and computation of kinematic and stellar parameters employed in this work. In Section~\ref{sec:results}, we elaborate on our separation of an indubitable set of pseudo and classical bulges with the identification and application of multiple stringent complimentary criteria. Following that, we compare the distribution of all structural and stellar parameters for the two bulge-types. We have also attempted to identify the most effective single structural indicator for both bulge-type and stellar activity in disc galaxies. In Section~\ref{sec:discussion}, we discuss our findings pertaining to the classification of bulges, bimodality of their properties and clues regarding the constitution of the green valley. We conform to a flat $\Lambda$-dominated Universe with $\Omega_{\Lambda}=0.714$, $\Omega_m=0.286$ and $H_o=69.6$ km sec$^{-1}$ Mpc$^{-1}$. All magnitudes are in the AB system.

\section{Data}
\label{sec:data}

\citet{Bottrelletal2019} performed bulge disc decomposition of all galaxies in the Stripe 82 region (16908 in total) in optical (u,g,r,i,z) bands using deep co-added SDSS images from \citet{Annisetal2014}. They demonstrated that deep images result in a more accurate determination of bulge and disc parameters compared to the computations based on SDSS Legacy images \citep{Simardetal2011}. In their work, only the decomposition performed in $r$ band has all fitting parameters kept free. For rest of the bands ({\it u,g,i,z}) either some or most of the parameters are held fixed according to their $r$ band values. We, thus, take their $r$ band catalogue for our sample selection and later for comparison with our mass-based morphological measurements. Using their $r$ band catalogue, we first select those galaxies which are determined to be 2-component (disc + free-bulge) systems, according to their analysis. 

We put a further cut on that sample by selecting only those galaxies which are in the 79 $deg^2$ contiguous imaging field of Herschel Stripe82 Survey \citep[HerS,][]{Vieroetal2014}. This ensures maximum overlap with all major multi-wavelength deep surveys in Stripe 82 including BOSS \citep{Eisensteinetal2011}, VLA-Stripe82 \citep{Hodgeetal2011}, HETDEX \citep{Hilletal2008}, SHELA \citep{Papovichetal2012}, SpIES \citep{Timlinetal2016}, HSC \citep{Miyazakietal2012} and VICS82 \citep{Geachetal2017}. This overlap is essential to perform mass-based morphological decomposition of galaxies and to obtain accurate estimates of their stellar parameters, dynamical parameters, dust content and gas content, which are crucial to trace the affect of structural transformation of galaxy on its stellar activity.

Out of the total sample of 16908 galaxies of \citet{Bottrelletal2019}, 1263 galaxies satisfy both the criteria, i.e., are determined to be 2-component (disc + free-bulge) systems and are present in the HerS field. We will perform the structural decomposition of this sample of 1263 galaxies in $K_s$ band. Selection of this band is crucial to trace the underlying mass distribution of galaxies in the most accurate manner. \citet{Blucketal2019} using 5 X 10$^5$ local ($z<0.2$) SDSS galaxies, have demonstrated that $B/T$ computed in all optical bands underestimates $B/T$ by mass. For $r$ band, the difference is more than 0.2, i.e., a component having 40\% of galaxy's full light might be designated to be accounting for only 20\% of that (see their Figure B1). They report that optical and mass-based structural fractions coincide closely for the $K_s$ band. This is because $K_s$ band accounts for all the light emitted by young and old stars and does not get affected by the brightness of young stellar populations \citep{Cowieetal1996,Bundyetal2006,Blucketal2019}. Since our work is focused on classifying bulges in an indubitable manner and examine the bimodality of their stellar properties, it is paramount to perform a mass (baryonic) based decomposition. In addition to that, our $K_s$ band images support a much better resolution (0.3") than the Sloan images (0.396") used by \citep{Bottrelletal2019}.

The $K_s$ band images are obtained from VICS82 survey, i.e., VISTA-CFHT Stripe 82 near-IR survey, which covers near-contiguous 150 $deg^2$ of Stripe 82 to an average depth of 21.9 mag in $J$ band and 21.4 mag in $K_s$ band \citep{Geachetal2017}. This survey has been conducted using Canada-France Hawaii Telescope (CFHT) WIRCam instrument and Visible Infrared Survey Telescope for Astronomy (VISTA) VIRCAM instrument. All images have been processed including dark and flat-field correction, refined sky subtraction, distortion correction, quality control, astrometric and photometric calibration. Using their $K_s$ selected catalogue, we match the RA Dec of our sample of 1263 galaxies to 1" difference and obtain information regarding the tile and coordinates on which each galaxy is present. Using this information, we write a code which automatically selects the tile (out of 33 VIRCAM and 55 WIRCam tiles in $K_s$ band) on which a galaxy is present and extracts a 180" X 180" cut-out for that galaxy. This cutout size is reasonably large considering that out of 1263, only 6 sources have total extent (Petrosian radius) more than 30" and none exceeds 40". It is optimal for the running of {\it ellipse}, {\it GALFIT} and other algorithms followed in this work. In addition to that, large size ensures that sky dominates the fitting region which is also an important criteria for {\it GALFIT} to create an accurate sigma image. We, thus, obtain cut-outs of all 1263 galaxies in the $K_s$ band.

\subsection{Sky variation and masking}

Accurate measurement of sky background variation is essential for robust fitting of galaxy images \citep{Pengetal2010}. We have estimated this variation separately for each of the VICS82 survey tiles from which our sample galaxies' cut-outs are obtained. The tiles have already been through rigorous background modification during pre-processing, which includes subtraction of the running sky, ``destriping" of the images in both directions, removal of background gradients, visual inspection of images to removes defects, patterns, residuals, in addition to fitting and removal of a fifth-degree polynomial surface \citep{Geachetal2017}. 

To compute the sky variation, we first run Source Extractor \citep{BertinandArnouts1996} with a low detection threshold (0.8 times relative to the background RMS) and broad filter (Gaussian with FWHM of 5 pixels) to mask all possible sources present on the tile along with their faint outskirts \citep{AkhlaghiandIchikawa2015}. We further grow each mask region according to the masked source size to rule out the possibility of contamination from extended diffused light. On the non-masked (sky) pixels, we perform 3-sigma clipping using standard {\it Python} algorithm to obtain the statistical measures. Since the background has already been through rigorous procedures, the mean and median of the sky background are found to be zero across all tiles. The standard deviation is found be similar to 2nd decimal place for all tiles of a given telescope (CFHT and VISTA) and band ($K_s$), confirming the robustness of computation. For CFHT $K_s$ band images, the value ranges from 23.90 to 24.17 mag/arcsec$^2$ and for VISTA $K_s$ band images, the value ranges from 23.80 to 23.98 mag/arcsec$^2$. These background variation values will be given as an input to {\it GALFIT} to aid the algorithm to obtain the most accurate bulge and disc parameters.

The mask (or segmentation-map) images generated for each cut-out in this process are also a critical input to the fitting algorithms employed in this work. These images are first given as an input to a custom-made code which unmasks the central source, i.e., the galaxy of interest, in each cut-out. We visually examined all modified mask images and corrected for a few cases (less than 5\%) in which some part of the galaxy of interest got masked or some neighbouring source did not get masked properly. The final mask images are given as an input to the isophotal fitting algorithm ({\it IRAF ellipse}) which is used to generate input values for {\it GALFIT} and later to {\it GALFIT} as well.

\subsection{Initial input parameters}

We will perform the bulge-disc decomposition on all galaxy images using {\it GALFIT} which is an algorithm that uses the Levenberg-Marquardt (LM) technique on galaxy images to find the best fitting structural parameters in a flexible and fast manner \citep{Pengetal2002,Pengetal2010}. To obtain the initial input parameters for {\it GALFIT}, we perform isophotal fitting on all galaxy images using the {\it ellipse} task of {\it IRAF}. This task finds best fitting isophotes for a galaxy at successively increasing radii, using LM technique, varying the intensity, central coordinates, ellipticity and position angle \citep{Jedrzejewski1987}. The cutouts and their corresponding masks are given as an input. We use the resulting table of isophotal values to create surface brightness profile for each galaxy according to their magnitude zero-point and plate-scale. These profiles are given as an input to a custom written code which extracts initial input parameters for {\it GALFIT}, i.e., total radius, magnitude, half light radius, ellipticity and position angle, for each of the 1263 galaxies. The total radius is defined by the outer-most fitting isophote and the total magnitude is according to the total flux contained in this radius. The half light radius is taken as that radius which contains half of the total flux, while ellipticity and position angle are obtained by computing the mode of all values across the profile. Other than generating the input values, the isophotal profiles will also be utilized after running {\it GALFIT} to examine if the output parameters generated by {\it GALFIT} are describing the galaxy well or significant residual is left. 

\subsection{Sigma image and PSF}

A vital requirement for {\it GALFIT}'s fitting process is the sigma image which informs the algorithm regarding the standard deviation of flux at each pixel of the input image \citep{Pengetal2002}. {\it GALFIT} can create an accurate sigma image internally if information regarding the units of the input image, flux conversion factor, gain of the telescope, exposure time involved, number of frames combined to create the input image and read noise of the detector is known to the user. We assimilated all such information from survey documentation, headers of the large CFHT/VISTA tiles and image analysis. While CFHT images are in micro-Janskys, VISTA images are in counts. Using the appropriate flux conversion factor, exposure time and gain of the telescope, we create the ``gain" parameter in the header of each cut-out (or input image) in such a manner that its multiplication with the image units will yield the total number of electrons for each pixel. We also create the ``rdnoise" parameter in each cut-out's header. These header parameters will be utilized by {\it GALFIT} to generate sigma image. 

Another important requirement for {\it GALFIT}'s fitting process is the point spread function (PSF) which the algorithm convolves with the model to mimic the effect of the telescope and filter on the actual flux distribution. The VICS82 survey provides an average PSF for each telescope (CFHT and VISTA) and band ($J$ and $K_s$) by evaluating the FWHM of $10^4$ point sources extracted from randomly selected image tiles belonging to each telescope and band. In the creation of the median stack, which includes the normalization of each source to its peak flux, they ensured that all selected $10^4$ point sources ($CLASS\_STAR>0.95$) are bright ($14<K_s<15$ mag) and unsaturated. They exhibited the quality of the generated PSF by employing it to derive aperture corrections for their photometric analysis. We followed the same process to create our own PSF, for each CFHT and VISTA $K_s$ band tile, selecting some $\sim100$ point sources using {\it DAOPHOT PSF} task of IRAF and compared the PSF thus generated with the one provided by the survey. Although we selected a far fewer number ($\sim100$) of point sources, the variation of the PSF across tiles and its difference with the survey PSF was minute and did not have any affect on the fitting parameters within the error range. This could possibly be due to small variation in seeing across the survey, where, CFHT reports a seeing of 0.96"($\sigma\sim\pm$0.10") and VISTA reports a seeing of 0.82"($\sigma\sim\pm$0.13") in $K_s$ band. Considering the insensitivity of the fitting to the PSF variation, we use the PSF provided by VICS82 survey, throughout this work, to rule-out any biases originating from the creation of a PSF in a non-rigorous and/or non-uniform manner for any tile.

\subsection{Fitting 1 and 2-component models}

To run {\it GALFIT} in a batch, i.e., on the full sample of 1263 galaxies, a customized input file for each galaxy is required. This file stores all information pertaining to the running of the algorithm on that galaxy, i.e., name of the cut-out, its mask file, PSF file, fitting box, convolution box, magnitude zeropoint, plate scale, initial input parameters, etc. Using the initial input parameters, obtained earlier through isophotal fitting, we write a code to generate such files for all galaxies in an automated manner. The dimension of the fitting box is twice the diameter of the central source and that of the convolution box ranges from 40-60 times the PSF-FWHM of the image. Using the customized files, we first fit all galaxies with a single S\'ersic component,

\begin{equation} 
I_{sersic}(r) = I_e(r_e) \exp [-b_n((\frac{r}{r_e})^{1/n} - 1)],
\end{equation}

\noindent where $r_e$ is the half light radius, $I_e(r_e)$ is the intensity at that radius, $n$ is the S\'ersic index and $b_n$ is a constant dependent on $n$. Out of 1263 galaxies, while 946 converged well ($1<\chi^2_\nu<2$), 317 did not converge. ``Did not converge" means that although {\it GALFIT} did not crash but one or more final parameter values were marked with asterisk signifying that those parameter values are non physical. We re-fitted those 317 galaxies fixing their S\'ersic index to range of values ($n=1.0,2.5,4.0$) and selecting the best fit, following which, 287 converged and 30 galaxies still did not converge. Examining the images of these 30 galaxies, we found that while 10 of them are poorly imaged, the rest do not have any clear signs that could explain their lack of convergence.

We consolidate the output parameters, obtained from 1-component fitting, of all galaxies and use them to create input files for 2-component fitting. In 2-component fitting, S\'ersic component (Equation~1) is simultaneously fitted with an exponential (or disc) component. The exponential component is the special case of the S\'ersic component with $n=1$, such that,

\begin{equation} 
I_{disc}(r) = I_o \exp (-\frac{r}{r_d}),
\end{equation}

\noindent where $I_o$ is the intensity at the centre of the disc and $r_d$ is the scale length of the disc. Thus, $I_{total}=I_{sersic}+I_{disc}$ is simultaneously fitted for the full sample. Out of 1263 galaxies, while 605 converged well ($1<\chi^2_\nu<2$), 658 did not converge. We re-fitted those 658 with a range of fixed S\'ersic indices ($n_b=1.0,2.5,4.0$) and selected the best fit. Out of those 658, 337 converged, however, 321 did not converge. We tried to re-fit them by fixing various parameters including ellipticity and position angle of the disc (derived from isophotal fitting), however, they still did not converge. 

\begin{figure*}
\centering
\mbox{\includegraphics[trim=75 0 75 0,clip,width=100mm]{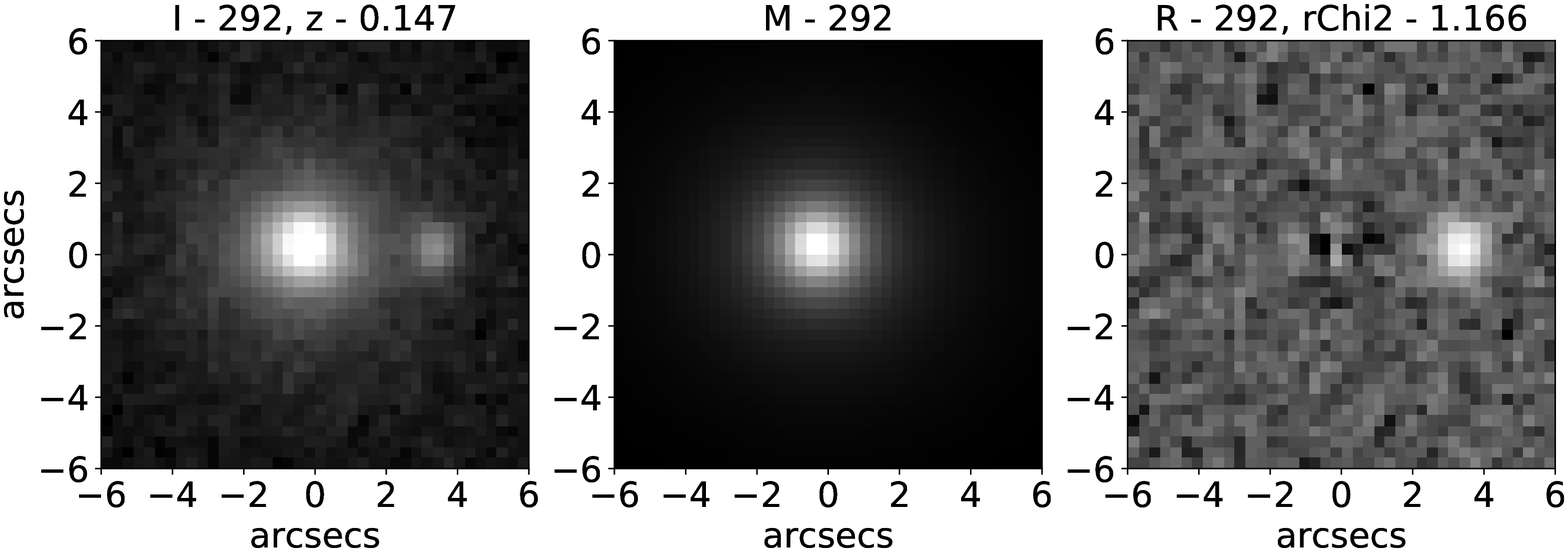}}
\mbox{\includegraphics[width=40mm]{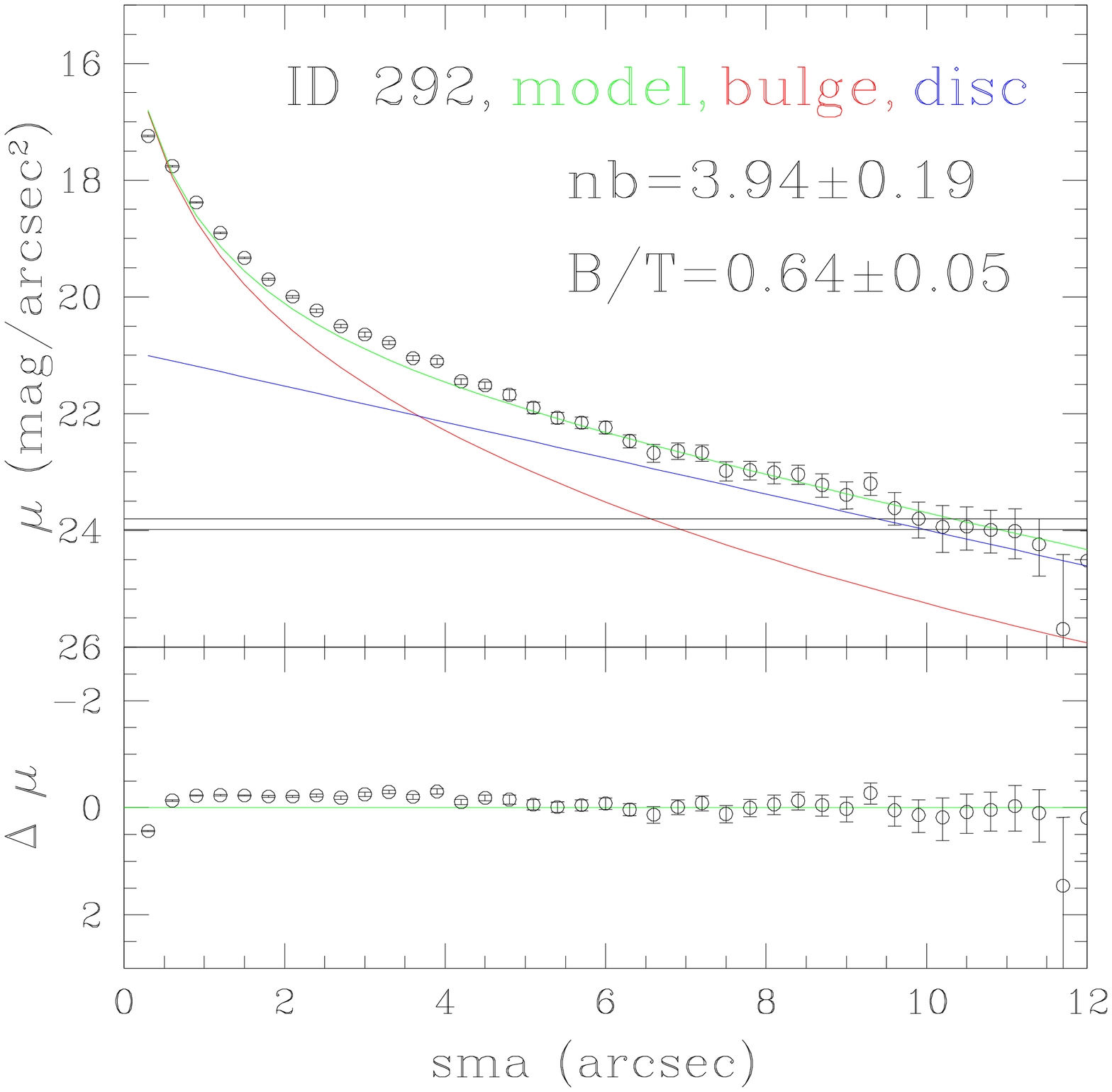}}\\
\mbox{\includegraphics[trim=75 0 75 0,clip,width=100mm]{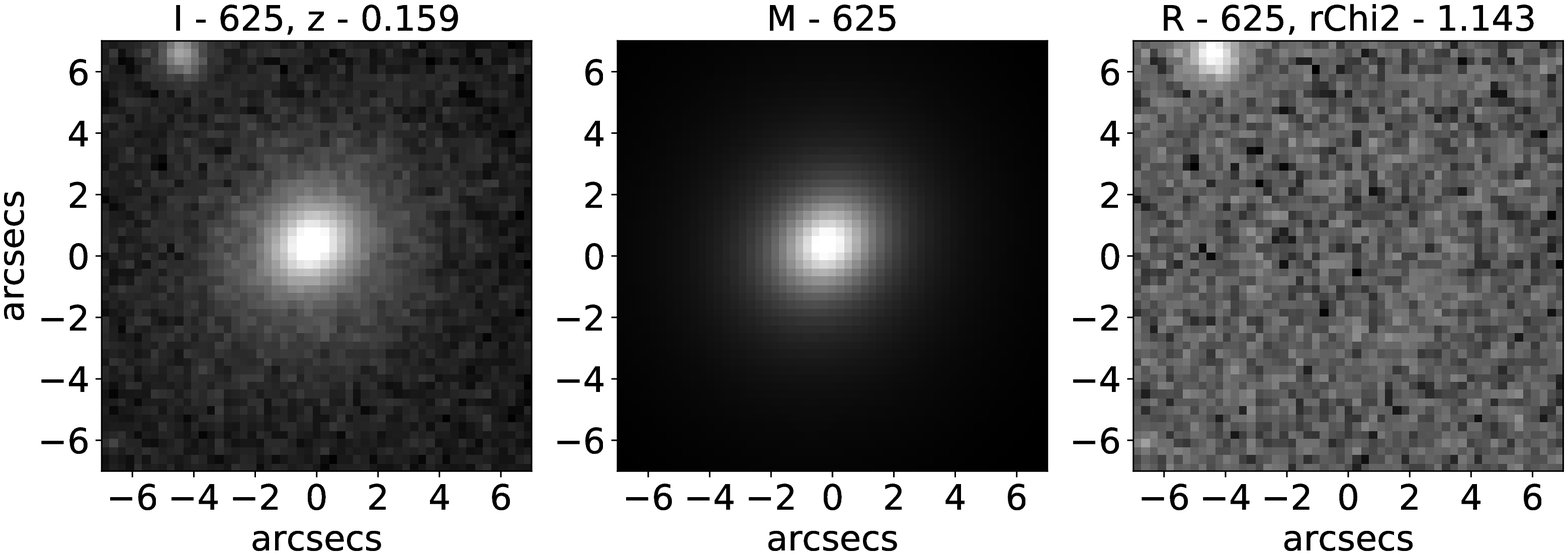}}
\mbox{\includegraphics[width=40mm]{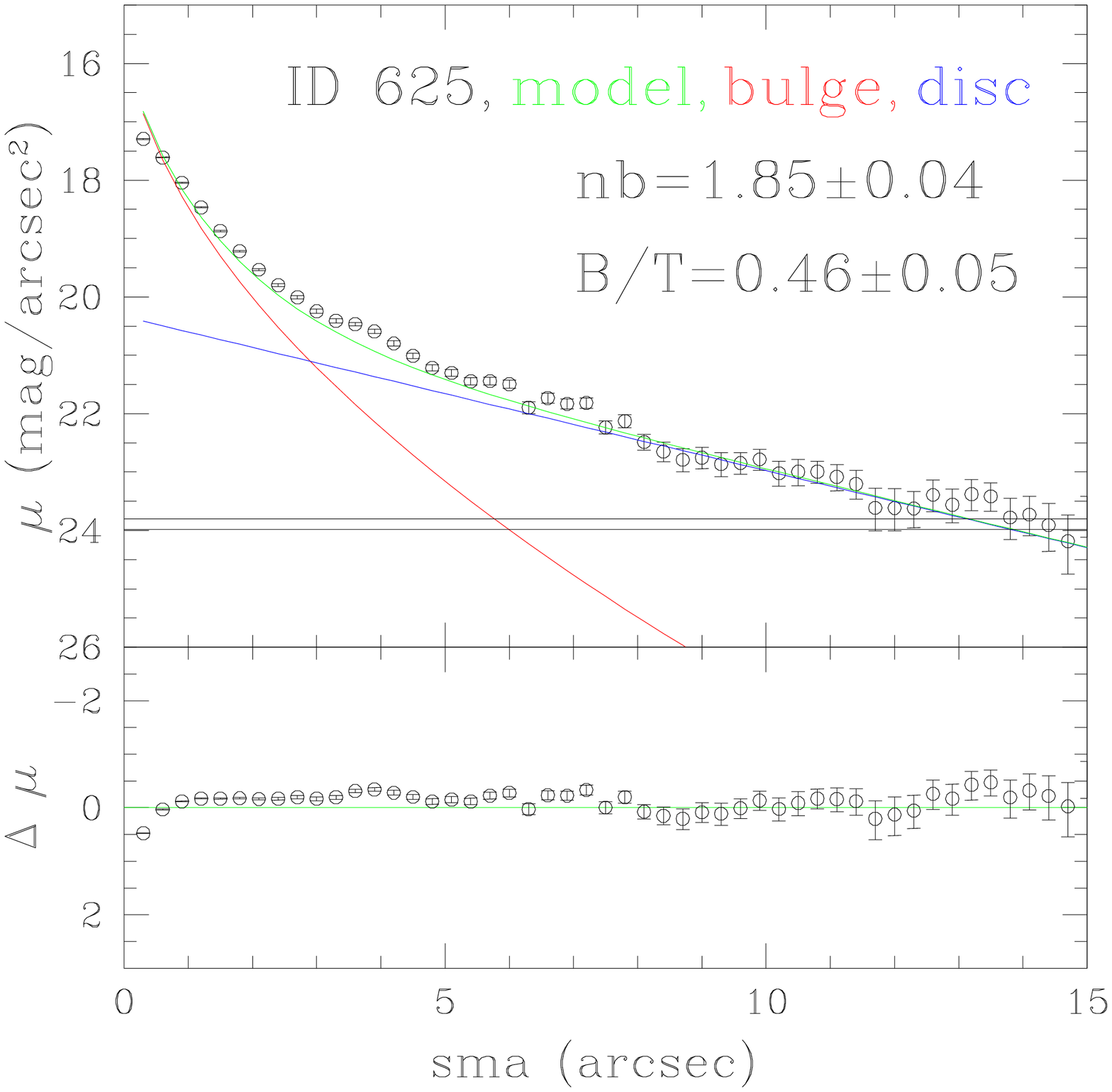}}\\
\mbox{\includegraphics[trim=75 0 75 0,clip,width=100mm]{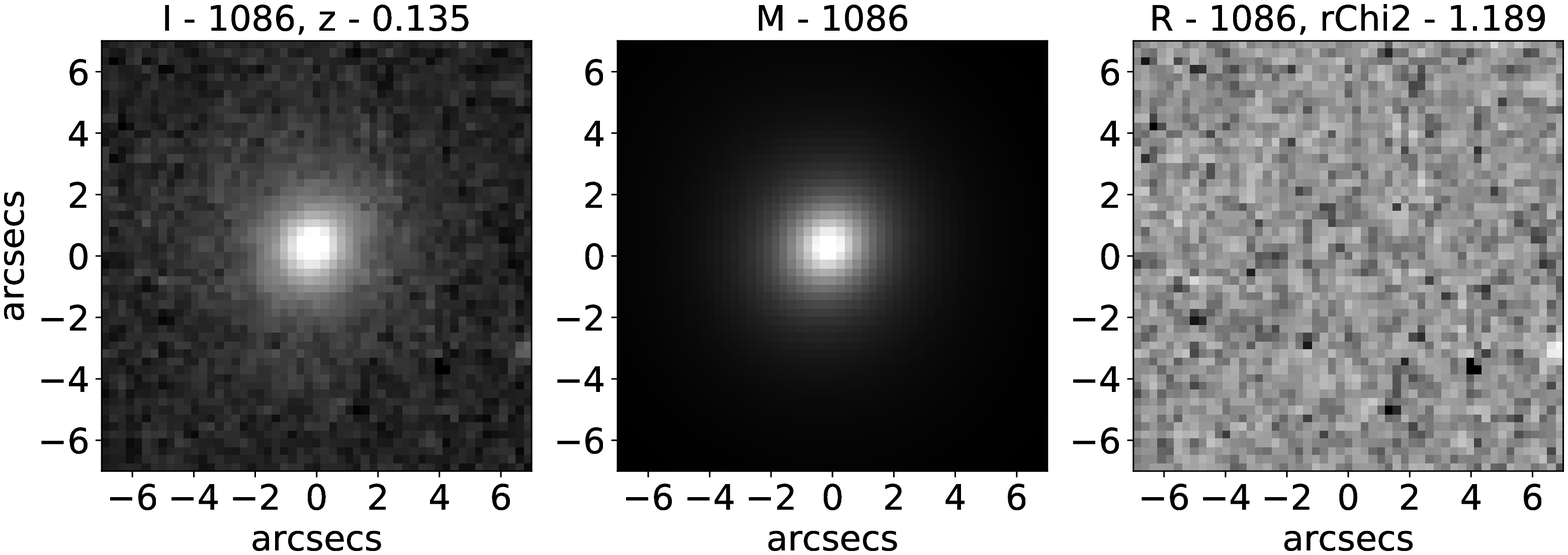}}
\mbox{\includegraphics[width=40mm]{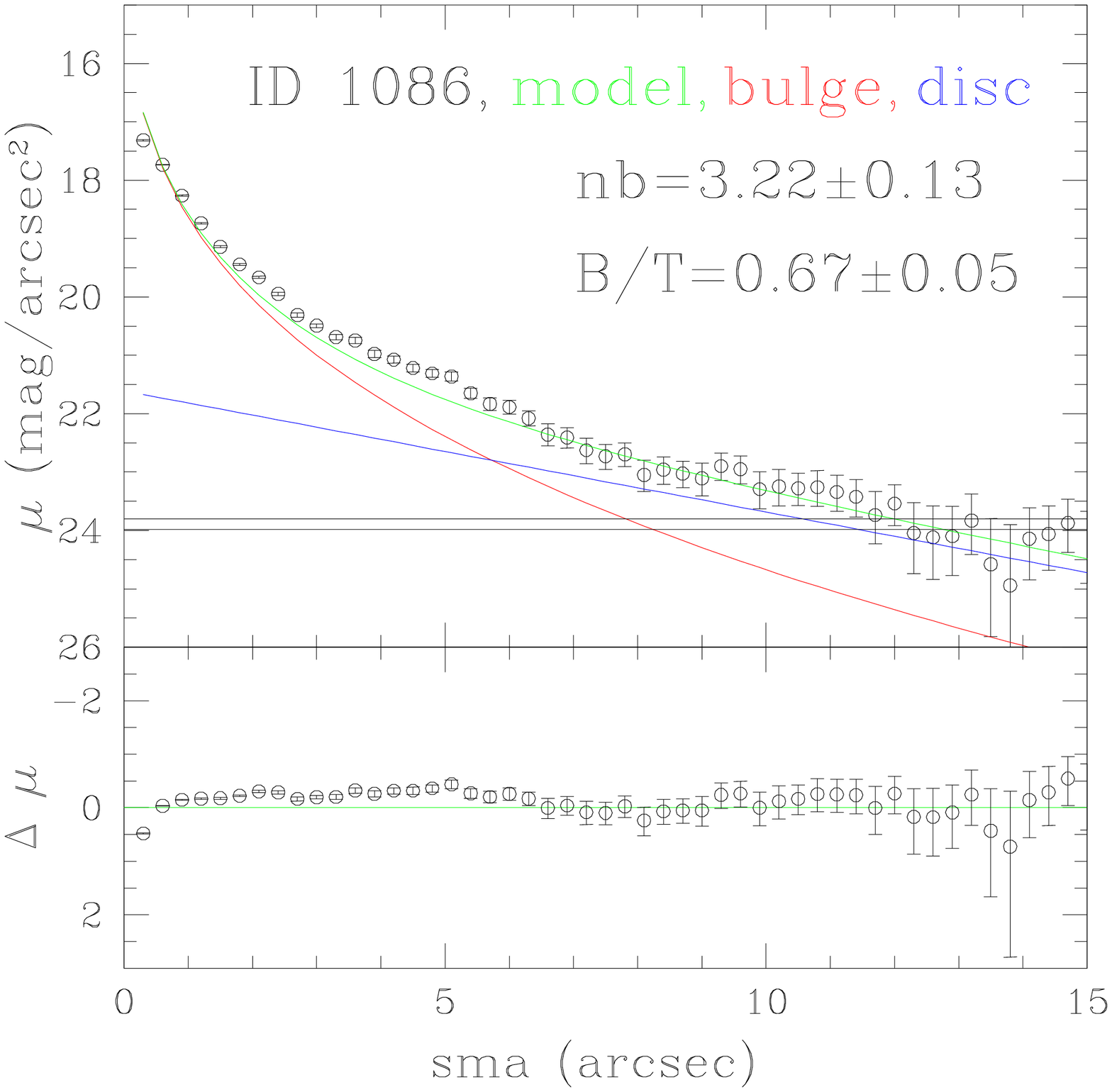}}\\
\mbox{\includegraphics[trim=75 0 75 0,clip,width=100mm]{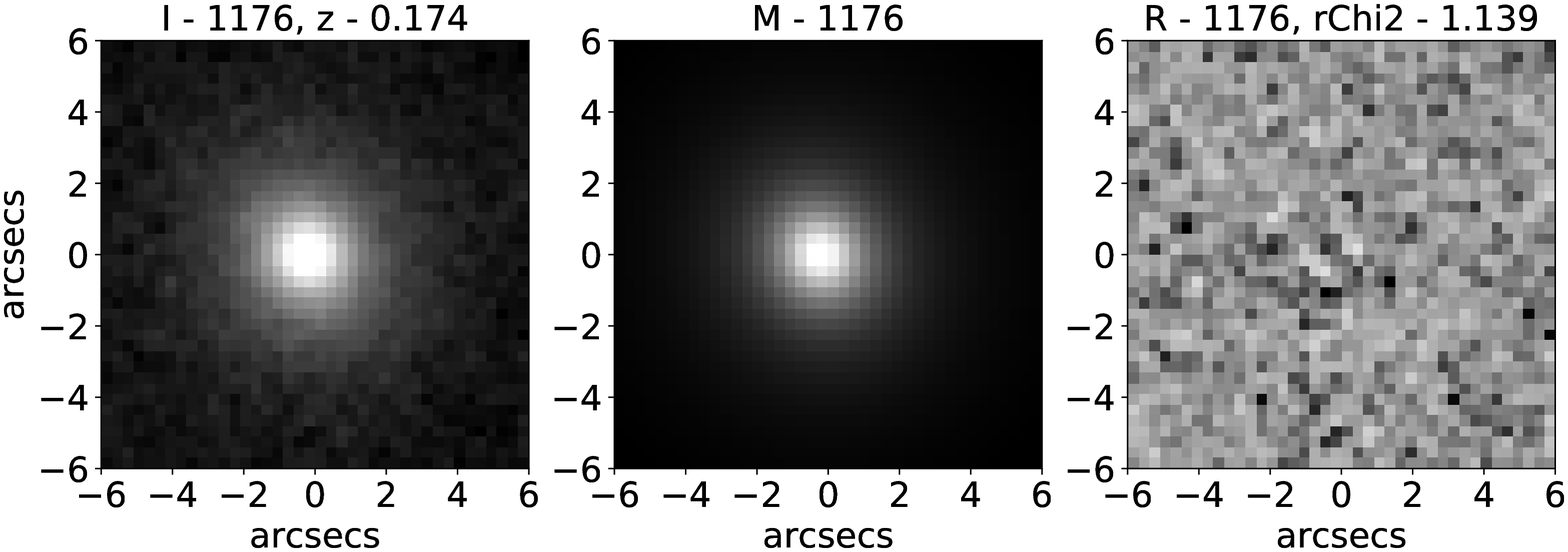}}
\mbox{\includegraphics[width=40mm]{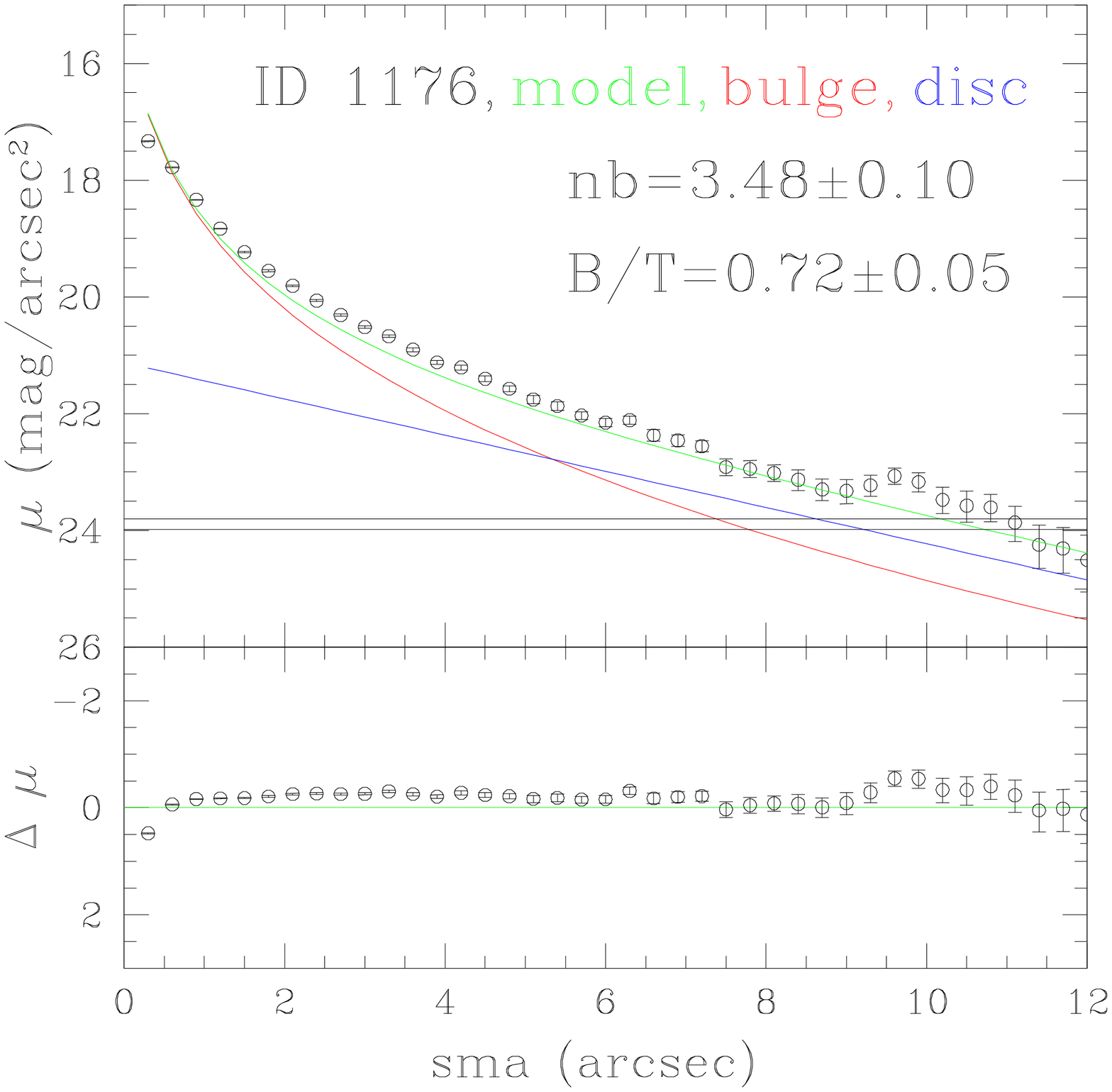}}\\
\mbox{\includegraphics[trim=75 0 75 0,clip,width=100mm]{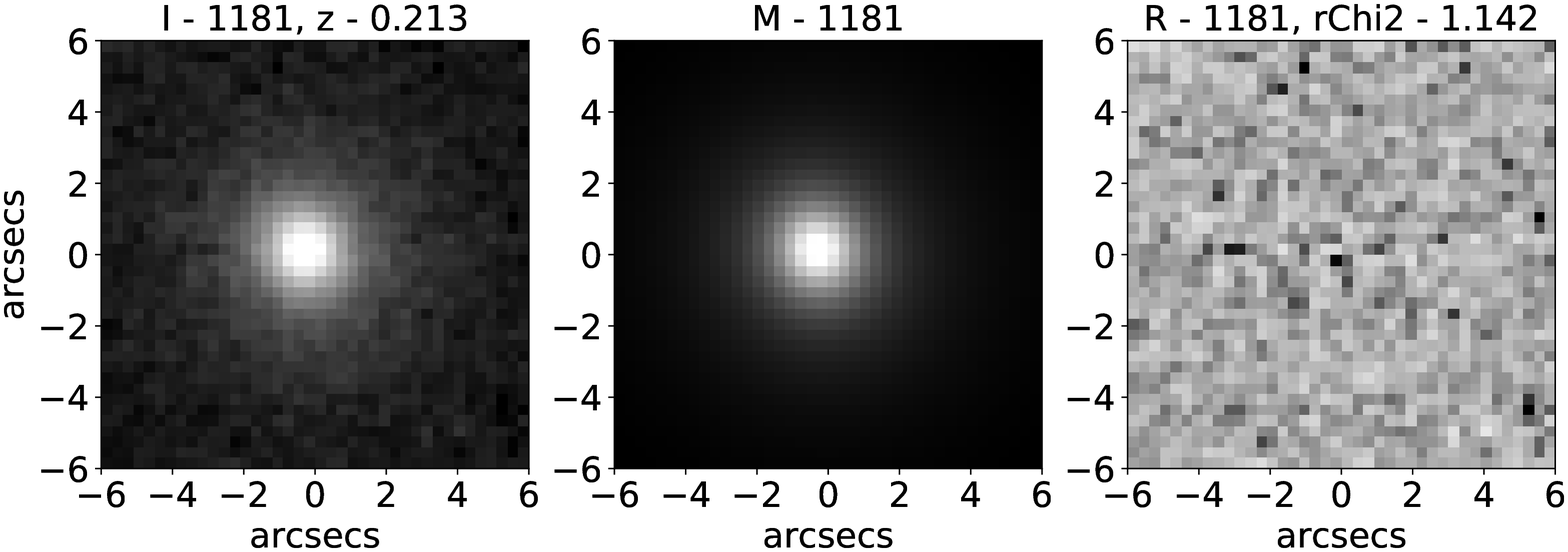}}
\mbox{\includegraphics[width=40mm]{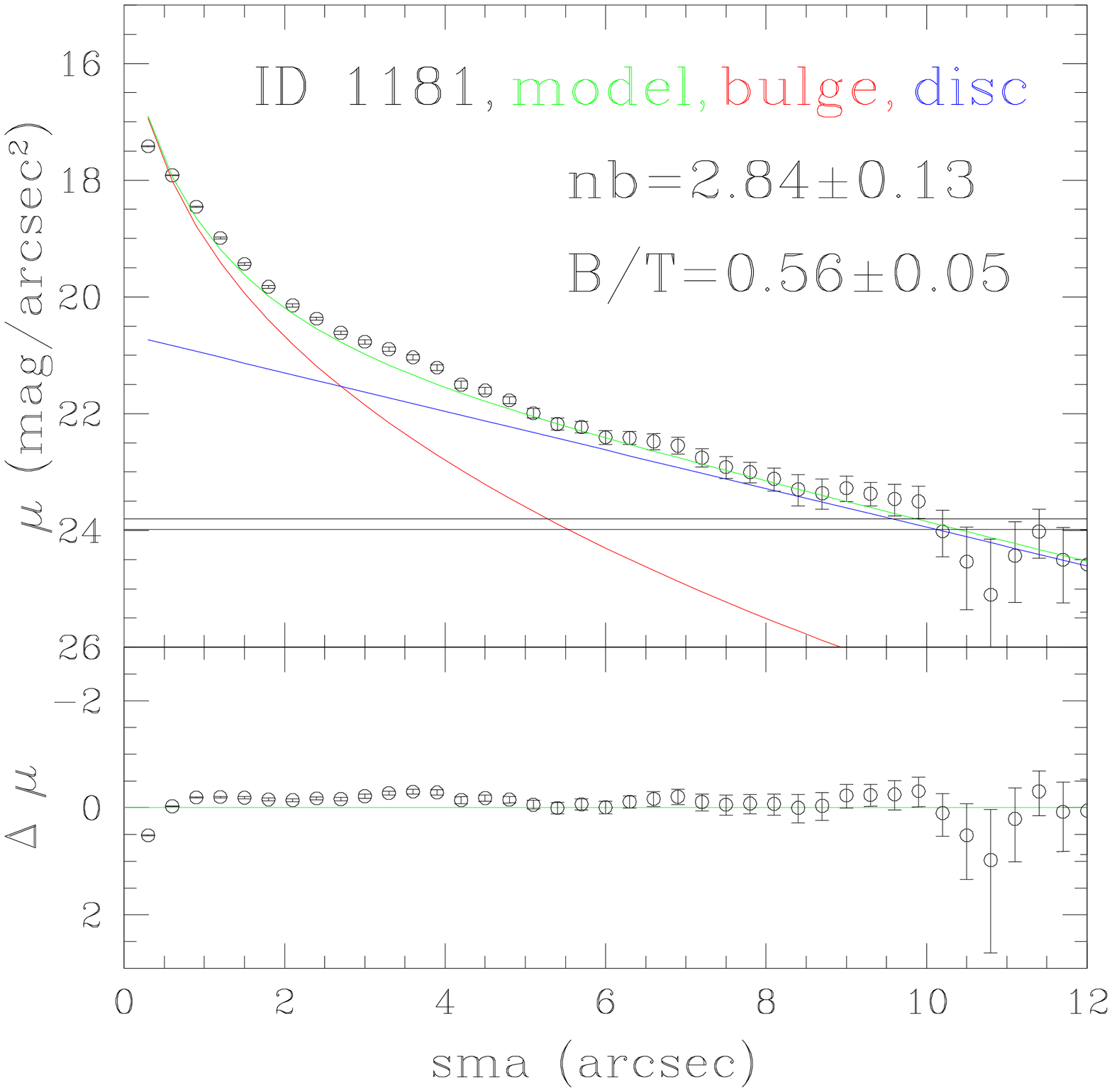}}
\caption{{\bf Fitting 2-components:} Each row depicts the fitting of a particular source in $K_s$ band. In a row, first three columns show the actual image of the source, its {\it GALFIT} model and its residual, respectively. While the redshift of the source has been mentioned at the top of actual image, the $\chi^2_\nu$ of the fitting has been mentioned at the top of residual image. The fourth column compares the model image's profile (green solid line) with the observed profile of the actual image (black solid points) obtained through isophotal analysis. Profiles of the bulge (red solid line) and the disc (blue solid line) component of the model image have also been marked. The difference of the model and observed profile is shown in the bottom panel of the plot.}
\label{fittingplota}
\end{figure*}

\begin{figure*}
\centering
\mbox{\includegraphics[trim=65 0 65 0,clip,width=100mm]{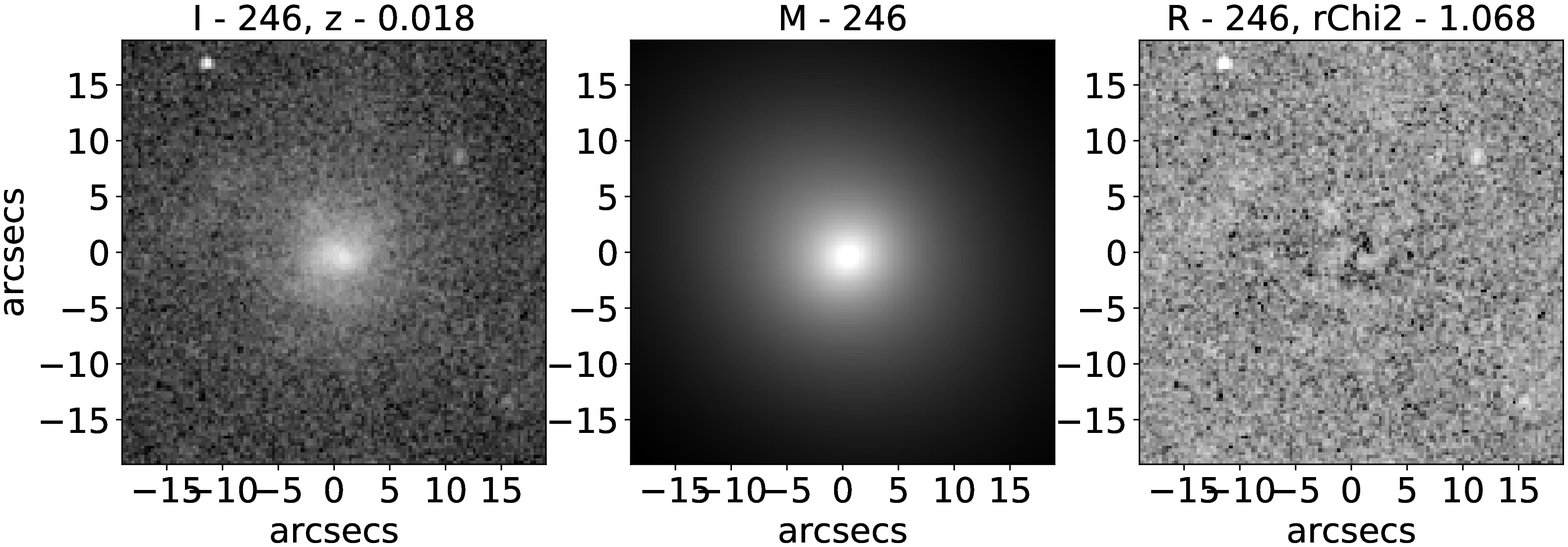}}
\mbox{\includegraphics[width=40mm]{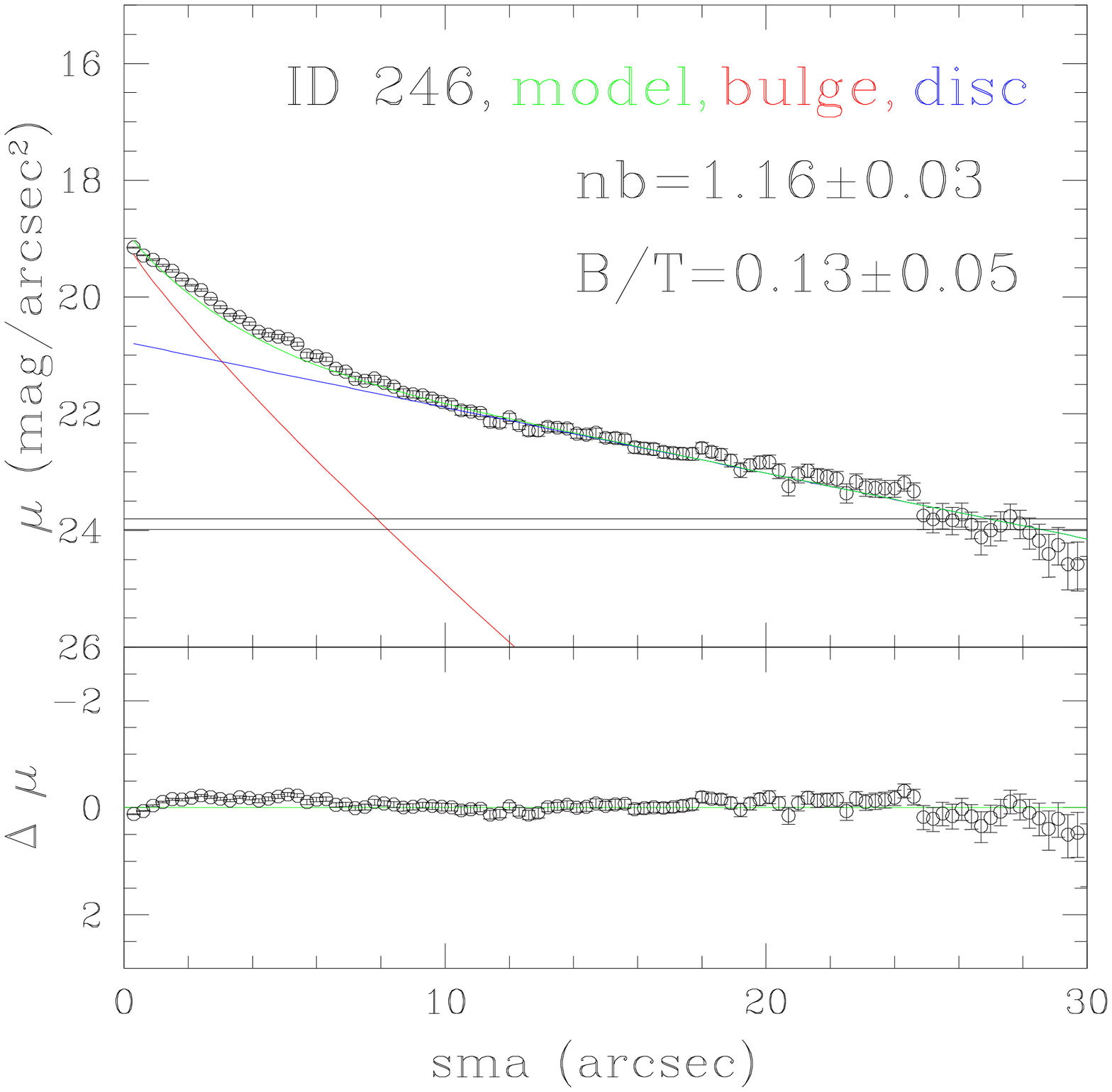}}\\
\mbox{\includegraphics[trim=75 0 75 0,clip,width=100mm]{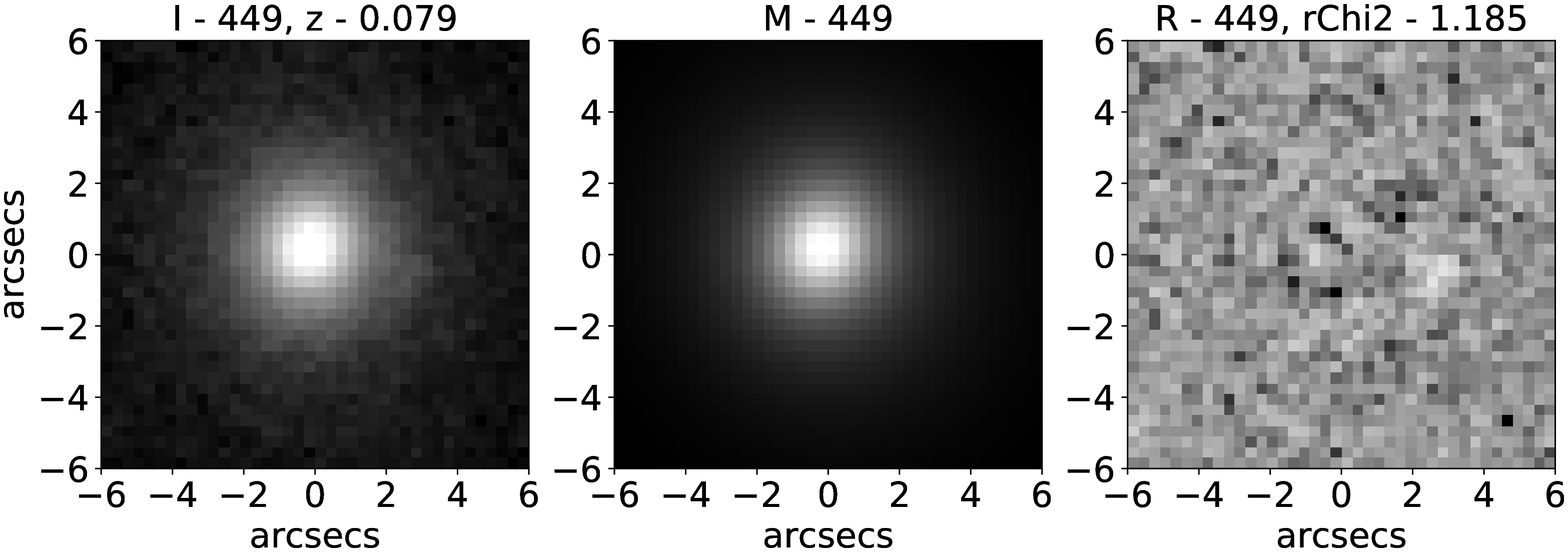}}
\mbox{\includegraphics[width=40mm]{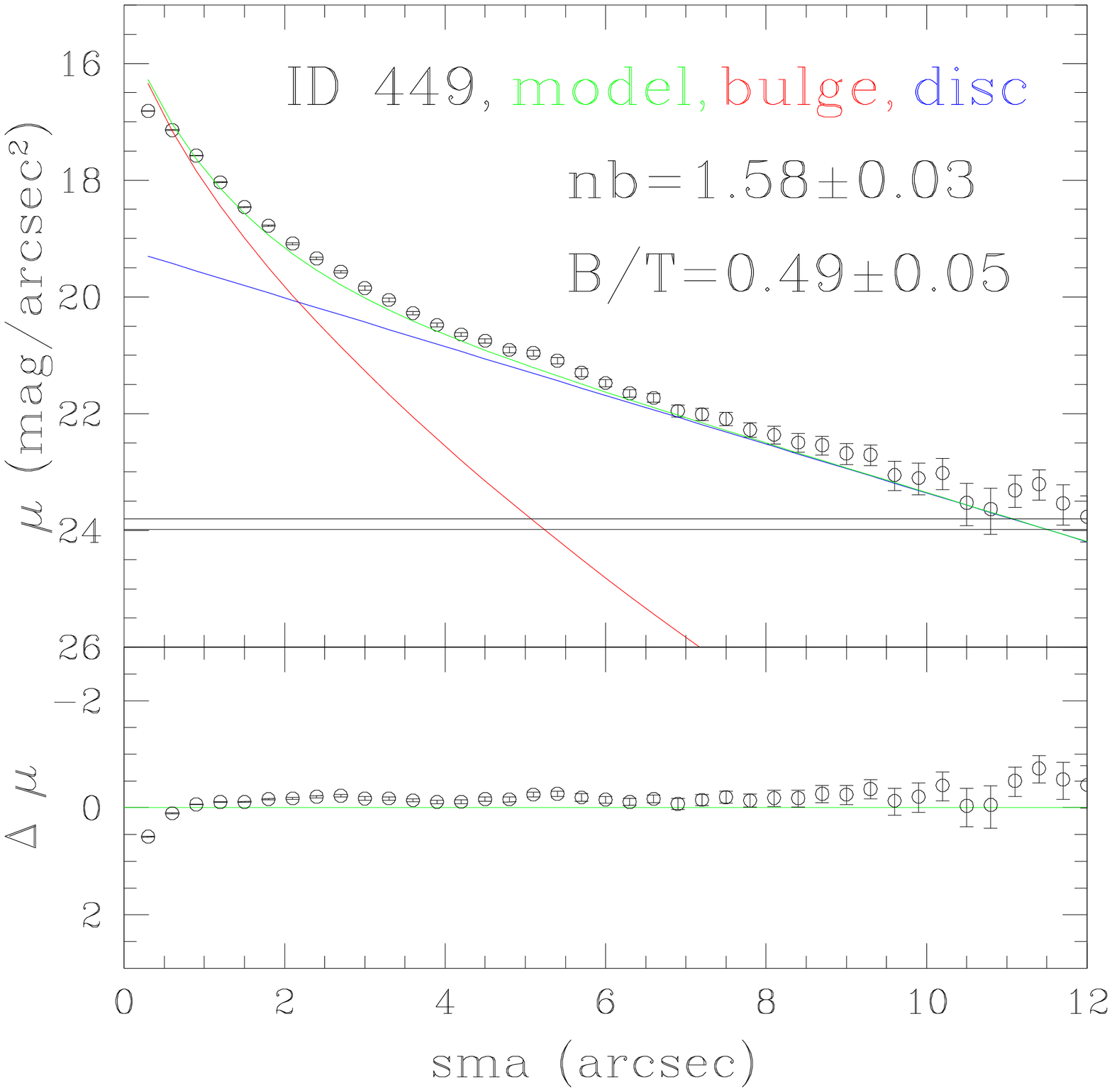}}\\
\mbox{\includegraphics[trim=75 0 75 0,clip,width=100mm]{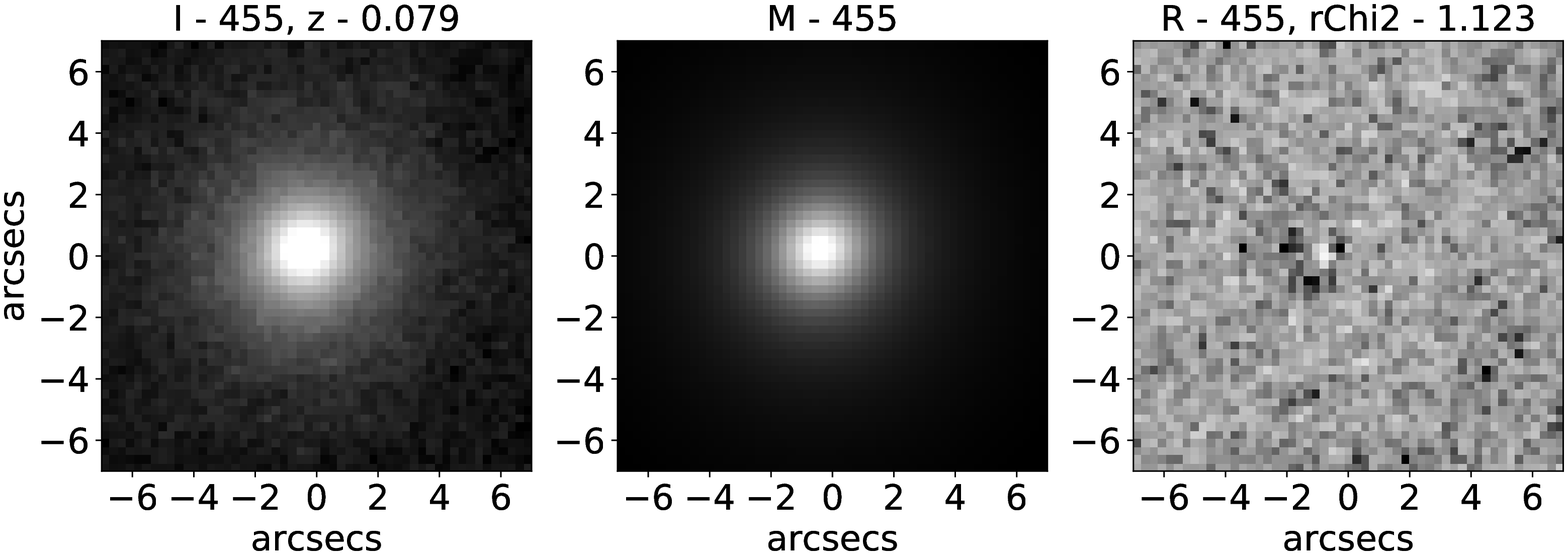}}
\mbox{\includegraphics[width=40mm]{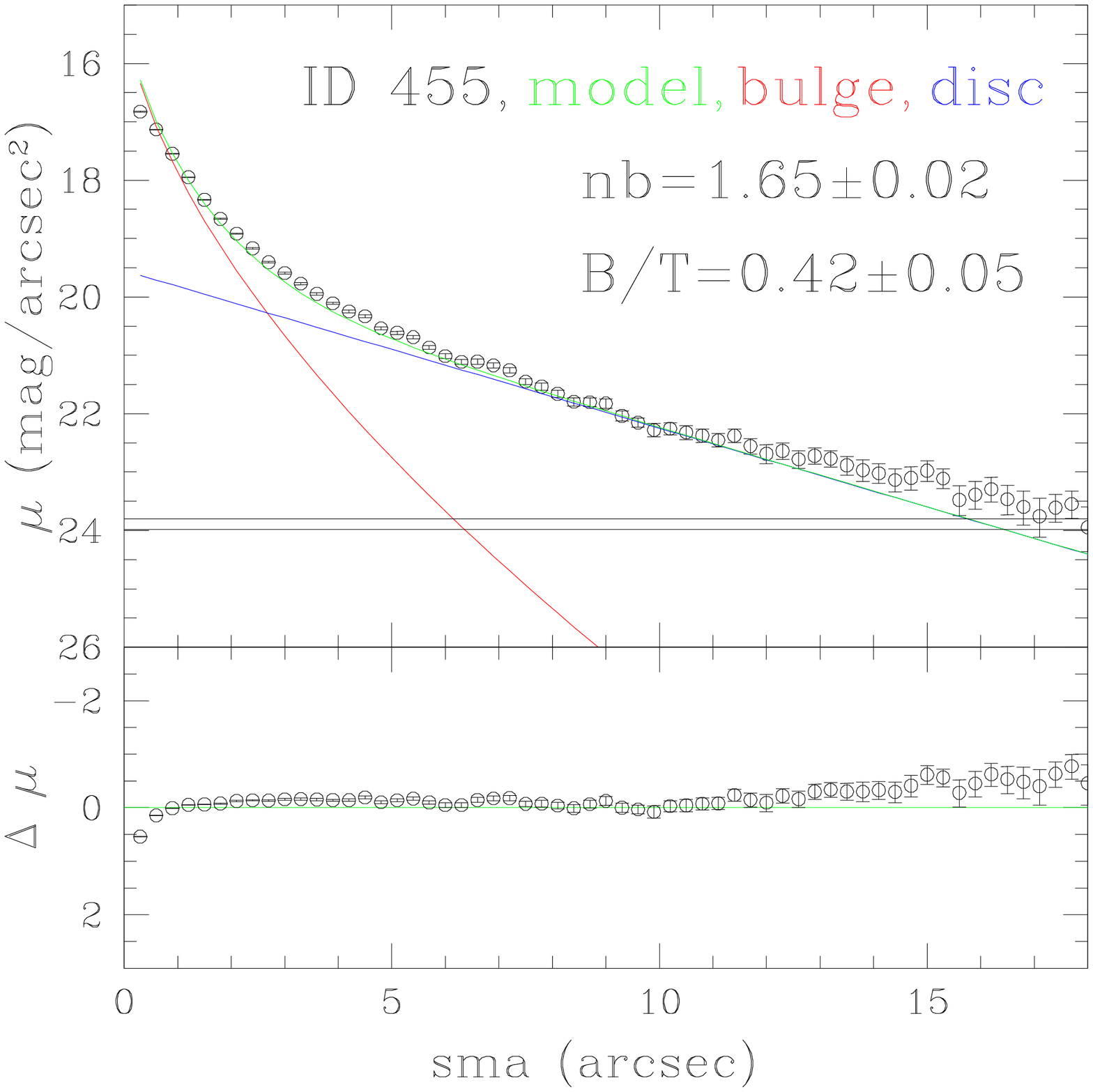}}\\
\mbox{\includegraphics[trim=75 0 75 0,clip,width=100mm]{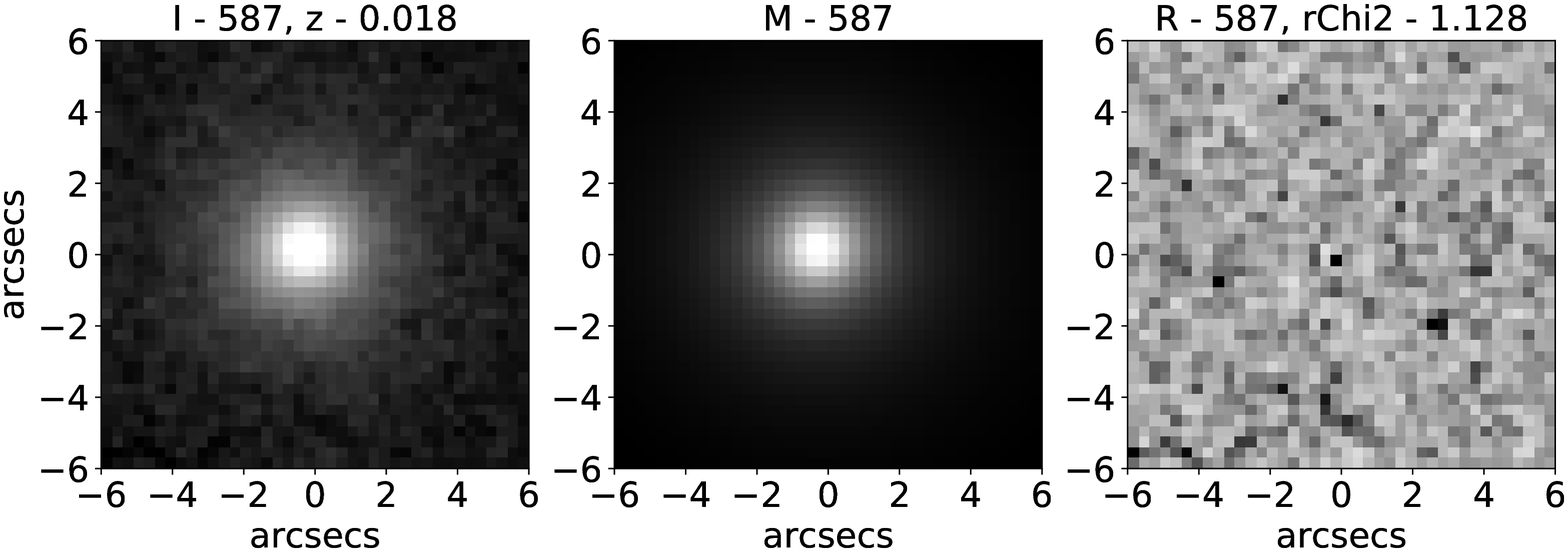}}
\mbox{\includegraphics[width=40mm]{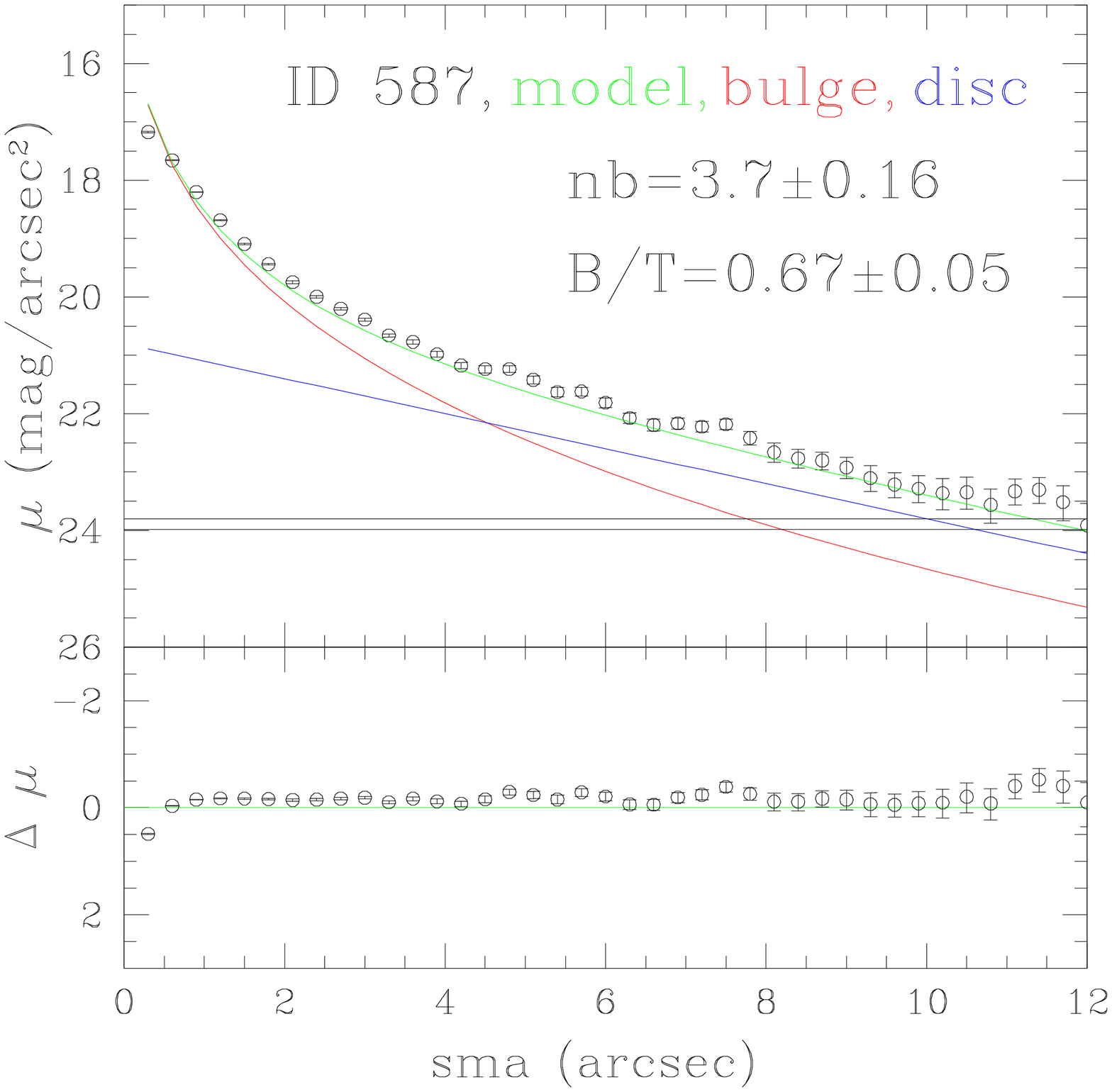}}\\
\mbox{\includegraphics[trim=75 0 75 0,clip,width=100mm]{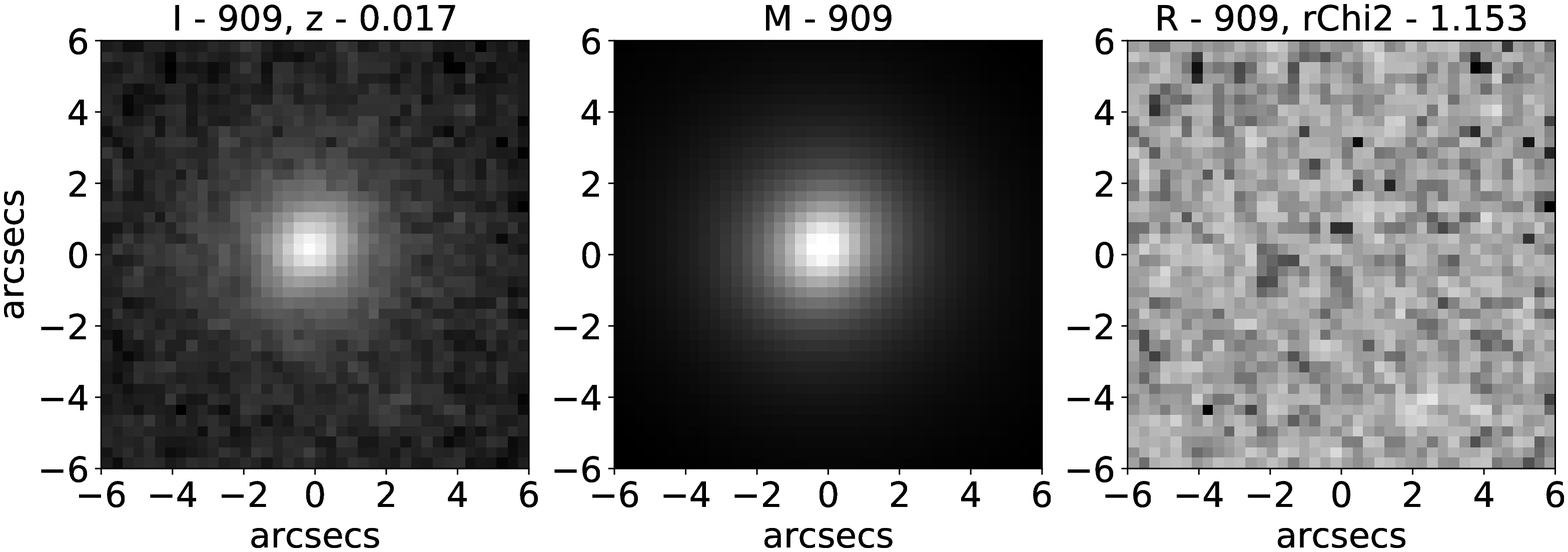}}
\mbox{\includegraphics[width=40mm]{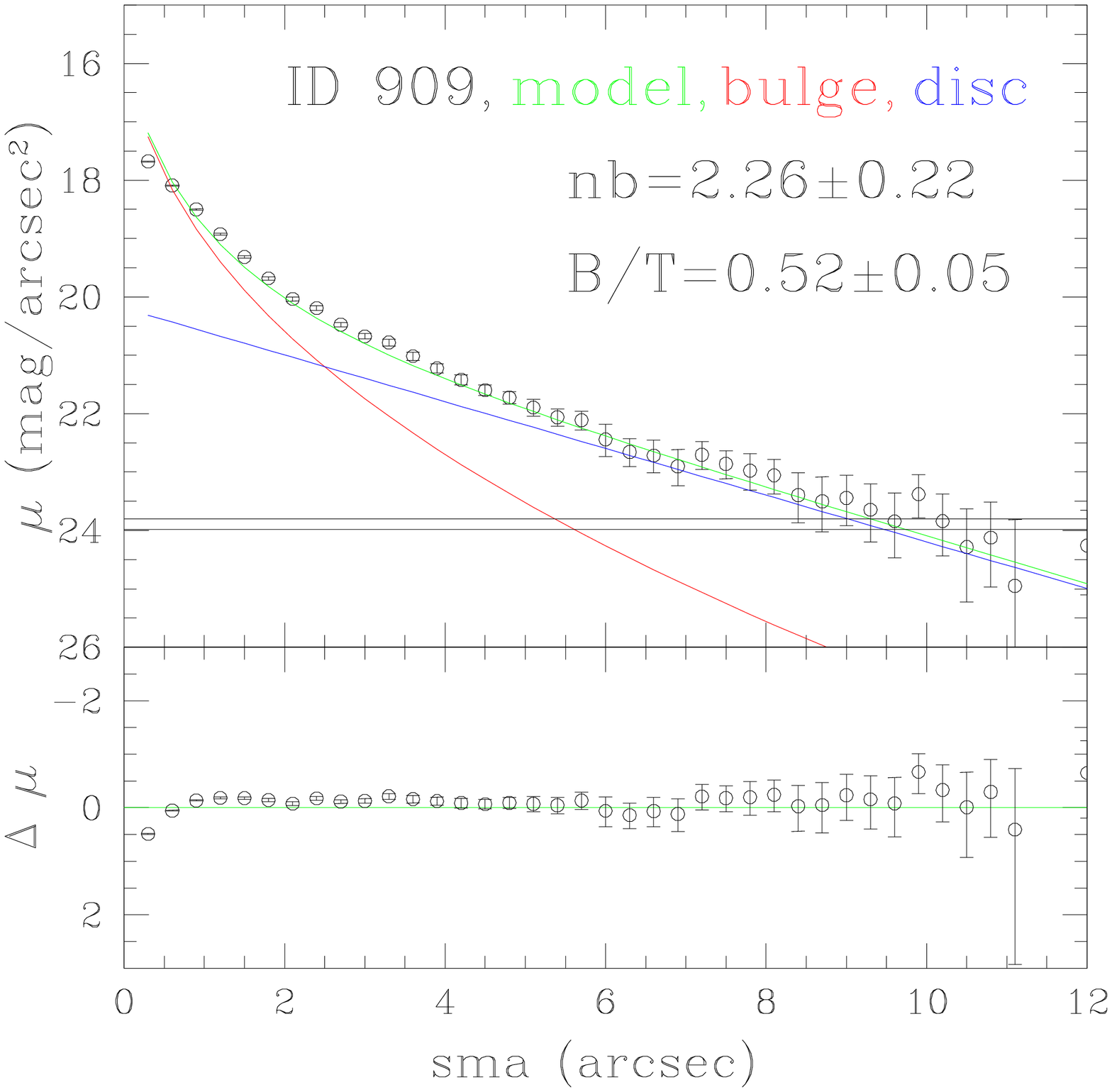}}\\
\mbox{\includegraphics[trim=75 0 75 0,clip,width=100mm]{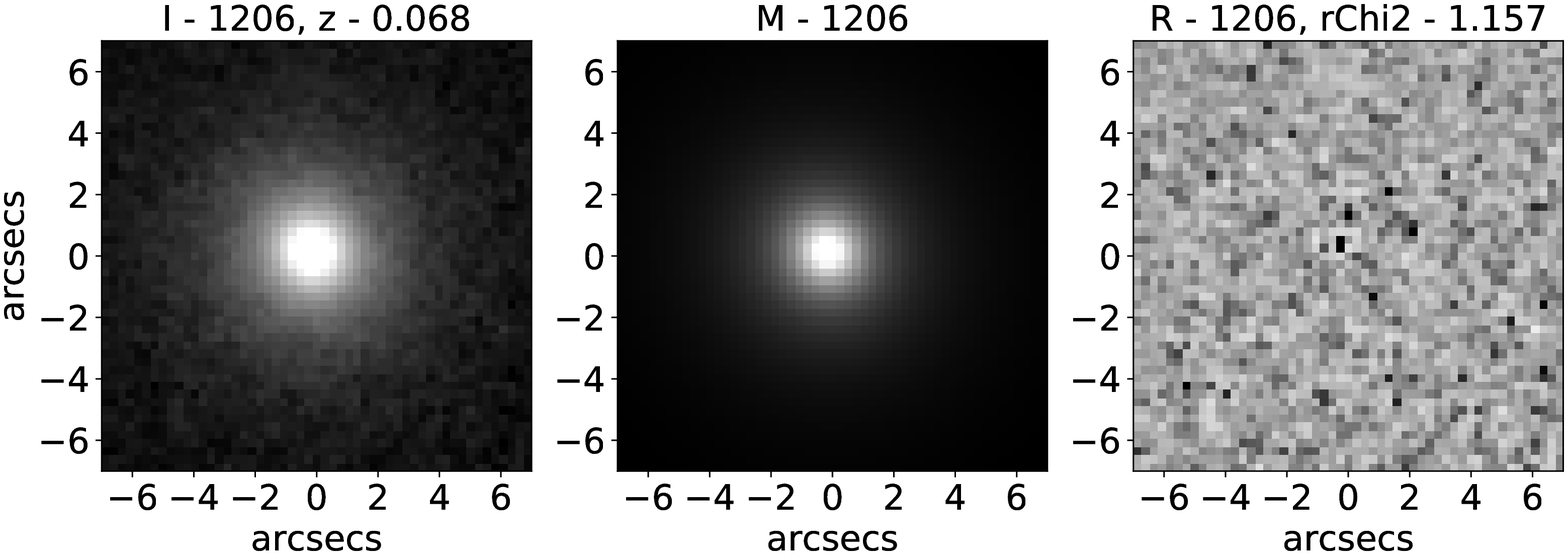}}
\mbox{\includegraphics[width=40mm]{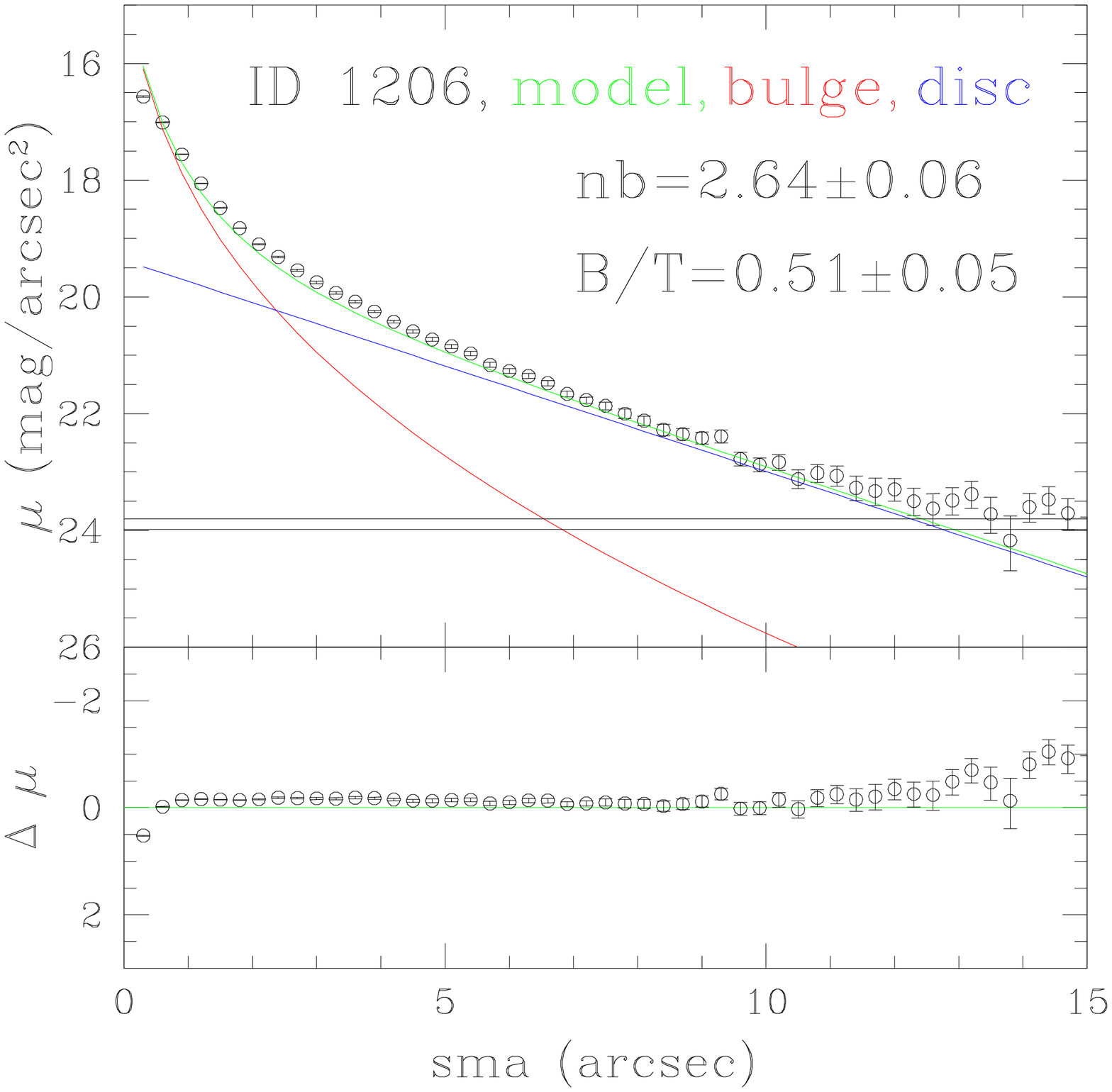}}
\caption{{\bf Fitting 2-components:} While the description is same as Fig.~\ref{fittingplota}, this figure is focused on smaller redshift ($z<0.1$) sources.} 
\label{fittingplotb}
\end{figure*}

Thus, for 2-component (bulge+disc) fitting, 942 galaxies converged. Fig.~\ref{fittingplota} and Fig.~\ref{fittingplotb} depict the fitting of a few sources in $K_s$ band along with a comparison of the best-fit parameters with the observed intensity profile for each source. The first three columns exhibit the real, model and residual images produced by {\it GALFIT} for the best fit parameters. The fourth column shows the bulge, disc and total (bulge+disc) intensity profiles that were generated using the best-fit model parameters, along with the galaxy's observed isophotal intensity profile. Computation and analysis of the residual of the observed and model profiles reveals that galaxies fitted with a free bulge S\'ersic index fit significantly better than those for which it was held fixed suggesting that it is a crucial factor in accurate determination of bulge parameters.

\begin{table*}
\begin{minipage}{170mm}
\caption{Bulge-disc decomposition values of all the 1263 sources analysed in $K_s$ band$^*$}
\begin{tabular}{@{}lllllllll@{}}
\hline
ID & $m_b$ & $r_{eb}$ & $n_b$ & $ar_b$ & PA$_b$ & $m_d$ & $r_d$ & $ar_d$\\
 & mag & arcsec & & & & mag & arcsec & \\
1 & 2 & 3 & 4 & 5 & 6 & 7 & 8 & 9\\
\hline
310 & 13.00($\pm$0.01) & 5.05($\pm$0.25) & 5.58($\pm$0.03) & 0.72($\pm$0.01) & 12.45($\pm$0.16) & 14.23($\pm$0.01) & 6.64($\pm$0.33) & 0.30($\pm$0.00)\\
312 & 15.92($\pm$0.02) & 0.50($\pm$0.02) & 1.69($\pm$0.07) & 0.55($\pm$0.01) & -79.20($\pm$0.49) & 14.90($\pm$0.01) & 1.68($\pm$0.08) & 0.72($\pm$0.01)\\
313 & 16.11($\pm$0.01) & 1.51($\pm$0.07) & 1.44($\pm$0.03) & 0.63($\pm$0.01) & -50.19($\pm$0.83) & 17.01($\pm$0.12) & 12.49($\pm$0.62) & 0.35($\pm$0.04)\\
314 & 15.13($\pm$0.04) & 1.22($\pm$0.06) & 3.64($\pm$0.12) & 0.72($\pm$0.01) & 87.99($\pm$0.63) & 16.46($\pm$0.10) & 4.21($\pm$0.21) & 0.50($\pm$0.03)\\
315 & 16.76($\pm$0.06) & 0.83($\pm$0.04) & 5.12($\pm$0.51) & 0.41($\pm$0.01) & -81.05($\pm$0.74) & 15.40($\pm$0.01) & 1.91($\pm$0.09) & 0.90($\pm$0.01)\\
317 & 14.74($\pm$0.03) & 2.03($\pm$0.10) & 5.79($\pm$0.13) & 0.82($\pm$0.01) & -86.05($\pm$0.59) & 15.73($\pm$0.04) & 5.04($\pm$0.25) & 0.56($\pm$0.01)\\
318 & 15.91($\pm$0.09) & 1.56($\pm$0.08) & 5.83($\pm$0.44) & 0.93($\pm$0.01) & -63.32($\pm$4.56) & 17.45($\pm$0.20) & 2.83($\pm$0.14) & 0.72($\pm$0.04)\\
320 & 17.21($\pm$0.14) & 0.76($\pm$0.03) & 4.72($\pm$0.92) & 0.28($\pm$0.02) & -51.80($\pm$1.31) & 15.82($\pm$0.03) & 2.16($\pm$0.11) & 0.53($\pm$0.01)\\
321 & 18.39($\pm$0.10) & 0.76($\pm$0.04) & 0.92($\pm$0.15) & 0.63($\pm$0.03) & -36.76($\pm$4.94) & 15.84($\pm$0.01) & 2.35($\pm$0.12) & 0.45($\pm$0.00)\\
322 & 15.21($\pm$0.01) & 1.85($\pm$0.09) & 5.99($\pm$0.09) & 0.78($\pm$0.01) & 18.55($\pm$0.80) & 17.23($\pm$0.05) & 1.87($\pm$0.09) & 0.98($\pm$0.05)\\
326 & 15.86($\pm$0.01) & 0.66($\pm$0.03) & 1.00($\pm$0.02) & 0.54($\pm$0.01) & 86.28($\pm$0.36) & 13.13($\pm$0.01) & 4.53($\pm$0.23) & 0.32($\pm$0.00)\\
327 & 17.69($\pm$0.38) & 0.48($\pm$0.02) & 5.59($\pm$3.66) & 0.56($\pm$0.04) & -72.17($\pm$4.01) & 16.24($\pm$0.08) & 1.70($\pm$0.08) & 0.77($\pm$0.02)\\
328 & 15.50($\pm$0.04) & 0.52($\pm$0.03) & 4.61($\pm$0.26) & 0.72($\pm$0.01) & 79.67($\pm$0.64) & 15.78($\pm$0.04) & 1.54($\pm$0.08) & 0.78($\pm$0.01)\\
330 & 16.42($\pm$0.08) & 0.47($\pm$0.02) & 3.20($\pm$0.38) & 0.48($\pm$0.01) & 30.61($\pm$0.84) & 16.52($\pm$0.08) & 1.31($\pm$0.06) & 0.74($\pm$0.01)\\
\hline
\label{fullcatalog}
\end{tabular}
$^*${\footnotesize This table presents the values obtained through 2-component (bulge-disc) 2D fitting of galaxies. The first column is the unique ID we have given to the 1263 objects. The next five columns (2-6) depict the magnitude ($m_b$), effective radius ($r_{eb}$), S\'ersic index ($n_b$), axis ratio ($ar_b$) and position angle (PA$_b$) of the bulge component. The last three columns (7-9) are the magnitude ($m_d$), scale length ($r_d$) and axis ratio ($ar_d$) of the disc component. The full table has been made available.}
\end{minipage}
\end{table*}

\begin{table*}
\begin{minipage}{170mm}
\caption{S\'ersic fitting and non-parametric values of all the 1263 sources analysed in $K_s$ band$^*$}
\begin{tabular}{@{}lllllllll@{}}
\hline
ID & $m_g$ & $r_{eg}$ & $n_g$ & $ar_g$ & PA$_g$ & $r_P$ & $C$ & $A$\\
 & mag & arcsec & & & & arcsec & & \\
1 & 2 & 3 & 4 & 5 & 6 & 7 & 8 & 9\\
\hline
310 & 12.35($\pm$0.01) & 12.49($\pm$0.62) & 6.61($\pm$0.02) & 0.64($\pm$0.01) & 10.66($\pm$0.09) & 17.86($\pm$0.89) & 4.24($\pm$0.26) & 0.23($\pm$0.01)\\
312 & 14.41($\pm$0.01) & 2.12($\pm$0.10) & 3.62($\pm$0.03) & 0.87($\pm$0.01) & 67.58($\pm$1.06) & 7.64($\pm$0.38) & 3.50($\pm$0.48) & 0.27($\pm$0.01)\\
313 & 16.06($\pm$0.01) & 1.60($\pm$0.08) & 1.56($\pm$0.03) & 0.63($\pm$0.01) & -50.09($\pm$0.84) & 5.68($\pm$0.28) & 2.96($\pm$0.52) & 0.45($\pm$0.03)\\
314 & 14.79($\pm$0.01) & 2.05($\pm$0.10) & 4.83($\pm$0.07) & 0.70($\pm$0.01) & 87.14($\pm$0.55) & 7.77($\pm$0.38) & 3.94($\pm$0.53) & 0.31($\pm$0.02)\\
315 & 14.88($\pm$0.01) & 3.94($\pm$0.19) & 3.63($\pm$0.05) & 0.85($\pm$0.01) & -77.33($\pm$1.44) & 8.85($\pm$0.44) & 3.14($\pm$0.36) & 0.35($\pm$0.03)\\
317 & 14.01($\pm$0.01) & 7.64($\pm$0.38) & 8.80($\pm$0.09) & 0.79($\pm$0.01) & -89.52($\pm$0.50) & 11.46($\pm$0.57) & 4.35($\pm$0.40) & 0.19($\pm$0.02)\\
318 & 15.49($\pm$0.02) & 2.98($\pm$0.15) & 7.04($\pm$0.15) & 0.97($\pm$0.01) & -55.32($\pm$12.30) & 7.62($\pm$0.38) & 3.97($\pm$0.52) & 0.45($\pm$0.03)\\
320 & 15.32($\pm$0.02) & 4.12($\pm$0.21) & 3.47($\pm$0.08) & 0.53($\pm$0.01) & -33.31($\pm$0.59) & 9.51($\pm$0.47) & 3.34($\pm$0.38) & 0.48($\pm$0.01)\\
321 & 15.70($\pm$0.01) & 3.61($\pm$0.18) & 1.51($\pm$0.02) & 0.48($\pm$0.01) & -1.03($\pm$0.45) & 9.21($\pm$0.46) & 2.94($\pm$0.34) & 0.49($\pm$0.06)\\
322 & 15.21($\pm$0.01) & 1.86($\pm$0.09) & 6.01($\pm$0.10) & 0.78($\pm$0.01) & 18.54($\pm$0.80) & 6.32($\pm$0.31) & 3.71($\pm$0.60) & 0.2($\pm$0.01)\\
326 & 12.88($\pm$0.01) & 8.23($\pm$0.41) & 2.33($\pm$0.01) & 0.36($\pm$0.01) & -84.20($\pm$0.05) & 16.43($\pm$0.82) & 3.24($\pm$0.21) & 0.28($\pm$0.01)\\
327 & 15.79($\pm$0.02) & 3.16($\pm$0.16) & 3.35($\pm$0.08) & 0.73($\pm$0.01) & -74.26($\pm$1.38) & 7.34($\pm$0.36) & 3.17($\pm$0.41) & 0.42($\pm$0.06)\\
328 & 14.68($\pm$0.01) & 1.68($\pm$0.08) & 6.23($\pm$0.07) & 0.88($\pm$0.01) & 72.38($\pm$1.12) & 6.56($\pm$0.32) & 3.76($\pm$0.60) & 0.34($\pm$0.01)\\
330 & 15.61($\pm$0.01) & 1.30($\pm$0.06) & 4.67($\pm$0.11) & 0.63($\pm$0.01) & 23.57($\pm$0.62) & 5.39($\pm$0.26) & 3.53($\pm$0.64) & 0.44($\pm$0.02)\\
\hline
\label{table1comp}
\end{tabular}
$^*${\footnotesize This table presents the values obtained through single component (S\'ersic) 2D fitting and non-parametric fitting of galaxies. The first column is the unique ID we have given to the 1263 objects. The next five columns (2-6) are the magnitude ($m_g$), effective radius ($r_{eg}$), S\'ersic index ($n_g$), axis ratio ($ar_g$) and position angle (PA$_g$) of the galaxy. The last three columns (7-9) are the Petrosian radius ($r_P$), concentration ($C$) and asymmetry index ($A$) of the galaxy. The full table has been made available with this work.}
\end{minipage}
\end{table*}

\begin{table*}
\begin{minipage}{170mm}
\caption{Intrinsic values of all the 1263 sources analysed in $K_s$ band with kinematic and stellar parameters$^*$}
\begin{tabular}{@{}lllllllll@{}}
\hline
ID & $M_b$ & $R_{eb}$ & $\langle\mu_{eb}\rangle$ & $M_d$ & $R_d$ & $\log(M_*)$ & $\log($SFR$)$ & $\sigma_o$\\
 & mag & kpc & mag/arcsec$^2$ & mag & kpc & [$M_{\odot}$] & [$M_{\odot}/yr$] & km/s\\
1 & 2 & 3 & 4 & 5 & 6 & 7 & 8 & 9\\
\hline
310 & -23.32($\pm$0.69) & 4.15($\pm$0.20) & 18.33($\pm$0.55) & -22.09($\pm$0.66) & 5.45($\pm$0.27) & 11.13($\pm$0.01) & -1.65($\pm$0.43) & 260.3($\pm$4.2)\\
312 & -21.26($\pm$0.63) & 0.59($\pm$0.02) & 16.17($\pm$0.48) & -22.28($\pm$0.66) & 1.98($\pm$0.09) & 10.65($\pm$0.02) & 0.29($\pm$0.08) & 133.4($\pm$6.0)\\
313 & -23.99($\pm$0.71) & 5.25($\pm$0.26) & 18.16($\pm$0.54) & -23.09($\pm$0.69) & 43.30($\pm$2.16) & 11.13($\pm$0.04) & 1.09($\pm$0.08) & 168.8($\pm$16.7)\\
314 & -23.98($\pm$0.71) & 3.02($\pm$0.15) & 16.98($\pm$0.50) & -22.65($\pm$0.67) & 10.44($\pm$0.52) & 11.38($\pm$0.06) & -0.18($\pm$0.47) & 260.8($\pm$5.1)\\
315 & -23.21($\pm$0.69) & 2.77($\pm$0.13) & 17.57($\pm$0.52) & -24.57($\pm$0.73) & 6.36($\pm$0.31) & 11.42($\pm$0.02) & 0.81($\pm$0.08) & 173.5($\pm$11.6)\\
317 & -24.48($\pm$0.73) & 5.23($\pm$0.26) & 17.68($\pm$0.53) & -23.49($\pm$0.70) & 12.99($\pm$0.64) & 11.58($\pm$0.07) & -0.69($\pm$0.66) & 301.7($\pm$9.6)\\
318 & -23.58($\pm$0.70) & 4.43($\pm$0.22) & 18.21($\pm$0.54) & -22.04($\pm$0.66) & 8.01($\pm$0.40) & 11.13($\pm$0.03) & -1.15($\pm$0.81) & 199.3($\pm$12.6)\\
320 & -22.50($\pm$0.67) & 2.33($\pm$0.11) & 17.90($\pm$0.53) & -23.89($\pm$0.71) & 6.61($\pm$0.33) & 11.46($\pm$0.02) & 0.47($\pm$0.13) & 187.5($\pm$13.7)\\
321 & -18.79($\pm$0.56) & 0.90($\pm$0.04) & 19.55($\pm$0.58) & -21.34($\pm$0.64) & 2.77($\pm$0.13) & 10.15($\pm$0.04) & 0.13($\pm$0.06) & 28.1($\pm$27.3)\\
322 & -23.99($\pm$0.71) & 4.74($\pm$0.23) & 17.95($\pm$0.53) & -21.97($\pm$0.65) & 4.79($\pm$0.23) & 11.11($\pm$0.01) & -0.46($\pm$0.42) & 262.8($\pm$12.0)\\
326 & -19.72($\pm$0.59) & 0.39($\pm$0.02) & 16.82($\pm$0.50) & -22.45($\pm$0.67) & 2.71($\pm$0.13) & 10.73($\pm$0.02) & 0.37($\pm$0.03) & 108.3($\pm$5.5)\\
327 & -21.30($\pm$0.63) & 1.15($\pm$0.05) & 17.57($\pm$0.52) & -22.75($\pm$0.68) & 4.05($\pm$0.20) & 10.99($\pm$0.01) & -0.08($\pm$0.09) & 92.5($\pm$13.5)\\
328 & -21.92($\pm$0.65) & 0.68($\pm$0.03) & 15.81($\pm$0.47) & -21.64($\pm$0.64) & 2.00($\pm$0.10) & 10.64($\pm$0.02) & -1.40($\pm$0.56) & 134.7($\pm$5.0)\\
330 & -22.70($\pm$0.68) & 1.17($\pm$0.05) & 16.20($\pm$0.48) & -22.60($\pm$0.67) & 3.26($\pm$0.16) & 11.05($\pm$0.01) & -0.25($\pm$0.06) & 238.6($\pm$12.2)\\
\hline
\label{tableintrinsic}
\end{tabular}
$^*${\footnotesize This table presents the intrinsic (or absolute) values of the bulge-disc component obtained according to their redshift, adopted cosmology and K-correction. The first column is the unique ID we have given to the 1263 objects. The next three columns (2-4) are the magnitude ($M_b$), effective radius ($R_{eb}$) and average surface brightness ($\mu_{eb}$) inside that radius for the bulge component. The following two columns (5-6) are the magnitude ($M_d$) and scale length ($R_d$) of the disc component. The last three columns (7-9) are the total stellar mass ($M_*$), star formation rate (SFR) and central velocity dispersion ($\sigma_o$) of the galaxy. The full table has been made available with this work.}
\end{minipage}
\end{table*}

\subsection{Non-parametric measures} 

In addition to parametric measures, we compute non-parametric measures for all galaxies, mainly their Petrosian radius, Concentration and Asymmetry. Non-parametric measures, by definition, are not constrained by any functional form and are thus considered least biased measures of galaxies' structure \citep{Conselice2014}. The algorithm for measuring Petrosian radius of a galaxy is based on the computation of the ratio of intensity at successively increasing radii to the intensity inside those radii. When the ratio $\eta(r)$ falls to an empirically determined fraction (0.2 in this work),

\begin{equation}
\eta(r_p)=\frac{I(r_p)}{\langle I(<r_p)\rangle}=0.2,
\end{equation}

\noindent at some radius $r_p$, then Petrosian radius ($r_P$) is given as $1.5\times r_p$ where 1.5 is again an empirically determined multiple. By ``empirically determined" we mean that these values (0.2 and 1.5) were found to provide most accurate estimates for a representative sample of galaxies \citep{Bershadyetal2000,Conselice2003,Lotzetal2004}. The algorithm thus works on extrapolation of intensity profile from the centre to the outskirts to obtain the total extent of the galaxy. For Concentration, first the flux inside the total (Petrosian) radius of the galaxy is computed. Then the algorithm finds those radii which contain 20\% and 80\% of the total flux to obtain,

\begin{equation}
C = 5 \log_{10}(\frac{r_{80}}{r_{20}}),
\end{equation}

\noindent where $C$ is the concentration index of the galaxy and 5 is an empirically determined multiple \citep{Bershadyetal2000,Grahametal2005}. For Asymmetry, galaxy image is rotated by $180^o$ about its ``centre of symmetry" (a central point, found iteratively, where asymmetry is minimum) and subtracted from the main image. Flux thus obtained from the subtracted image is normalized to obtain the asymmetry index ($A$) of the galaxy. Our code for computing the three measures ($r_P$, $C$ and $A$) ran smoothly for 1253 out of the total sample of $1263$ galaxies. For the rest $10$ galaxies, iterative algorithm to find the centre of the galaxy did not converge.

\subsection{Stellar parameters}

Stellar parameters, i.e., stellar masses ($M_*$) and star formation rates (SFR) are taken from GALEX-SDSS-WISE Legacy Catalogue 2 \citep{Salimetal2018}. They follow a Bayesian approach to SED fitting on the combination of GALEX and SDSS data of all galaxies (0.7 million) with $z<0.3$. Their most important modification over previous works \citep{Salimetal2005,Salimetal2007,Salimetal2016} is the usage of IR luminosity to set constraints on dust emission which allows dust attenuation curve parameters to be fitted freely while creating models. Computation of ``true" IR luminosity is based on far-IR flux from Herschel \citep{Valianteetal2016} in addition to mid-IR flux from WISE \citep{Langetal2016}. An accurate estimation of the dust attenuation affect is a critical factor in the determination of stellar activity of galaxies. They demonstrate by comparing with other works that if far-IR is not included in the computation, SFRs get systematically over-estimated especially for quiescent galaxies.

Their project involves usage of Herschel ATLAS to calibrate computation of true IR luminosity for the full sample of $0.7$ million galaxies which have only mid-IR fluxes from WISE. Our sample being on the Herschel field, has full coverage of the calibration sample itself. Out of our total sample of $1263$ galaxies, there are ``good quality" $M_*$ and SFR estimates for 1205 galaxies. 

\subsection{Central velocity dispersion}

Stellar velocity dispersions for galaxies in our sample have been obtained from SDSS DR15 spectroscopic catalogue \citep{Boltonetal2012,Aguadoetal2019}. The SDSS spectra, since DR9, have been obtained using BOSS spectrograph \citep{Dawsonetal2013} which has a fibre diameter of 2" and wavelength coverage from $365$ to $1040$ nm. Each spectra is reduced by the spectroscopic pipeline ({\it idlspec2d}) which is refined with each data release. Stellar velocity dispersions are derived following a Principal Component Analysis (PCA) method where 24 eigenspectra from ELODIE stellar library \citep{PrugnielandSoubiran2001} are convolved and binned to match the instrumental resolution and constant-velocity pixel scale of reduced SDSS spectra. These template sets are redshifted, broadened by successively large velocity widths and modelled through least square fitting of linear combination of each trial broadening. Best-fit velocity dispersion value is thus determined through chi-square minimization and error on the value is determined from curvature of the chi-squared curve around global minimum. Based on the average S/N and instrumental resolution of SDSS spectra, velocity dispersion measurements below 70 km/s should be treated with caution \citep{Thomasetal2013}. In our work, we will use these measurements only for bulge classification, where, all galaxies with velocity dispersion below 90 km/s will be clubbed together. Thus even large ($\sim$30\%) errors on below resolution values will not affect our bulge classification or any of the presented results and inferences.  Out of our total sample of 1263 galaxies, reliable (based on data quality flags) velocity dispersion measurements are present for 1249 galaxies, i.e., almost the full sample. Following this, we apply aperture correction to ensure that all velocity dispersions are computed within same effective aperture which is chosen to be 1/8th of the half light radius of the bulge. We, thus, obtain central velocity dispersion ($\sigma_o$), using,

\begin{equation}
\sigma_o = (\frac{r_{eb}/8}{r_{ap}})^{-0.04} \sigma_{ap}
\end{equation}

\noindent where, $r_{eb}$ is the half light radius of the bulge in arcseconds, $r_{ap}$ is the radius of the aperture (1"), $\sigma_{ap}$ is the measured stellar velocity dispersion and $0.04$ is an empirically derived value \citep{Jorgensenetal1995,Teimooriniaetal2016}. Note that the corrections ($<$0.5\%) are insubstantial compared to the errors ($\sim$3-5\%) on velocity dispersion values.

\subsection{Consolidation and comparison}

In consolidation, we have obtained both parametric and non-parametric morphological measurements for our sample of 1263 galaxies in $K_s$ band. Based on the cosmological parameters adopted, redshift of the galaxies and corresponding K-corrections involved, we convert all measured apparent quantities to intrinsic ones in rest-frame $K_s$ band \citep[equations illustrated in][]{GrahamandDriver2005,Sachdeva2013,Sachdevaetal2015}. Thus, for the full galaxy and for the bulge and the disc component separately, we obtain their absolute magnitudes ($M_g$, $M_b$, $M_d$), intrinsic characteristic sizes ($R_{eg}$, $R_{eb}$, $R_d$) in Kpc, S\'ersic indices ($n_g$, $n_b$) along with other defining parameters, i.e., position angle, ellipticity, etc. Note that subscript `g' is for global, `b' is for bulge and `d' is for disc. In addition to that, we have acquired the latest and most accurate values of their stellar ($M_*$, SFR, sSFR) and kinematic ($\sigma_o$) parameters. In terms of parametric morphology, we have performed both single component (S\'ersic function) and 2-component (S\'ersic function + exponential function) fitting. For single component fitting, nearly all galaxies (barring 30) converged to physically meaningful parameters, where, 946 converged with free $n_g$ and 287 converged when it was held fixed. For 2-component fitting, 942 galaxies (75\% of the total) converged to physically meaningful parameters, where, 605 converged with free $n_b$ and 337 converged when it was held fixed.

Tables (or catalogues) consolidating parametric, non-parametric, kinematic and stellar measures, of all the 1263 galaxies in our sample, have been made available with this paper. This includes apparent as well as intrinsic measures in rest-frame $K_s$ band, along with errors. In Table~\ref{fullcatalog}, \ref{table1comp} and \ref{tableintrinsic}, we present a small glimpse of the full appended version. Each version also includes a detailed description of the columns in its beginning.

\begin{figure*}
\centering
\mbox{\includegraphics[width=45mm]{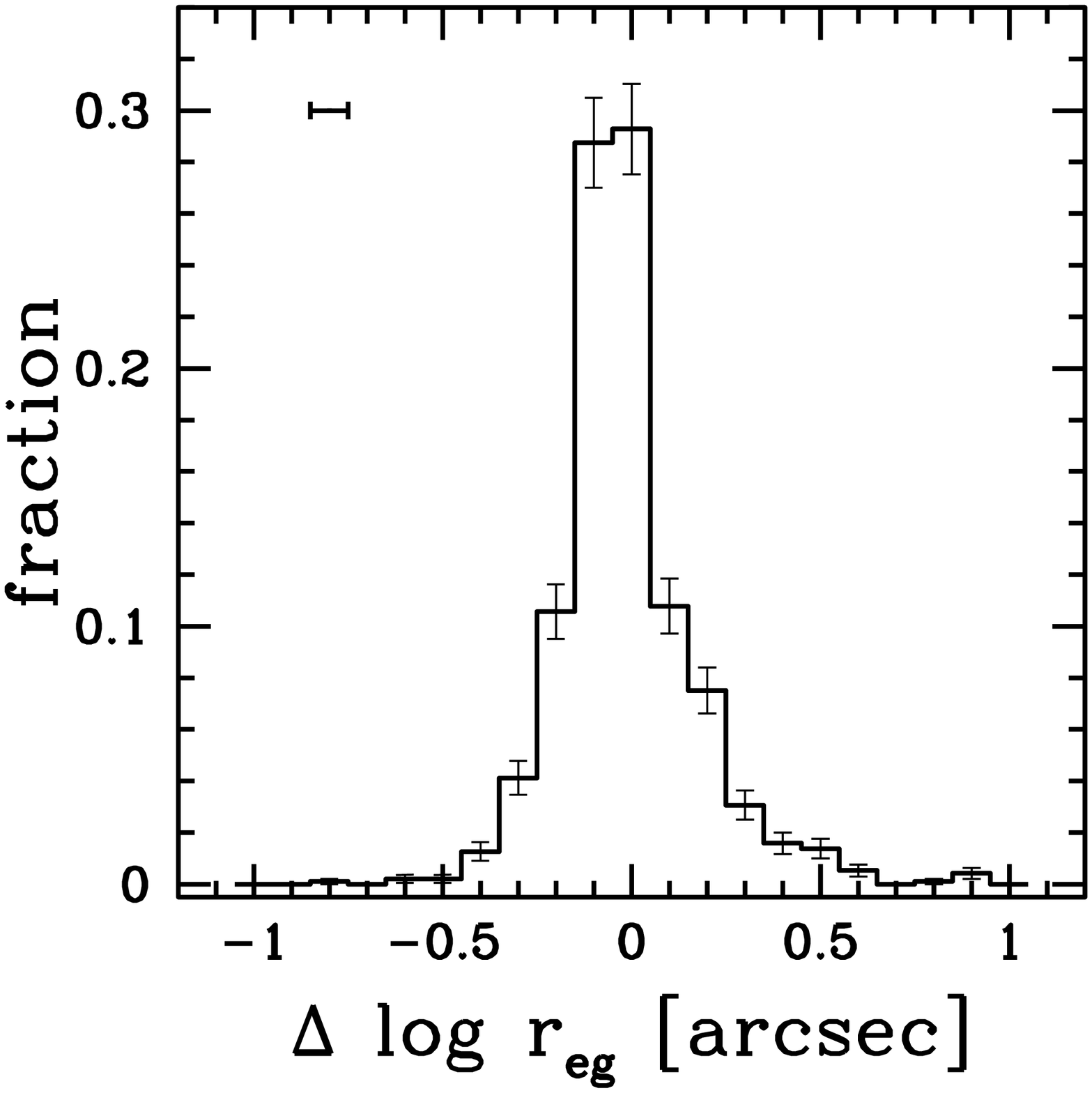}}
\mbox{\includegraphics[trim=30 0 0 0,clip,width=42.8mm]{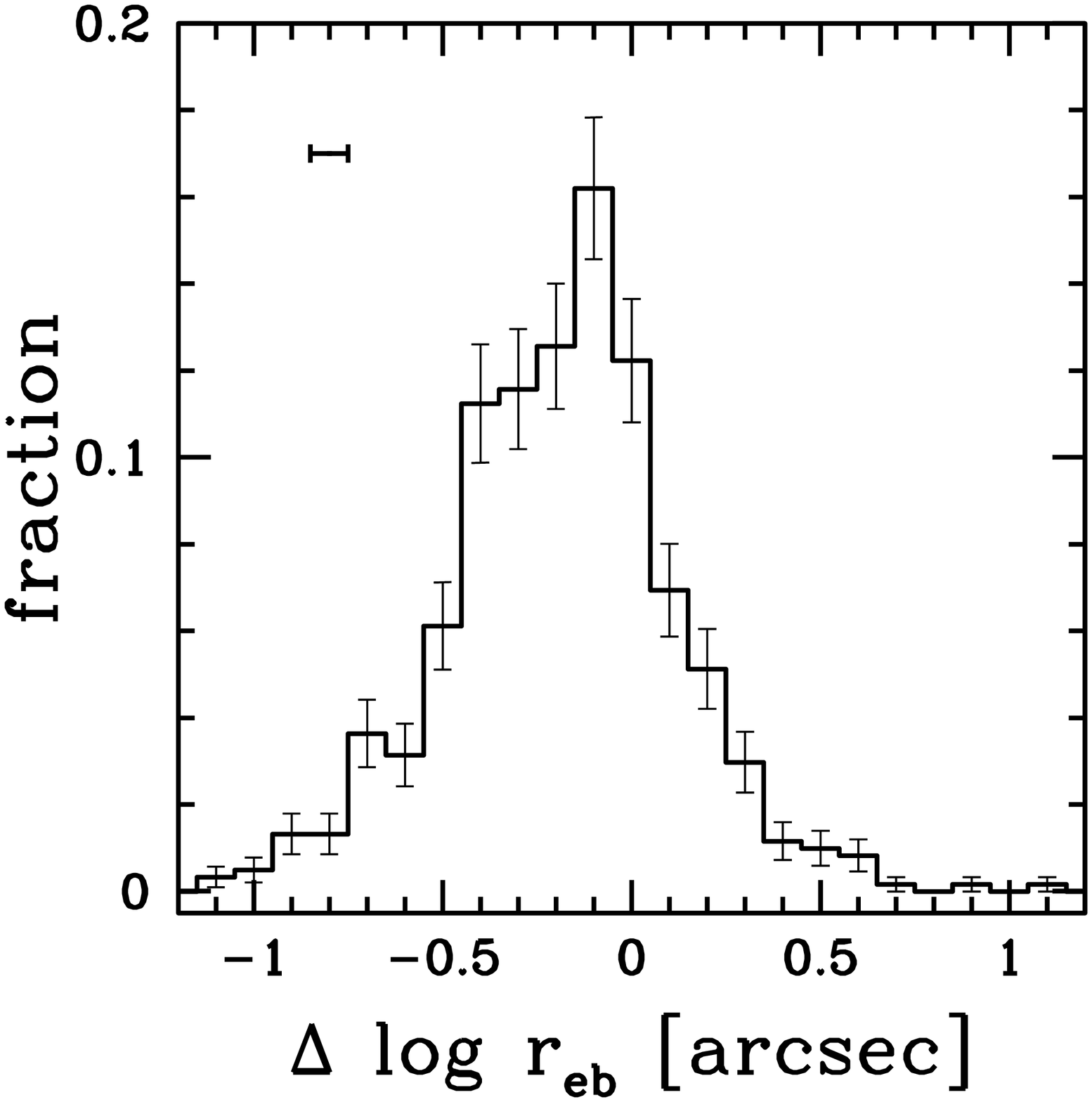}}
\mbox{\includegraphics[trim=30 0 0 0,clip,width=42.8mm]{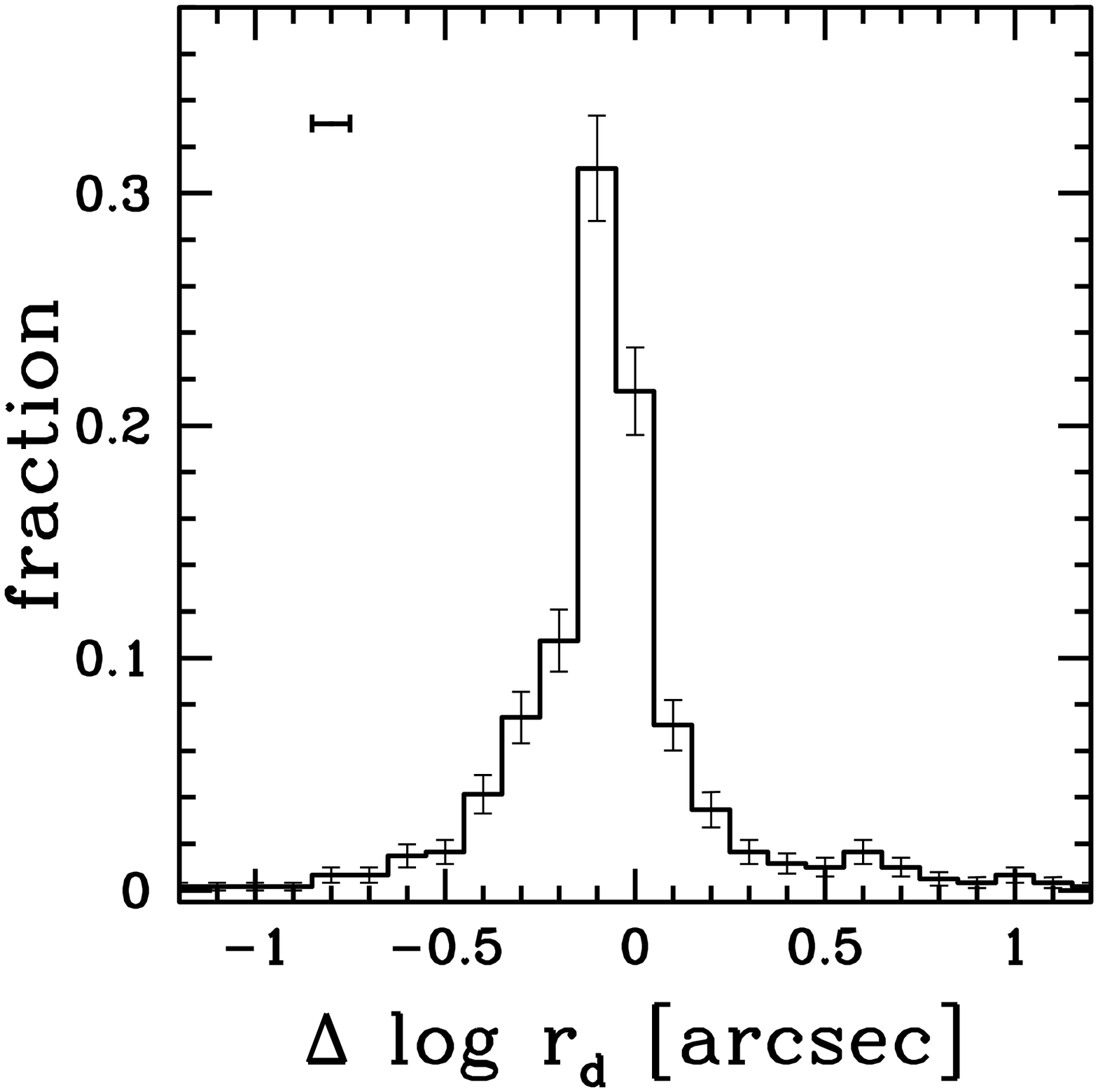}}
\mbox{\includegraphics[trim=30 0 0 0,clip,width=42.8mm]{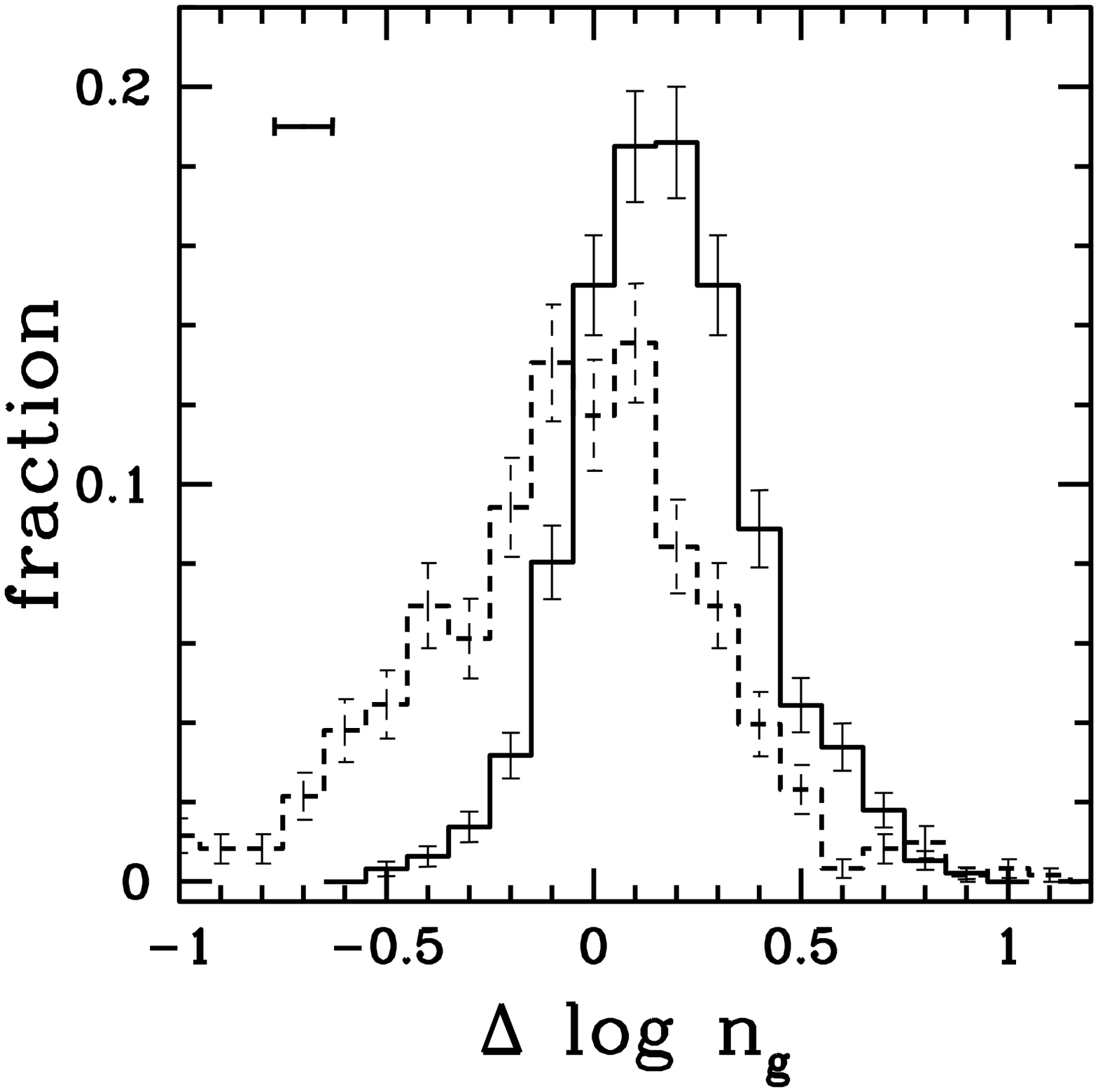}}\\
\mbox{\includegraphics[width=45mm]{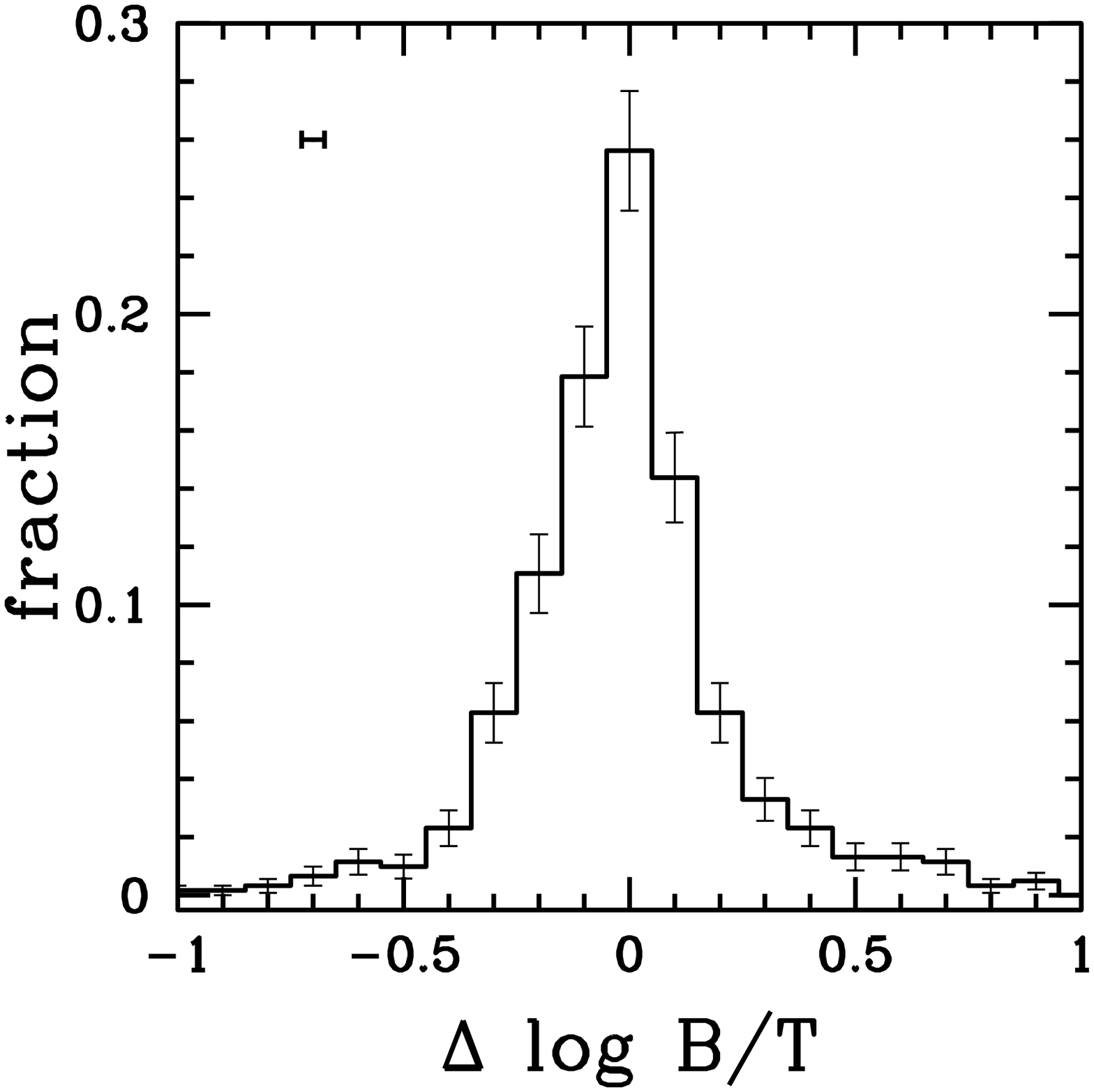}}
\mbox{\includegraphics[trim=30 0 0 0,clip,width=42.8mm]{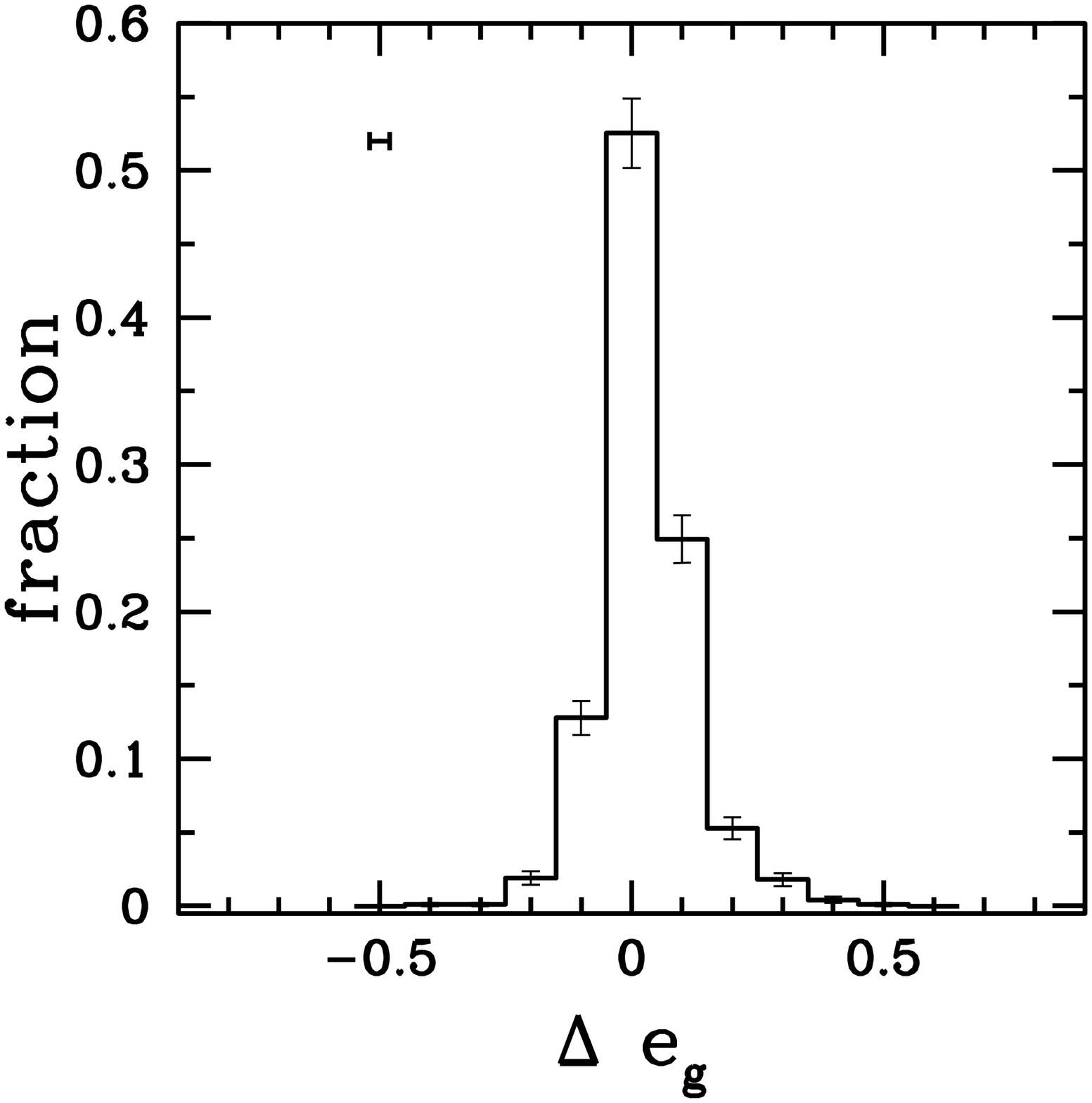}}
\mbox{\includegraphics[trim=30 0 0 0,clip,width=42.8mm]{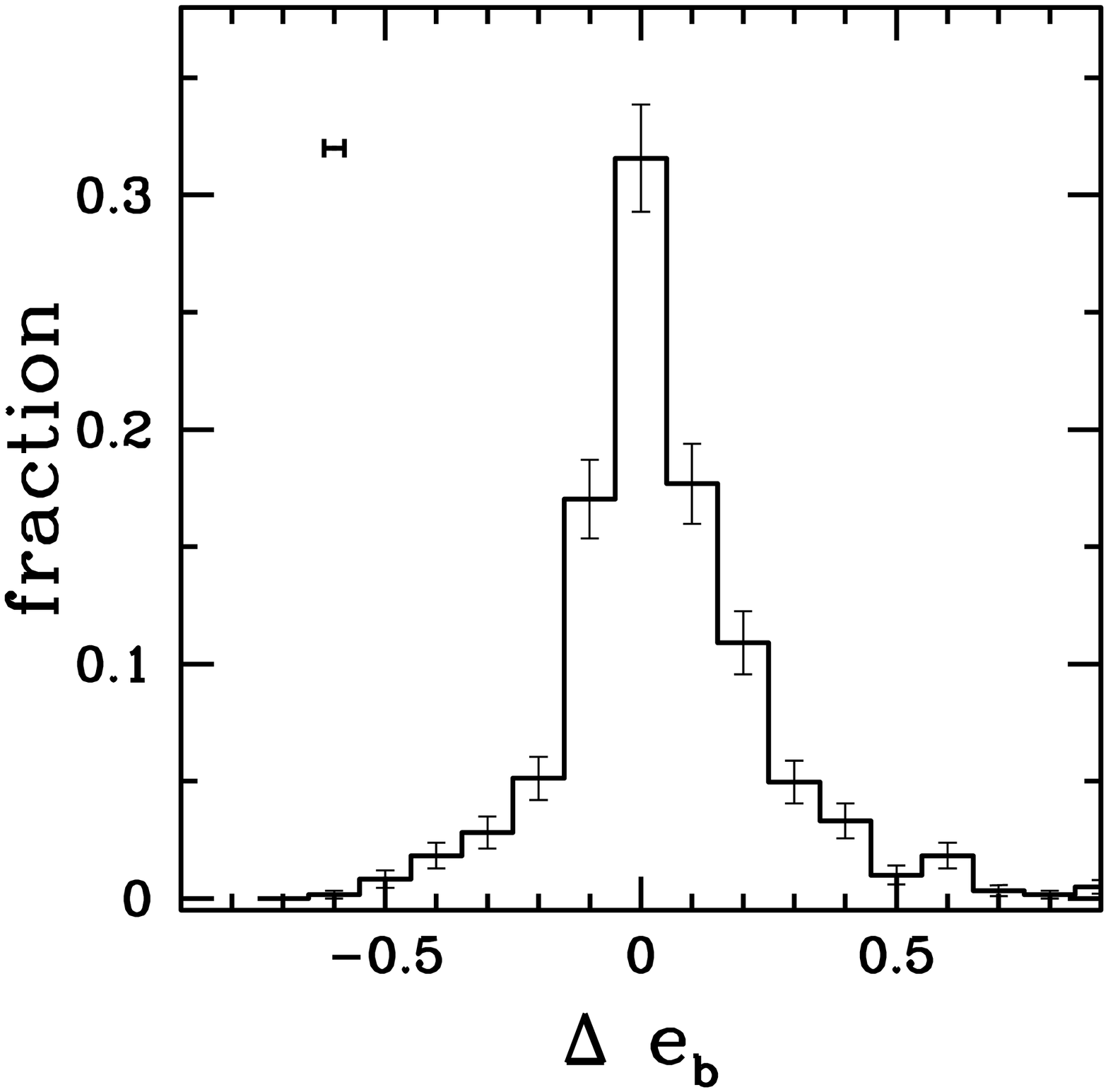}}
\mbox{\includegraphics[trim=30 0 0 0,clip,width=42.8mm]{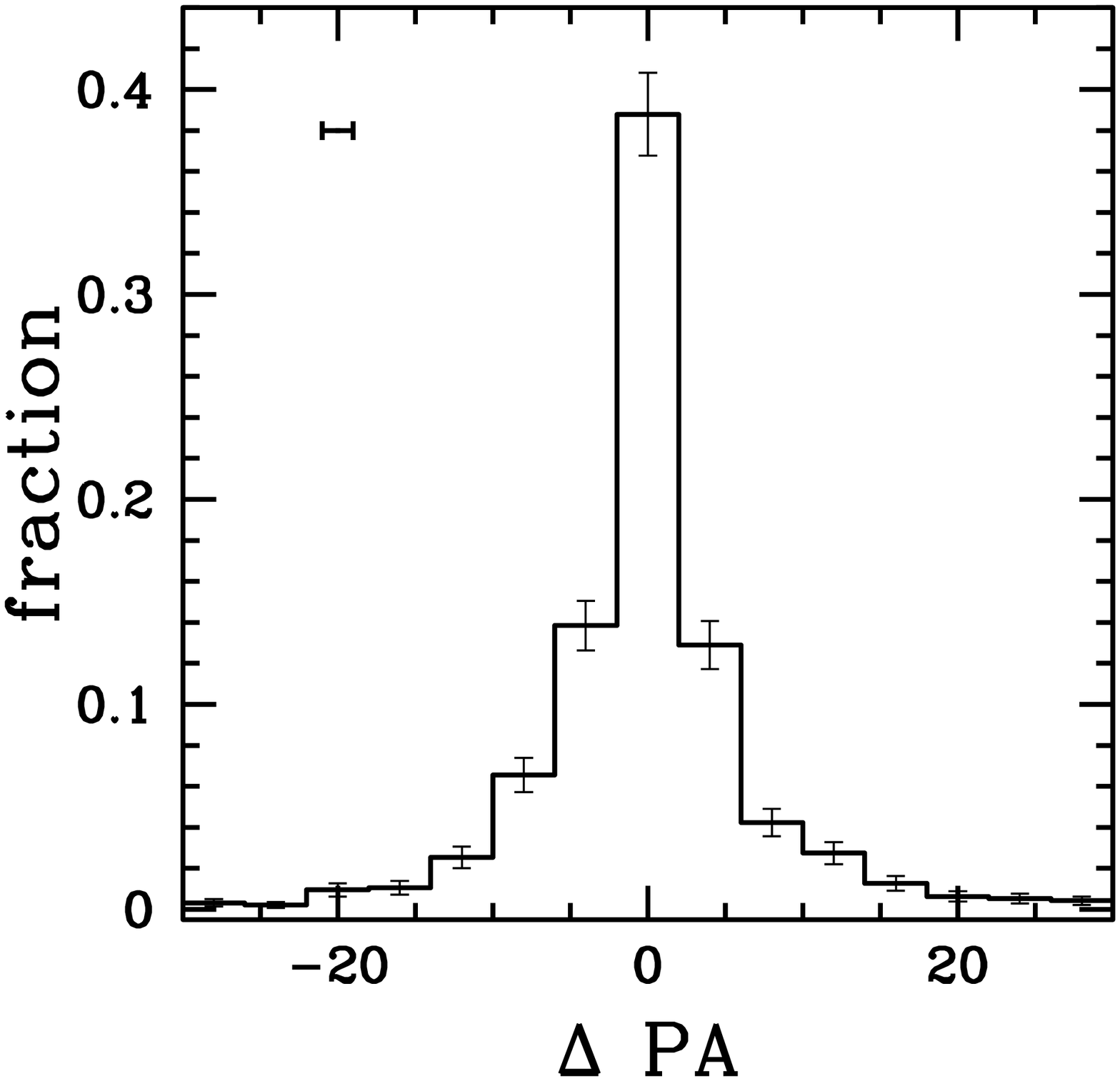}}
\caption{{\bf Band comparison:} The histograms show the distribution of the difference of parameter values computed by us in $K_s$ band and those computed by \citet{Bottrelletal2019} in the $r$ - band. The first three plots in the first row depict the distribution of the difference in the global effective radius ($r_{eg}$), bulge effective radius ($r_{eb}$) and disc scale length ($r_d$). The fourth plot shows the difference in the S\'ersic index values of the two bands, both for the global (solid line) and the bulge (dashed line). In the second row, difference in the bulge-to-total ratio ($B/T$), global ellipticity ($e_g$), bulge ellipticity ($e_b$) and global position angle (PA$_g$) for the two bands is shown. Average error-bars are marked in each plot.} 
\label{bulgefix}
\end{figure*}

In Fig.~\ref{bulgefix}, we compare the distribution of the parameters obtained in $K_s$ band with those obtained in the optical ($r$ band) by \citet{Bottrelletal2019}. Note that we compare only that sample which converged in our fitting process with S\'ersic index (whether global $n_g$ or bulge $n_b$) kept free so as to match the assumptions made by \citet{Bottrelletal2019} during their fitting of the same sample. We find that there is quite a reasonable match between the two bands confirming the accuracy of the decomposition carried out using two different methods, i.e., {\it GIM2D} and {\it GALFIT}. More than $80\%$ of all radii ($r_g$, $r_{eb}$, $r_d$) and most importantly the bulge-to-total ratio ($B/T$) are within 0.2 dex for the two bands. Even for a sensitive parameter like $n_b$, for more than $70\%$ the difference is within 0.2 dex. Ellipticity ($e_g$, $e_b$) and position angle (PA) of all ($>95\%$) parameters are within 0.1 dex for the two bands.

Variations observed between the two bands are negligible and can be attributed to the real differences between the two bands (e.g., galaxies and their components extend more in bluer bands than in redder ones, etc.), along with measurement uncertainties. Since our primary aim is to perform indubitable classification of bulges before proceeding to the analysis of their stellar properties, we will focus on this well matched sample of 605 galaxies for which 2-component morphological decomposition converged with free $n_b$. This is also necessary because $n_b$ is one of the key parameters whose performance as a differentiator of bulge type will be examined in this work. Henceforth, all the presented analysis is based on this sample.

\section{Results}
\label{sec:results}

\subsection{Performance of bulge morphology indicators}

Over the past few years, Kormendy relation (KR) - a projection of the fundamental plane exhibited by elliptical galaxies \citep{Kormendy1977} - has established itself as one of the most efficient classifiers of bulge morphology \citep{Gadotti2009,Sachdevaetal2017,Sachdevaetal2019,Gaoetal2020}. This is driven by the fact that classical bulges - being more dominated by dispersion than pseudo bulges - are found to lie within $\pm2\sigma$ boundaries of the KR followed by elliptical galaxies. Pseudo bulges, in contrast, are found to be low surface brightness outliers to KR. To perform an unambiguous classification of bulges in our sample, we will extract a single morphology indicator ($\Delta$$\langle\mu_{eb}\rangle$) from KR and employ that to investigate the performance of other morphology indicators. Following that, the indicator which best compliments $\Delta$$\langle\mu_{eb}\rangle$, will be used in conjunction with it to select an indubitable class of classical and pseudo bulges. 

\begin{figure}
\centering
\mbox{\includegraphics[width=55mm]{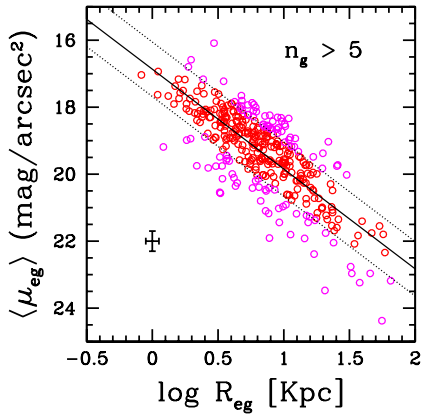}}
\caption{{\bf Kormendy relation:} The distribution of all galaxies with global S\'ersic index ($n_g$) more than 5.0 is shown on the Kormendy plane. These galaxies have also been confirmed visually to be lacking any disc component, i.e., have a high probability of being pure ellipticals. While the solid black line marks the relation followed by these sources, the two dotted lines mark the $\pm1\sigma$ boundary of the relation, i.e., confines $\sim66\%$ of the sources. We will use the $+1\sigma$ (lower dotted line) boundary as a separator of ``elliptical like" (EL) and ``non-elliptical like" (NEL) bulges. Average error-bar is marked.}
\label{ellipKP}
\end{figure}

To obtain KR for elliptical galaxies in rest-frame $K_s$ band, we select elliptical galaxies from the full sample according to their global S\'ersic index ($n_g$). Analysis of statistically large samples of galaxies, over the past two decades, has revealed that $n_g$ is an effective separator of late-type ($n_g<2.0-2.5$) and early-type ($n_g>2.5-3.0$) galaxies \citep{Shenetal2003,Ravindranathetal2004,Bardenetal2005}. In addition to that, studies focusing on early-type galaxies have stressed that higher is the $n_g$, lesser is the contamination from late-type counterparts \citep{Blakesleeetal2006,Retturaetal2006,vanderWel2008,Huangetal2013}. Based on these findings, we select those galaxies from the full sample that have $n_g>5$. There is a possibility that a high S\'ersic index cut may result in the selection of the most luminous of the ellipticals. However, we find that the relation observed by galaxies in the smaller range ($3.0<n_g<6.0$) matches with that obtained for our selected sample with $n_g>5$, ruling out this possibility. Other than that, we visually examine the selected sample to remove sources with any faint disc structure. Fig.~\ref{ellipKP} shows the distribution of our sample of ellipticals and the relation derived,

\begin{equation}
\langle\mu_{eg}\rangle = (2.98 \pm 0.13) \log(R_{eg}) + (16.86 \pm 0.11) 
\end{equation}

\noindent where $R_{eg}$ is the effective radius of the galaxy (in Kpc) and $\langle\mu_{eg}\rangle$ is the intrinsic average surface brightness inside it (in mag/arcsec$^2$). The fitting has been performed using the {\it fit} function of {\it Gnuplot} which implements non-linear least-square Marquardt-Levenberg algorithm. The $1\sigma$ scatter on $\langle\mu_{eg}\rangle$ is $\pm 0.81$ mag/arcsec$^2$. The scatter is larger than that observed by a recent work studying the placement of local ellipticals on the Kormendy plane in the optical \citep{Gaoetal2020}. This could be due to several factors, including a more stringent selection of ellipticals. To overcome this possibility, we adopt the $+1\sigma$ boundary, instead of $+2\sigma$ \citep{Neumannetal2017} or $+3\sigma$ \citep{Gadotti2009,Gaoetal2020} boundary, to ensure that only those bulges are eventually chosen to be classical which are most `elliptical like'. Fig.~\ref{ellipKP} also marks the $\pm1\sigma$ boundaries of the relation. Based on the $+1\sigma$ boundary, we define a quantity $\Delta$$\langle\mu_{eb}\rangle$ which marks the relative distance of the bulge from this boundary i.e.,

\begin{equation}
\Delta\langle\mu_{eb}\rangle = \langle\mu_{eb}\rangle - 2.98 \log(R_{eb}) - 17.68
\end{equation}

where $R_{eb}$ is the effective radius of the bulge (in Kpc) and $\langle\mu_{eb}\rangle$ is the intrinsic average surface brightness inside it (in mag/arcsec$^2$). The quantity $\Delta$$\langle\mu_{eb}\rangle$ works as a bulge morphology classifier which embodies KR, such that for bulges which lie on the $+1\sigma$ boundary of KR, this value is zero. More positive is this quantity more `non-elliptical like' (NEL) is the bulge and more negative is the quantity more `elliptical like' (EL) is the bulge. 

\begin{figure*}
\centering
\mbox{\includegraphics[width=48.5mm]{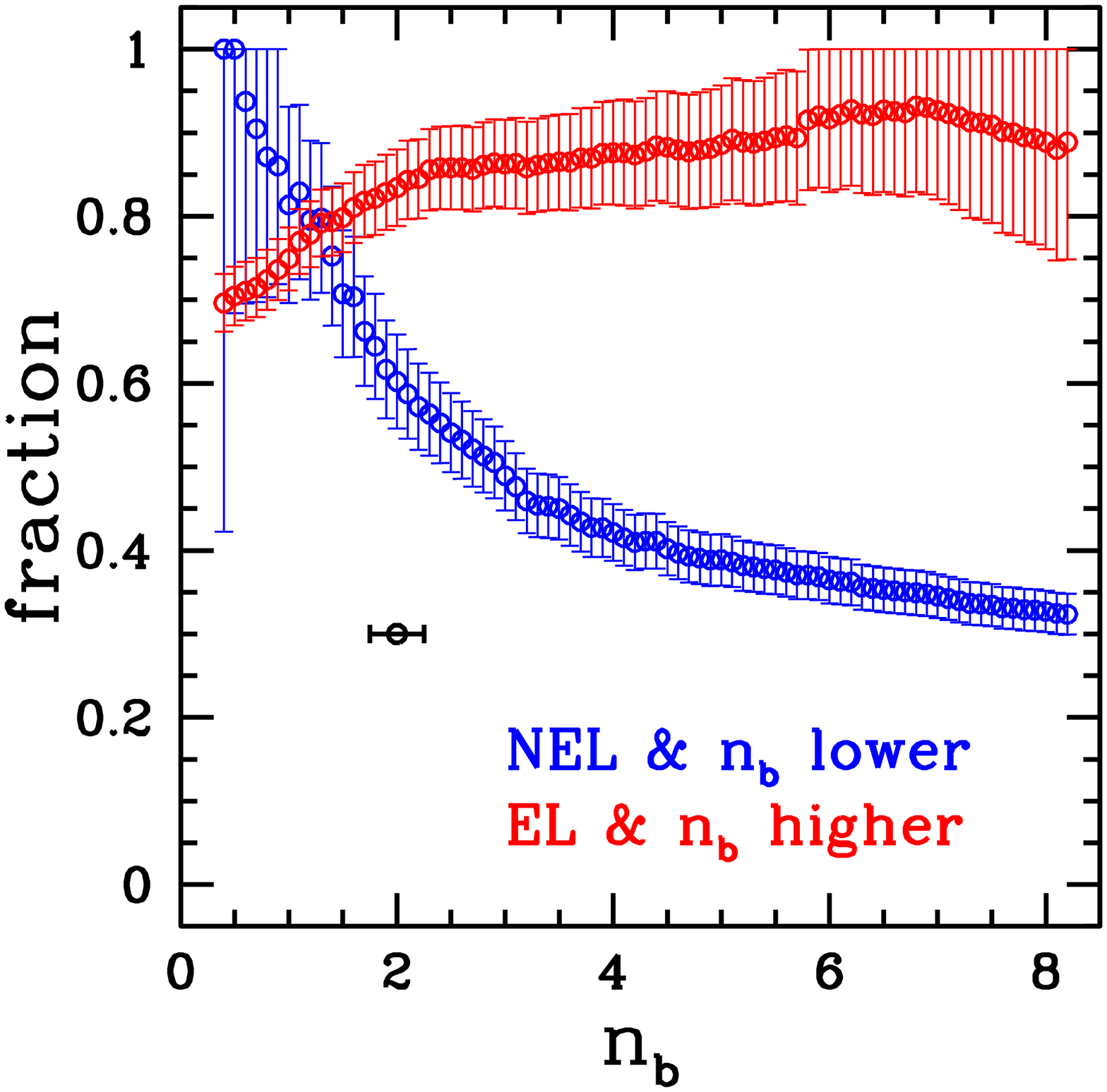}}
\mbox{\includegraphics[trim=75 0 0 0,clip,width=41.8mm]{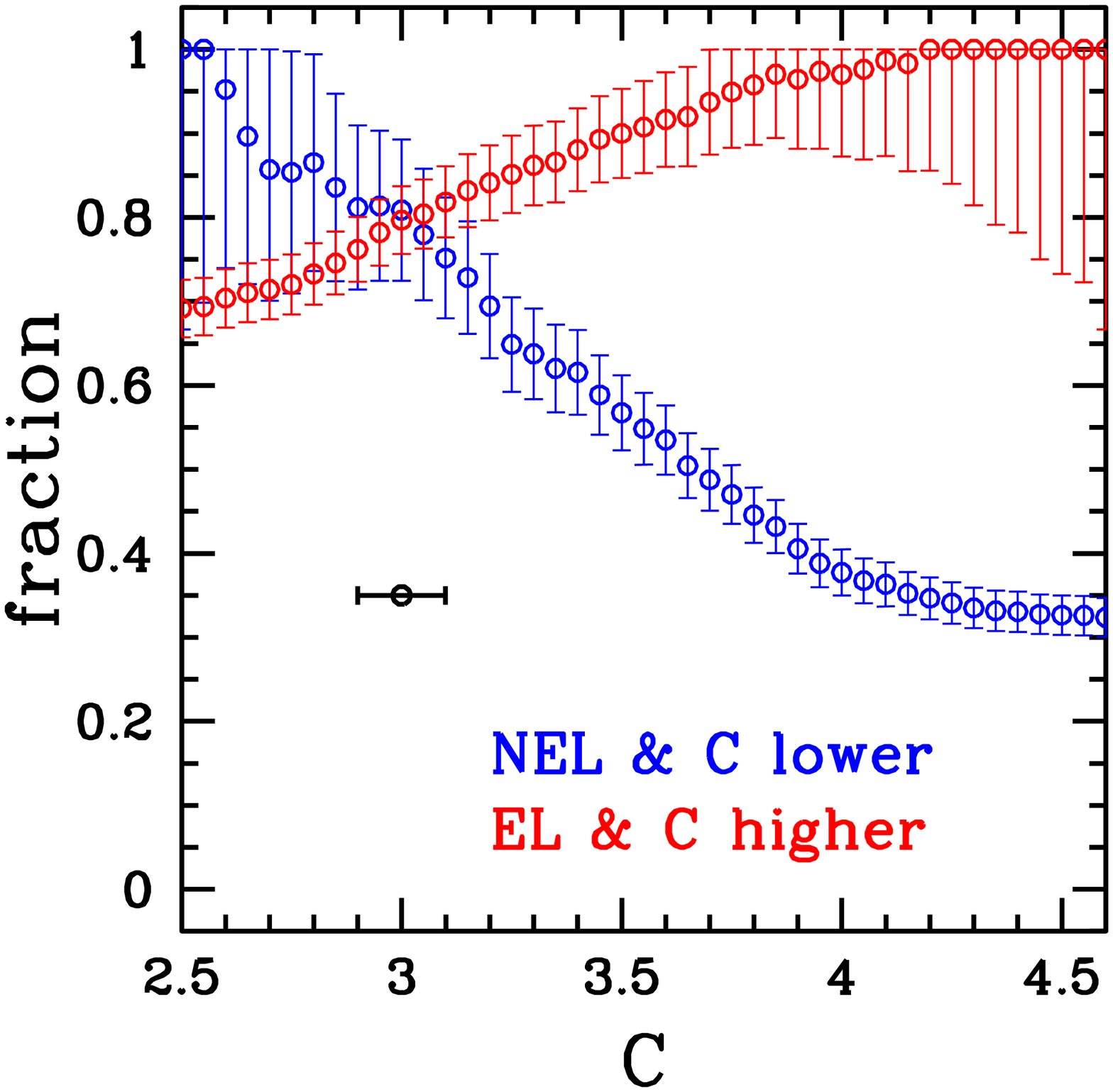}}
\mbox{\includegraphics[trim=75 0 0 0,clip,width=41.8mm]{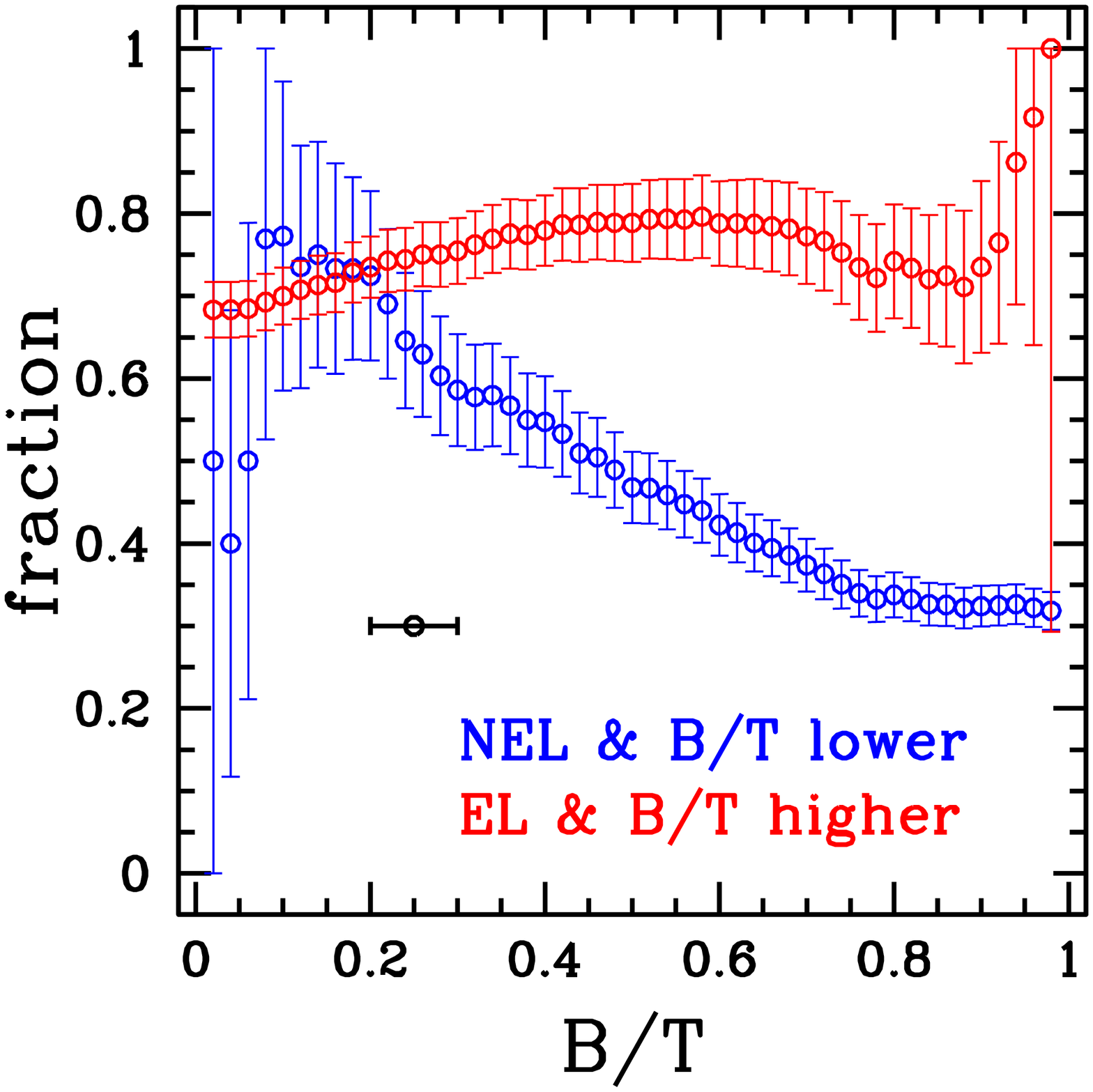}}
\mbox{\includegraphics[trim=75 0 0 0,clip,width=41.8mm]{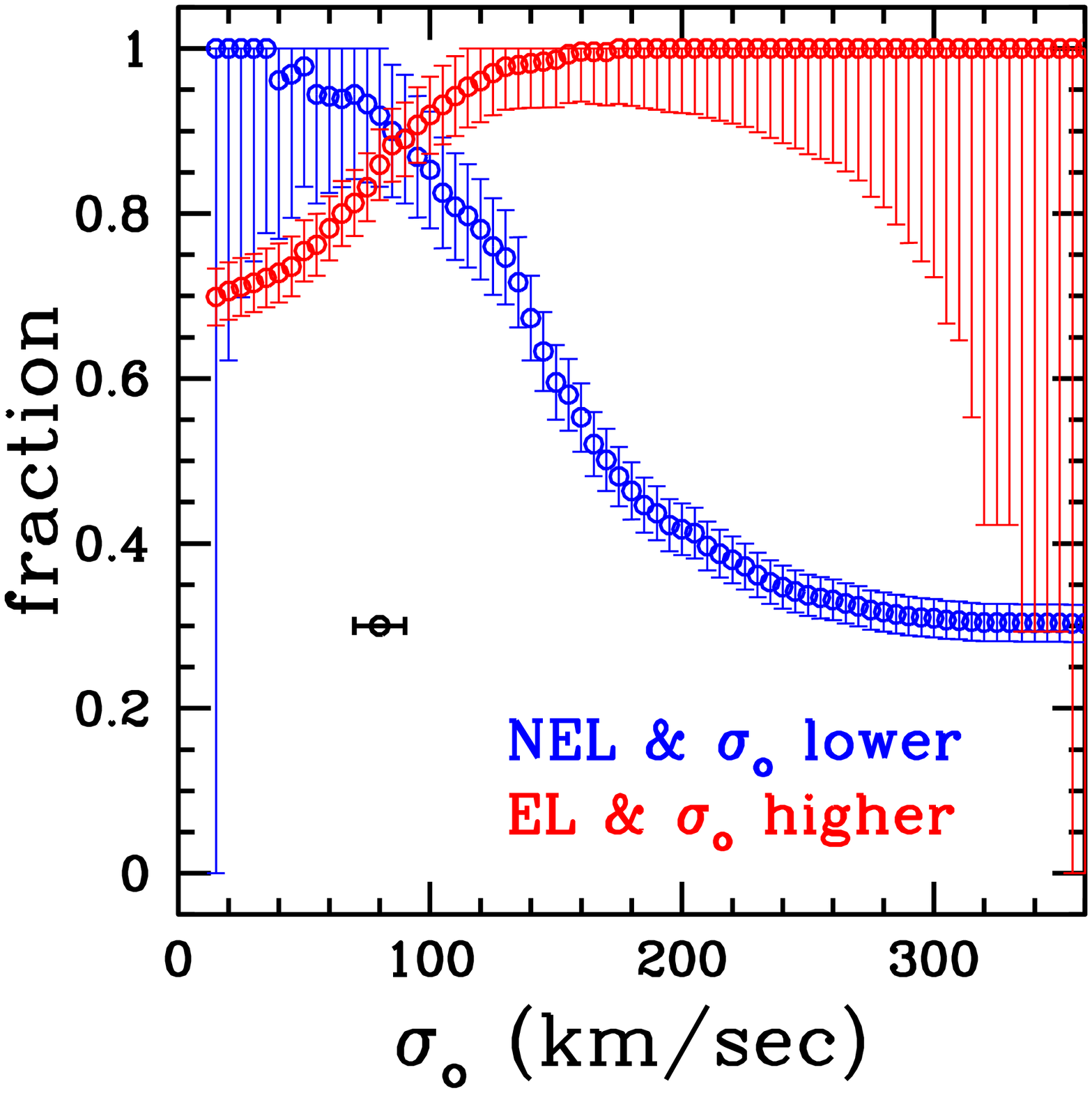}}
\caption{{\bf Complementarity to Kormendy relation:} The performance of four morphology indicators ($n_b$, $C$, $B/T$, $\sigma_o$), in separating ``elliptical like" (EL) and ``non-elliptical like" (NEL) bulges (defined according to Kormendy relation), is examined. Each point on the red curve marks the fraction of those galaxies which host EL bulges and have an indicator value higher than that point. On the contrary, each point on the blue curve marks the fraction of those galaxies which have NEL bulges and have an indicator value lower than that point. The efficiency of the indicator depends on the sharpness, smoothness and stability of the increase (decrease) in galaxy fraction exhibited by the red (blue) curve with successively increasing indicator value. Average error-bars are marked in each plot.}
\label{fracplots}
\end{figure*}

Now we employ this indicator to investigate the performance of other potential bulge morphology classifiers, i.e., bulge S\'ersic index ($n_b$), bulge-to-total light ratio ($B/T$), concentration index ($C$) and central velocity dispersion ($\sigma_o$). In Fig.~\ref{fracplots}, for successively increasing values of $n_b$, $B/T$, $C$ and $\sigma_o$, we trace the increase in the fraction of bulges which are EL (i.e., have $\Delta$$\langle\mu_{eb}\rangle$ $<0$) and decrease in the fraction of bulges which are NEL (i.e., have $\Delta$$\langle\mu_{eb}\rangle$ $>0$). Each point on the red curve marks the fraction of those galaxies which host EL bulges and have an indicator value higher than that point. On the contrary, each point on the blue curve marks the fraction of those galaxies which have NEL bulges and have an indicator value lower than that point (Fig.~\ref{fracplots}).

In the case of $n_b$, 80\% of the bulges with $n_b>1.3$ are EL and 80\% of the bulges with $n_b<1.3$ are NEL. In the case of $C$, 80\% of the galaxies with $C>3.0$ have EL bulges and 80\% of the galaxies with $C<3.0$ have NEL bulges. $C$ appears to be a more efficient classifier than $n_b$ since with the increasing values of $C$, rise in the fraction of EL bulges and decline in the fraction of NEL bulges is sharper than $n_b$ (Fig.~\ref{fracplots}). In contrast, $B/T$ does not appear to be a good classifier. The trends of increase in EL bulges and decrease in NEL bulges with the increasing values of $B/T$ are not as clear as in the case of $n_b$ and $C$.

Amongst the four morphology indicators, $\sigma_o$ turns out to be the most efficient classifier, where the two curves are most sharp, smooth and stable (Fig.~\ref{fracplots}). The transition occurs at $\sigma_o=90$ km/s, where 90\% of the galaxies with $\sigma_o>90$ km/s have EL bulges and 90\% of the galaxies with $\sigma_o<90$ km/s have NEL bulges. With increasing values of $\sigma_o$, rise in the fraction of EL bulges and decline in the fraction of NEL bulges is sharper and more stable than that observed for other morphology indicators. 

\subsection{Selection of pseudo, classical and ``ambiguous" bulges}

\begin{figure*}
\centering
\mbox{\includegraphics[width=49mm]{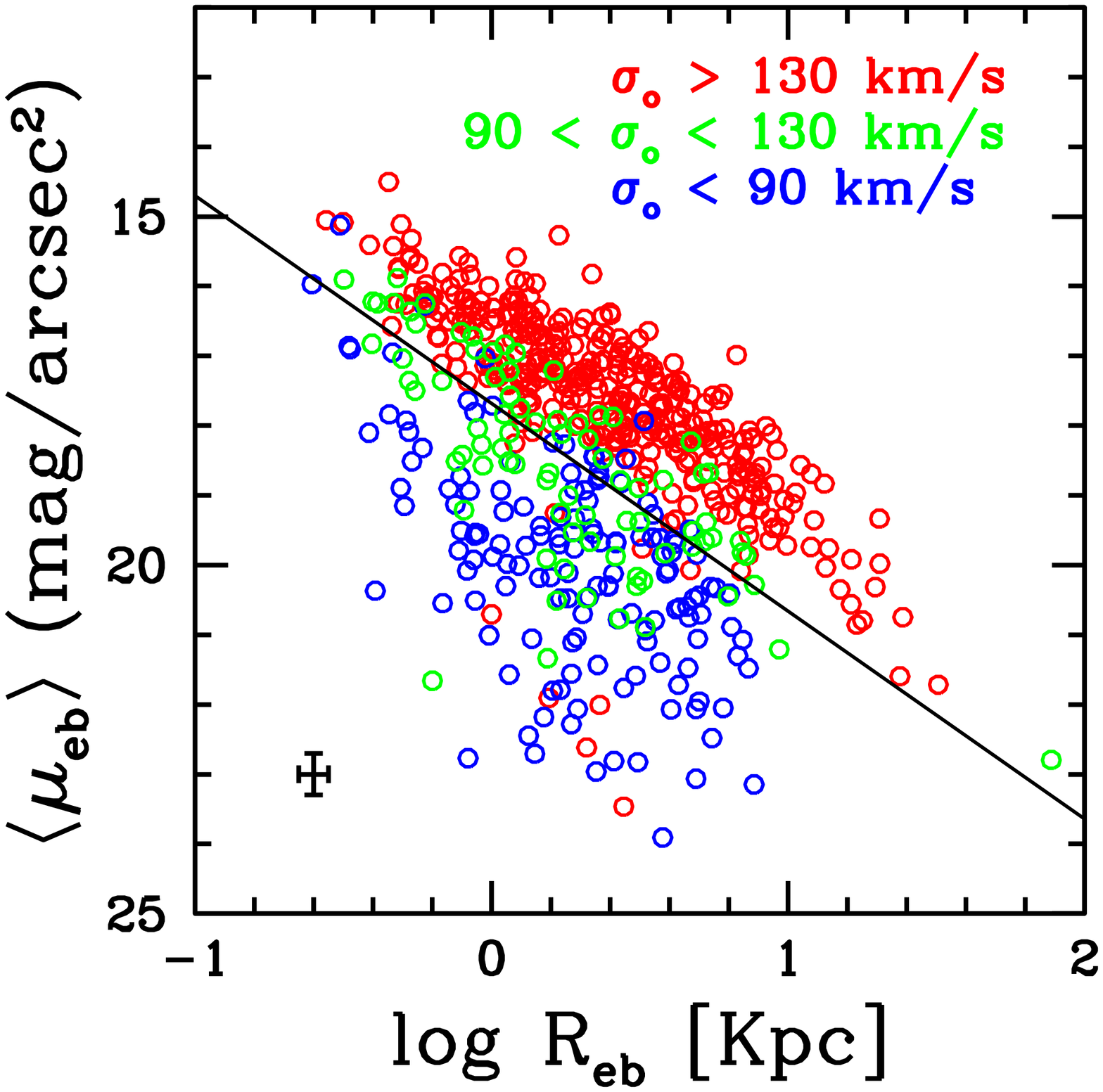}}
\mbox{\includegraphics[trim=83 0 0 0,clip,width=41.7mm]{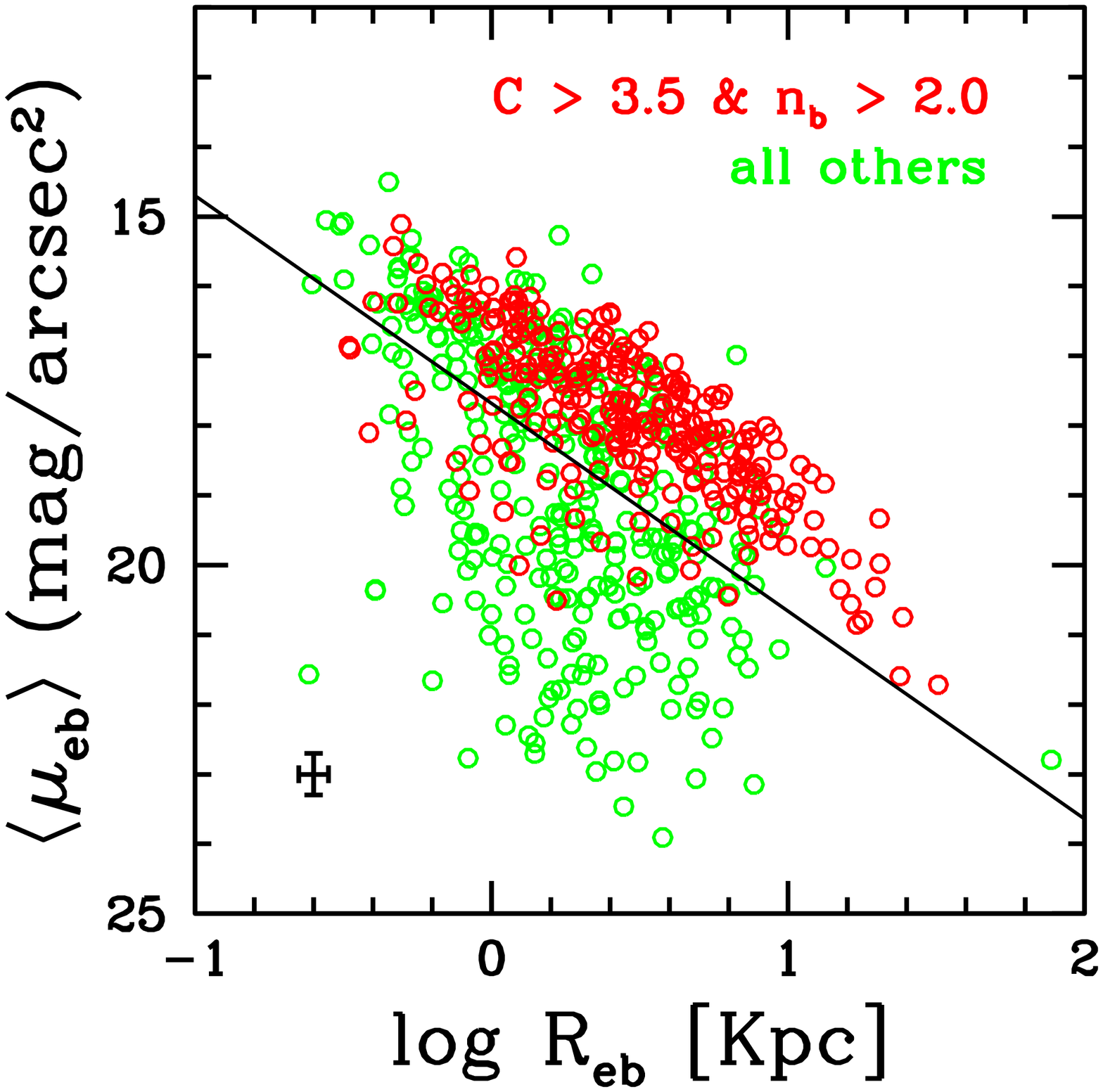}}
\mbox{\includegraphics[trim=83 0 0 0,clip,width=41.7mm]{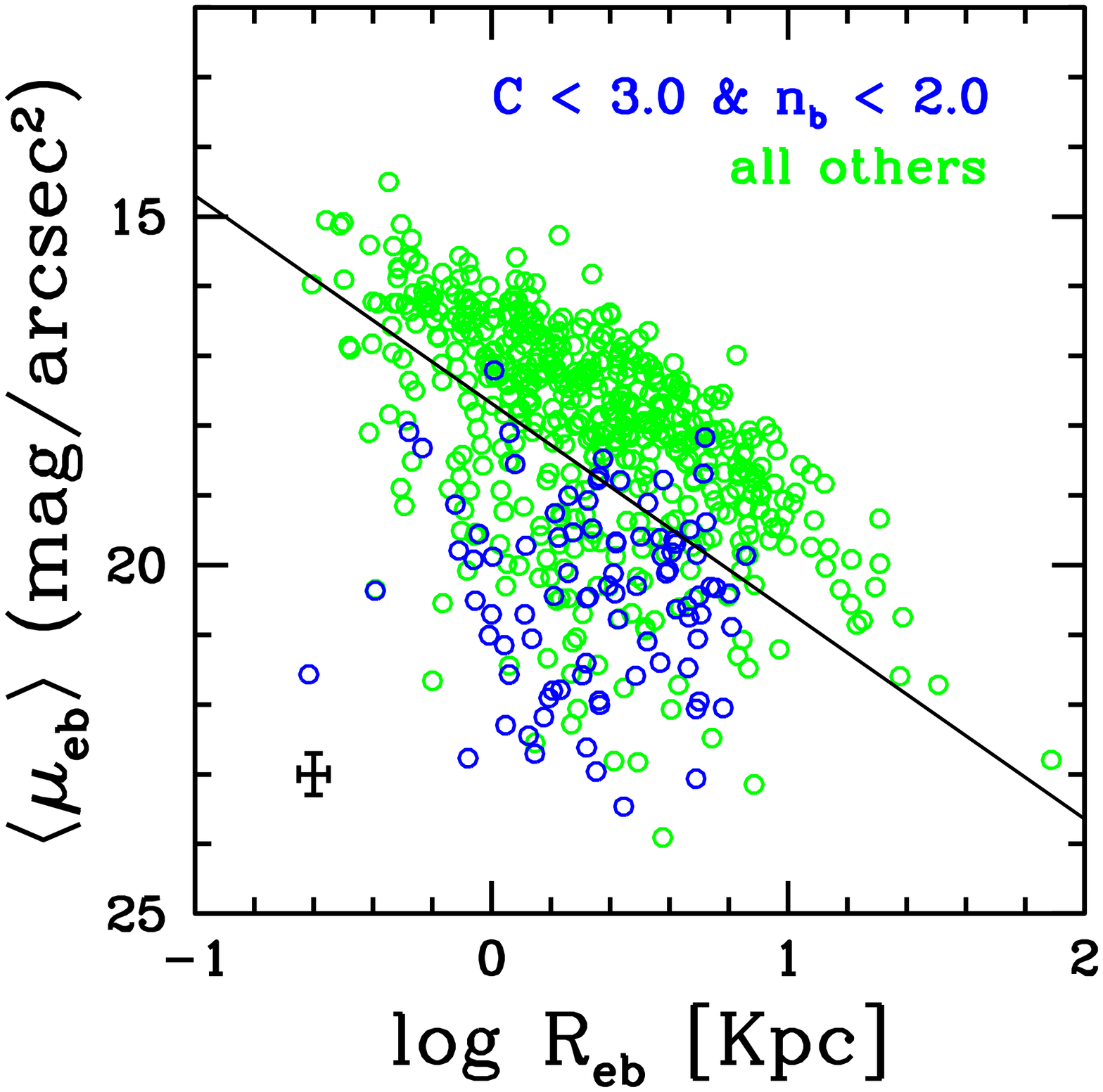}}
\mbox{\includegraphics[trim=83 0 0 0,clip,width=41.7mm]{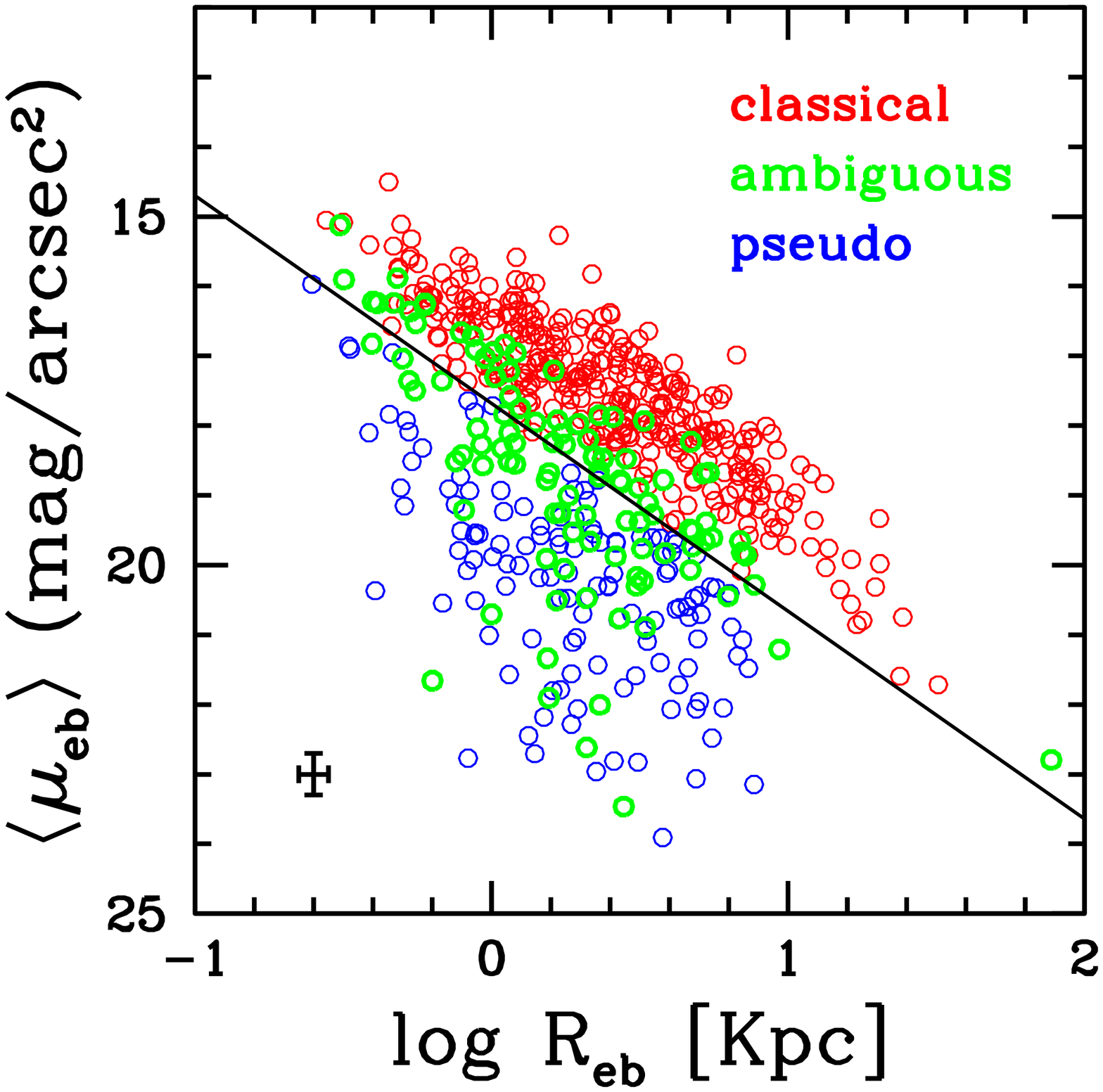}}
\caption{{\bf Efficient combinations and selection:} The first plot shows the placement of galaxies with different values of central velocity dispersion ($\sigma_o$) on the Kormendy plane. While $\sigma_o$ best compliments the Kormendy relation in bulge classification, in its absence the combination of concentration ($C$) and S\'ersic index of the bulge ($n_b$) turns out to be effective. The second and third plot depict the strategies for separating ``elliptical like" (red) and ``non-elliptical like" (blue) bulge disc galaxies from others (green) using combination of $C$ and $n_b$. In the fourth plot, the selection of pseudo (blue) and classical (red) bulges, as carried out in this work, based on the combination of Kormendy relation and $\sigma_o$, is shown. The bulges which could not be unambiguously classified, as either pseudo or classical, have been put in the ``ambiguous" (green) category. Average error-bars are marked in each plot.}
\label{combos}
\end{figure*}

The investigation of the performance of morphology indicators reveals that $\sigma_o$ best compliments KR in the separation of EL and NEL bulges. For a sufficiently high (low) value of $\sigma_o$, all bulges will be EL (NEL) based on KR, however, we also need to ensure that least fraction of bulges are left unclassified. In accordance to that, we find that more than 60\% of our total sample has $\sigma_o>130$ km/s and all ($\sim$98\%) of them are EL based on KR. For the rest 40\% of the sample, we find that more than 60\% has $\sigma_o<90$ km/s and nearly all ($\sim$90\%) of them are NEL based on KR. In Fig.~\ref{combos} the placement of galaxies in different ranges of $\sigma_o$ is shown on the Kormendy plane. It is the most efficient of classifiers because it leaves less than 20\% of the bulges as unclassified.

If $\sigma_o$ is not available, then the combination of $n_b$ and $C$ works best on KR for the separation. An analysis of all potentialities reveals that $(C>3.5 + n_b>2.0)$ and $(C<3.0 + n_b<2.0)$ are the best combinations for the selection of EL and NEL bulges, respectively (Fig.~\ref{combos}). However, note that even the combination of $n_b$ and $C$ is not as effective as $\sigma_o$.

Since in our work, $\sigma_o$ is available for almost the full sample (593 out of 605), we will employ it to classify bulges in conjunction with the Kormendy relation. Thus, those bulges which have $\Delta$$\langle\mu_{eb}\rangle$ $<0$ and $\sigma_o>130$ km/s are marked to be an indubitable class of classical bulges. Similarly, those bulges which have $\Delta$$\langle\mu_{eb}\rangle$ $>0$ and $\sigma_o<90$ km/s are marked to be an indubitable class of pseudo bulges. Out of the total sample of 593 galaxies, 353 galaxies are thus classified to be discs hosting classical bulges and 130 galaxies are classified to be disc hosting pseudo bulges. The rest of the bulges (110, $<$20\% of the total) are marked as ``ambiguous". Fig.~\ref{combos} also shows the placement of classical, pseudo and ambiguous bulges on the Kormendy plane. Note that this category ``ambiguous" is not the third bulge type. It only indicates that these bulges could not be unambiguously classified as either classical or pseudo.

It is certainly possible that some of the EL bulges, possibly having $\Delta$$\langle\mu_{eb}\rangle$ slightly greater than zero or $\sigma_o$ slightly less than 130 km/s, did not get included in the classical bulge sample. Similarly, it is also possible that some of the NEL bulges have been missed out of the pseudo bulge sample. However, strict criteria ensure that the two bulge-type samples have minimal contamination. Most importantly, the bulges which could not be indubitably classified into one type are saved from contaminating the sample of the other type by being grouped into the separate ambiguous bulge category.  

\begin{figure}
\centering
\mbox{\includegraphics[width=45mm]{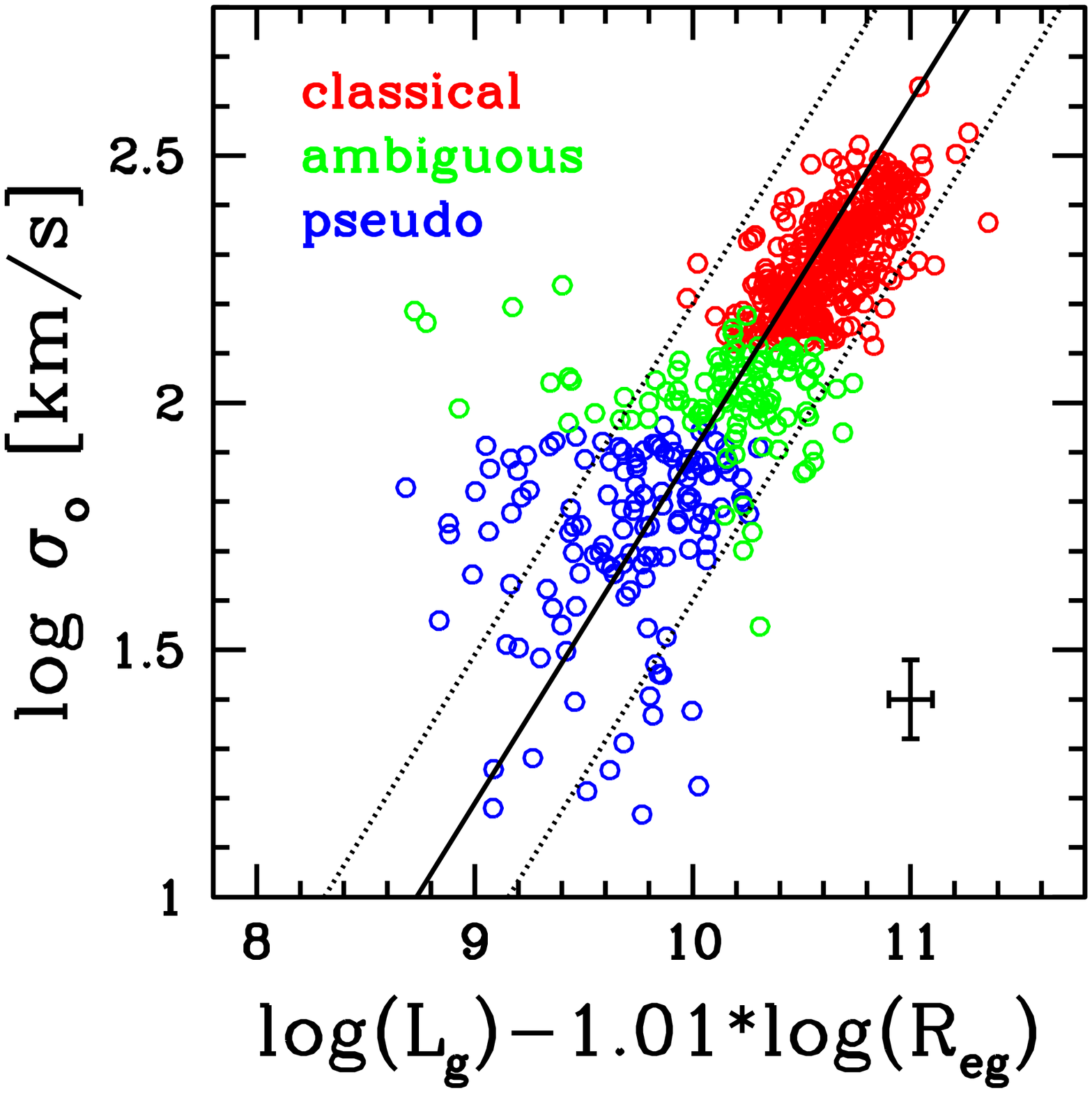}}
\mbox{\includegraphics[trim=94 0 0 0,clip,width=37.3mm]{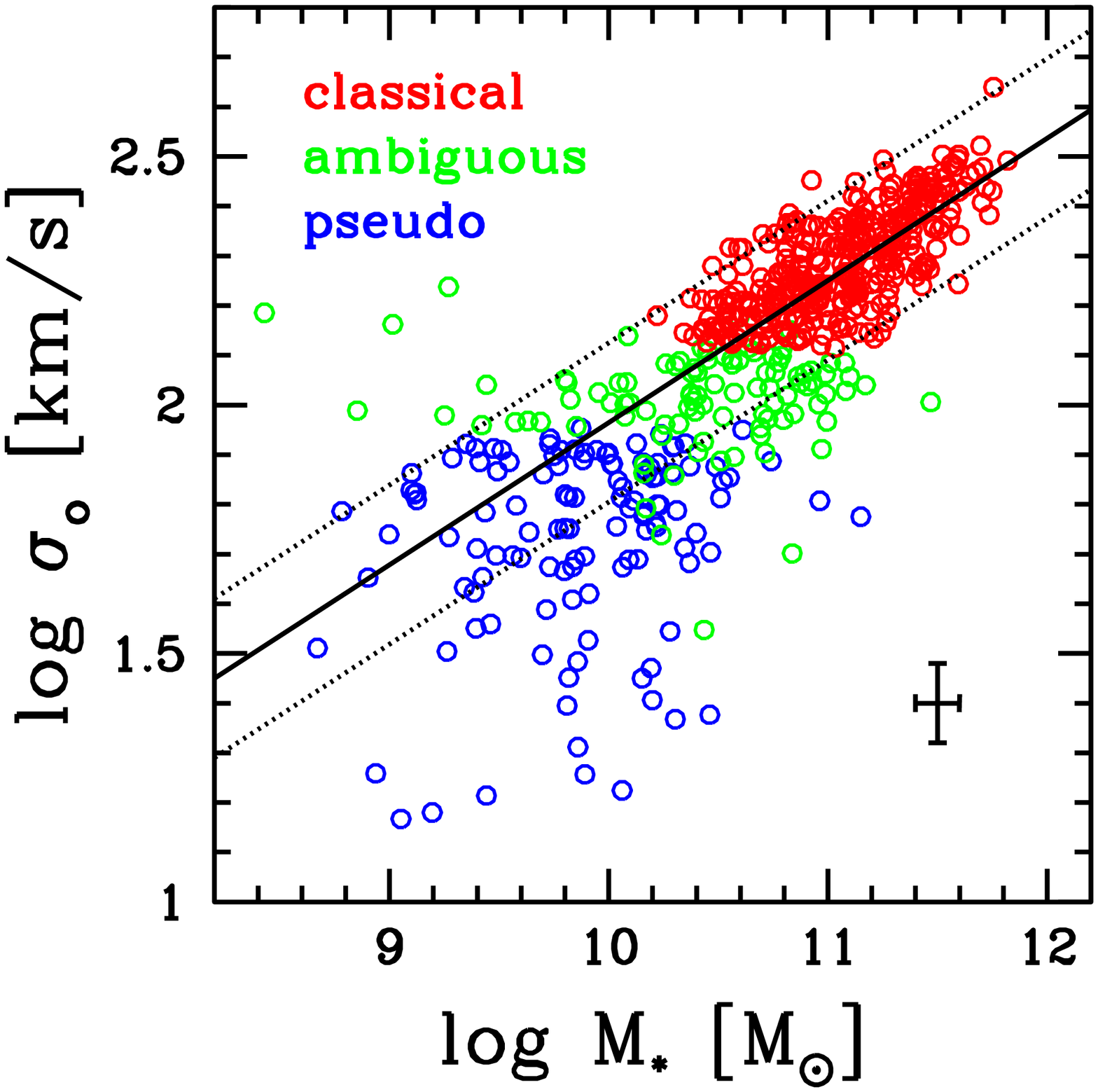}}
\caption{{\bf Fundamental and Faber-Jackson plane:} In the first plot, the placement of disc galaxies with pseudo (blue), classical (red) and ambiguous (green) bulges is shown on the Fundamental Plane. The plane, the relation (solid line) and the scatter (dotted lines) are from \citet{vandenBosch2016} analysis of BH-host galaxies. In the second plot, their placement is shown on the mass-based Faber-Jackson plane, where, total stellar mass of the galaxy ($M_*$) is plotted against its central velocity dispersion ($\sigma_o$). The relation (solid line) is from \citet{Gallazzietal2006} and the scatter (dotted lines) is as obtained by \citet{Corteseetal2014} and \citet{Aquino-Ortizetal2018} for their sample of early type galaxies. Average error-bars are marked in both plots.}
\label{bulgesFJ}
\end{figure}

In Fig.~\ref{bulgesFJ}, the placement of disc galaxies with pseudo, classical and ambiguous bulges has been depicted on the Fundamental \citep{Kormendy1977} and Faber-Jackson \citep{FaberandJackson1976} plane. The relation and the scatter of the Fundamental Plane is as reported by \citet{vandenBosch2016} for their analysis of BH-host galaxies in $K_s$ band. They found that irrespective of the bulge nature, all galaxies follow the Fundamental Plane in $K_s$ band because it efficiently reflects the total mass inside effective radius. They also noted that dominance of the bulge can increase the tightness of the relation. Distribution of our sample is consistent with their findings (Fig.~\ref{bulgesFJ}). The Faber-Jackson relation is as found by \citet{Gallazzietal2006}, whereas, the scatter lines are those reported by later works \citep{Corteseetal2014,Aquino-Ortizetal2018} for a more representative sample of early type galaxies. It can be seen that classical bulge hosting disc galaxies are the best adherents of the scaling relations found for elliptical galaxies. Although this is expected because the selection of these bulges is itself based on a scaling relation, this demonstrates the effectiveness of the Kormendy relation in bulge-type determination.

\subsection{Bimodality of structural and stellar properties}

\begin{figure*}
\centering
\mbox{\includegraphics[width=50mm]{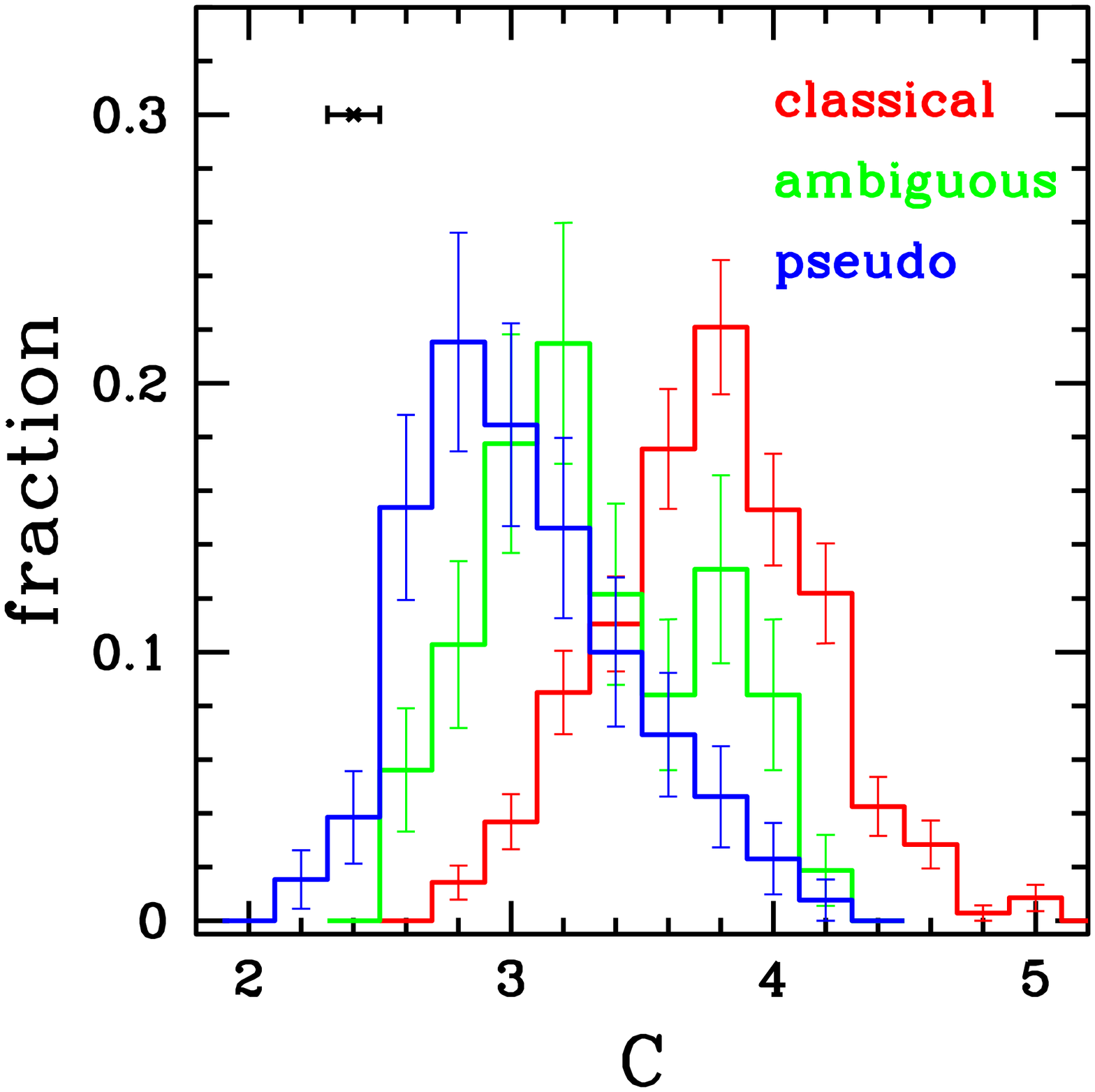}}
\mbox{\includegraphics[trim=30 0 0 0,clip,width=47.2mm]{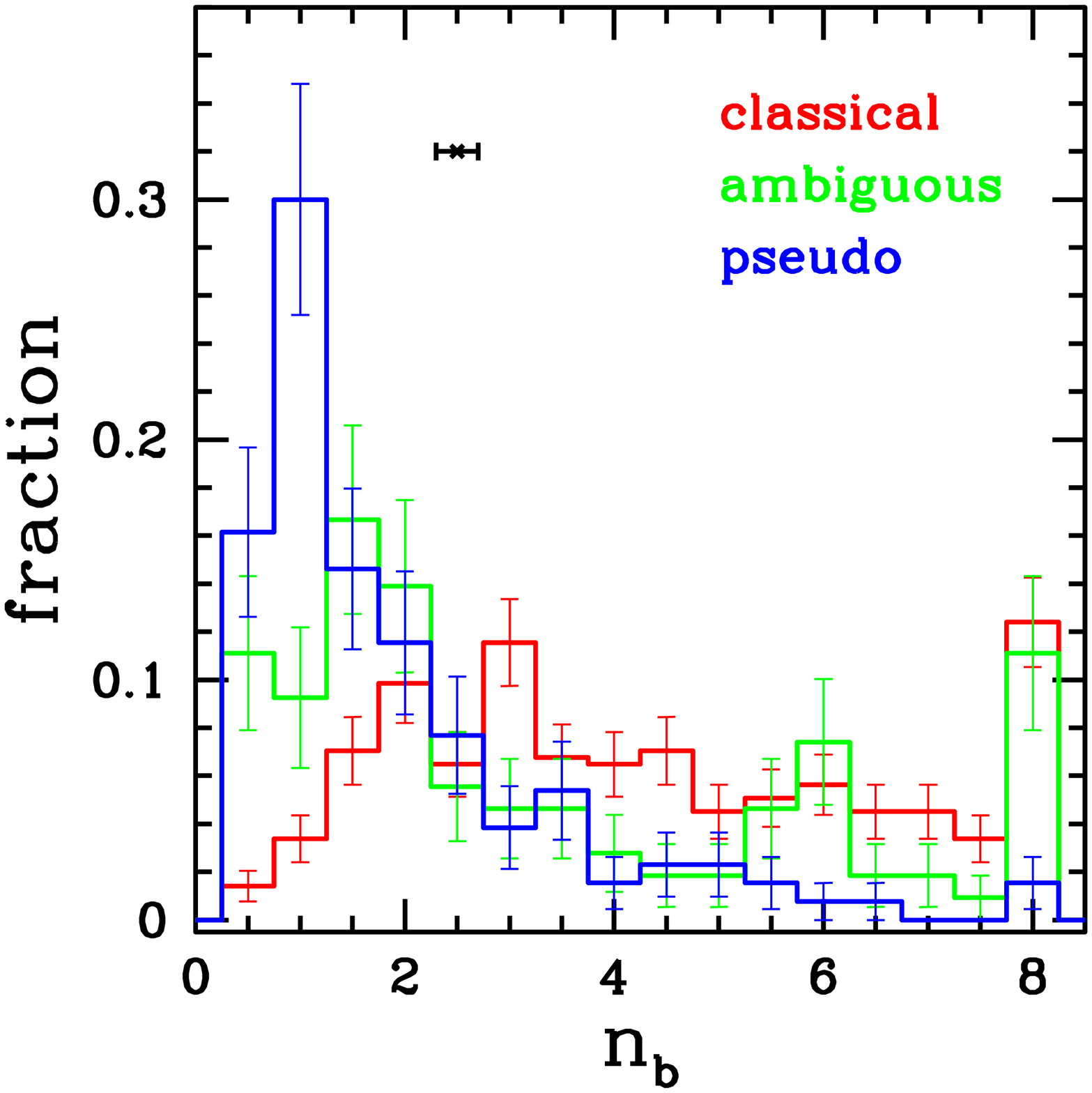}}
\mbox{\includegraphics[trim=30 0 0 0,clip,width=47.2mm]{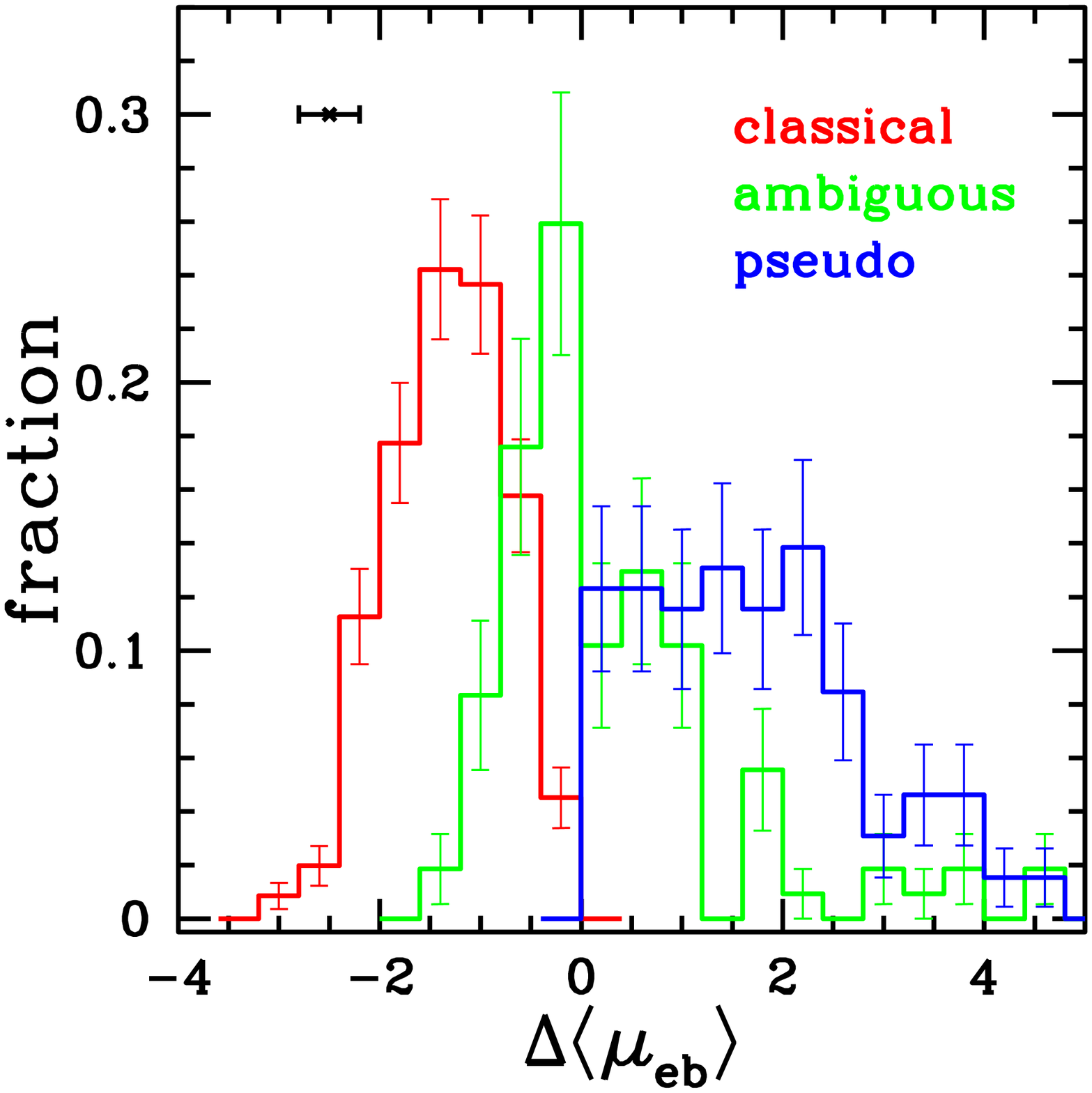}}\\
\mbox{\includegraphics[width=50mm]{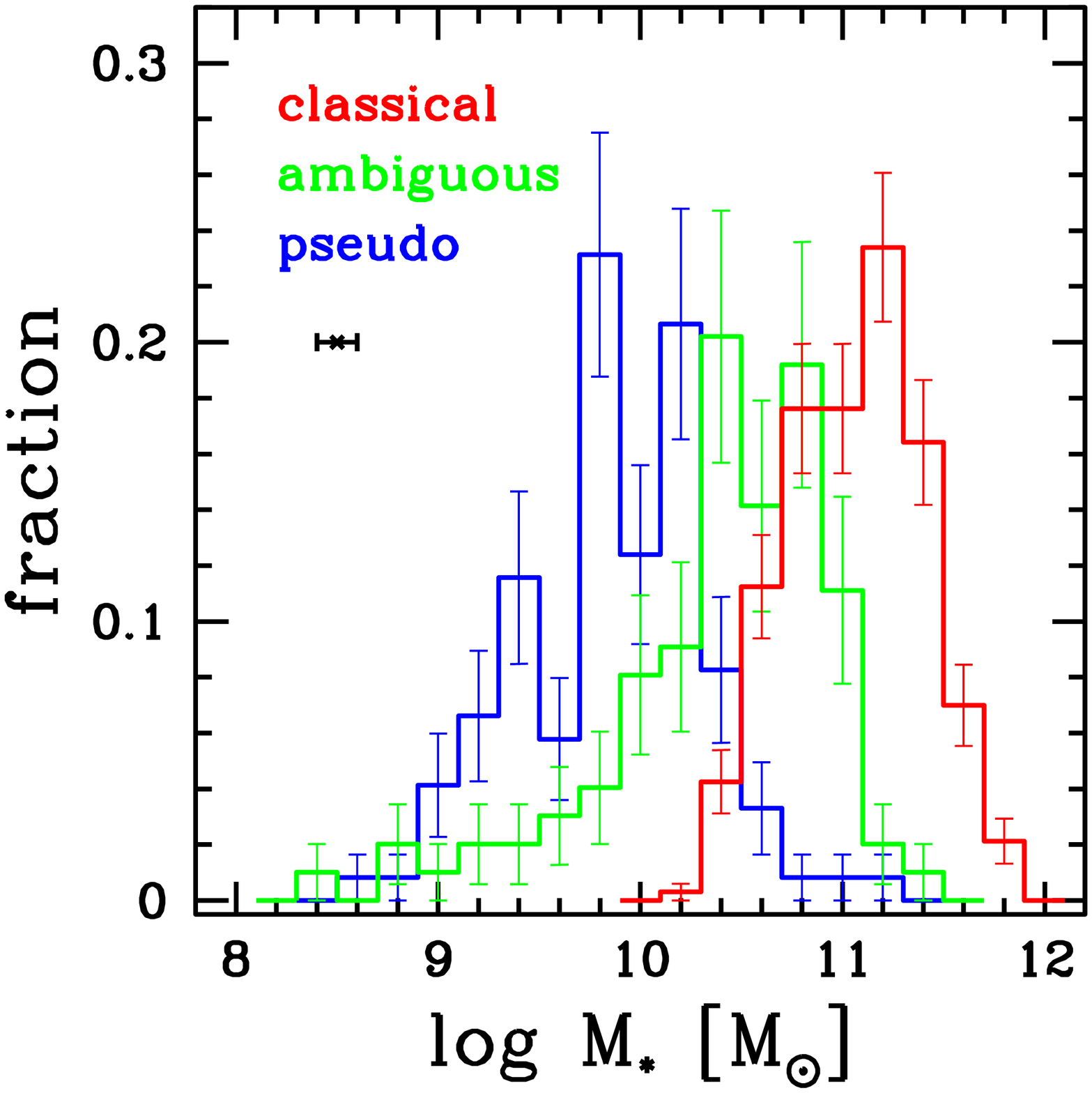}}
\mbox{\includegraphics[trim=30 0 0 0,clip,width=47.2mm]{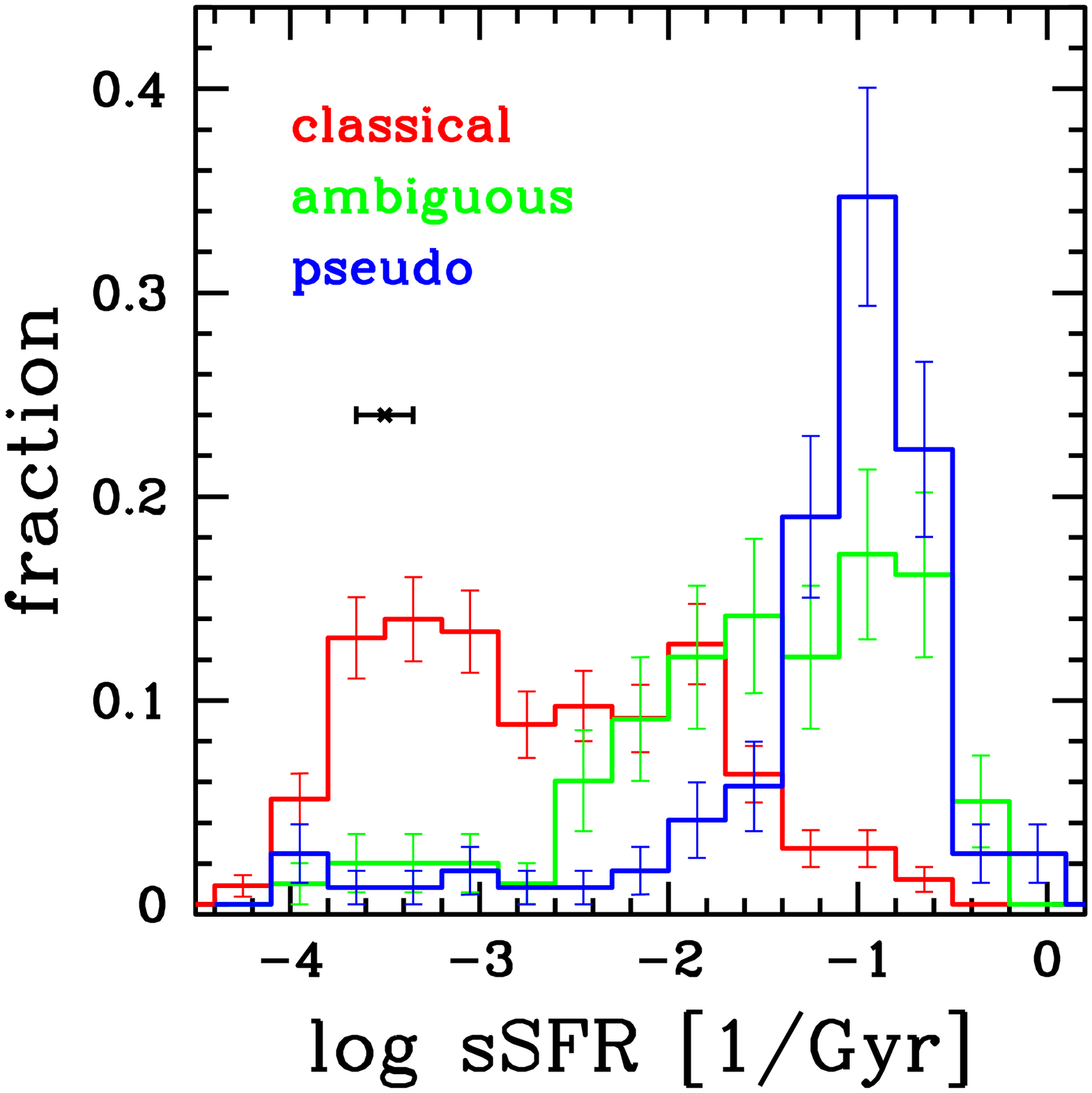}}
\mbox{\includegraphics[trim=30 0 0 0,clip,width=47.2mm]{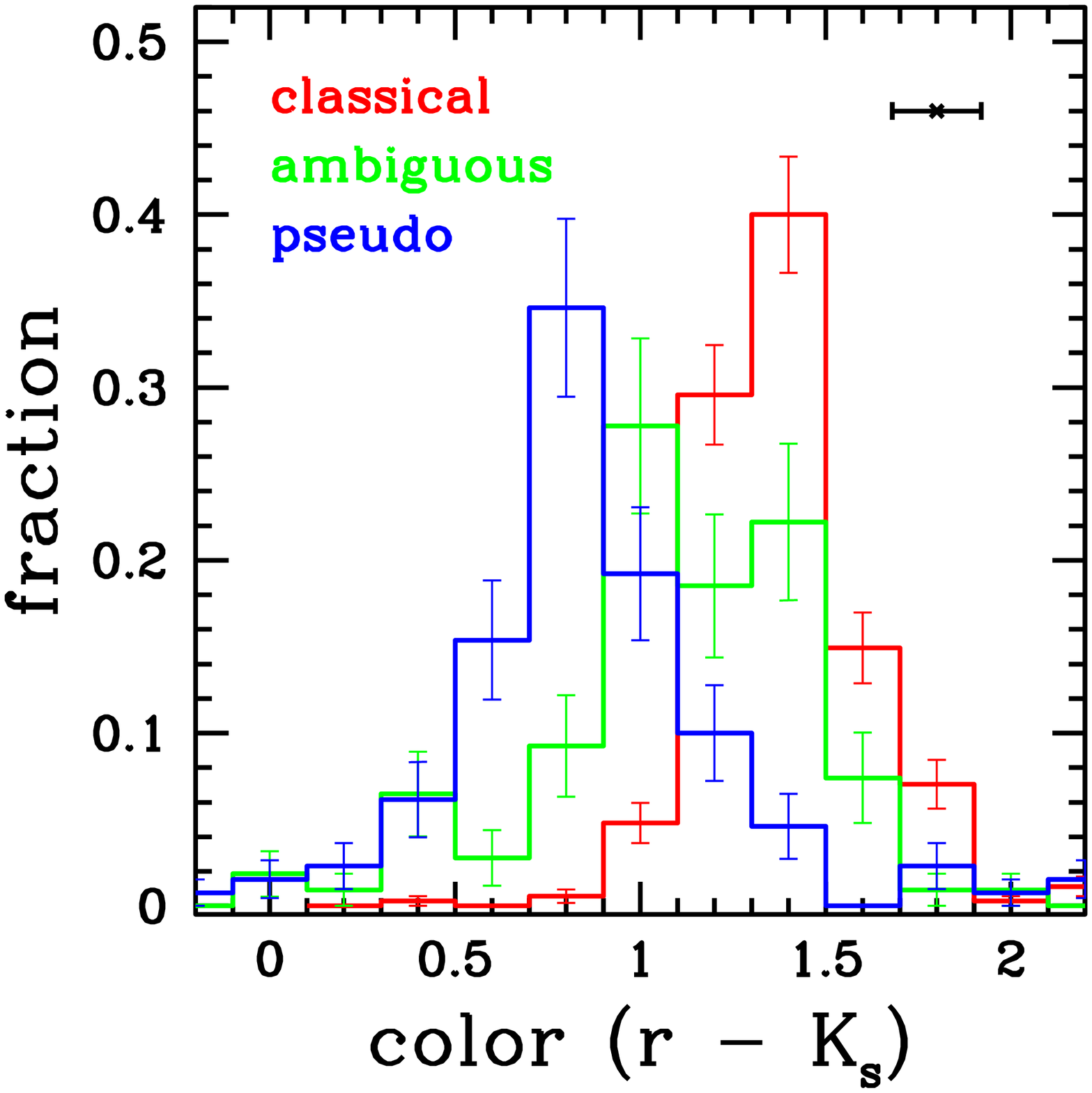}}
\caption{{\bf Bulge properties:} The distribution of galaxies with pseudo (blue), classical (red) and ambiguous (green) bulges is shown for different structural and stellar parameters. In the first row, the three parameters are concentration ($C$), S\'ersic index of the bulge ($n_b$) and relative average surface mass density within bulge effective radius ($\Delta$$\langle\mu_{eb}\rangle$). In the second row, the three parameters are total stellar mass ($M_*$), specific star formation rate (sSFR) and global colour ($r-K_s$) of the galaxy. Average error-bars are marked in each plot.}
\label{bulgeproperties}
\end{figure*}

In Fig.~\ref{bulgeproperties}, we analyse the distribution of the two bulge types, along with ambiguous bulges, with respect to structural indicators and stellar parameters. In the case of the concentration index ($C$), pseudo (PBD) and classical bulge disc (CBD) galaxies exhibit two well separated peaks. All CBDs ($\sim$98\%) have $C$ more than 3.0, whereas, more than 85\% of PBDs have $C$ less than 3.5. As found earlier, $n_b$ and $B/T$ are not as efficient differentiators of bulge types as $C$. Although more than 80\% of PBDs have $n_b$ less than 2.5, CBDs are quite uniformly distributed over the full range. This is consistent with the earlier and most recent findings \citep{Gadotti2009,Gaoetal2020}, discussed later.

Our focus in this work is to examine the distribution of these bulges in terms of their stellar parameters. We report the presence of clear bimodality in the case of total stellar mass ($M_*$), specific star formation rate (sSFR) and global colour ($r-K_s$) of the galaxy. In the case of $M_*$, $\log(M_*/M_{\odot})$$=$10.5 marks the middle point which critically separates the two bulge populations. All PBDs ($\sim95\%$) have $M_*$ lower than this value and all CBDs ($\sim95\%$) have $M_*$ higher than this value. ABDs neatly occupy a narrow region in the middle area, indicating that as in the case of $\Delta$$\langle\mu_{eb}\rangle$ and $\sigma_o$, sufficiently high or low value of $M_*$ can lead to unambiguous classification of bulge type. 

In the case of sSFR, all PBDs ($\sim$95\%) have log(sSFR/(1/Gyr))$<$-1.4 and CBDs have log(sSFR/(1/Gyr))$>$-1.6. Similarly, in the case of colour ($r-K_s$), the difference between the peaks of PBDs and CBDs is as large as 0.6 mag. Here, the middle point (1.1 mag) reports less than 10\% digressions. Our work, thus, demonstrates that {\it if bulges are indubitably classified to be pseudo and classical, their stellar properties will be markedly distinct}, i.e., not only all PBDs will be young and star forming, all CBDs will be old and quiescent with negligible ($<$5-10\%) digressions. The placement of ABDs in these plots suggests that they are dominantly in the middle region or the green valley. The next section will provide more evidence in support of this observation.

\subsection{Most effective single structural indicator}

Discs hosting different bulge types have been found to differ in their stellar properties. However, instead of determining the bulge type, it will be more efficient if a single morphological indicator can be applied for the separation of star forming or quiescent populations. Although $M_*$ (total stellar mass), $M_*/R_{eg}$, $M_*/R_{eg}^2$, $\sigma_o$, $n_b$, $\Sigma_1$ (surface mass density within 1 kpc), etc., many indicators have been suggested, they have not been proven to be ultimate demarker. The issue being that substantial amount of dispersion survives and the thresholds determined are at the most necessary, not sufficient, in separating star forming and quiescent populations.

\begin{figure*}
\centering
\mbox{\includegraphics[trim=0 85 0 0,clip,width=50mm]{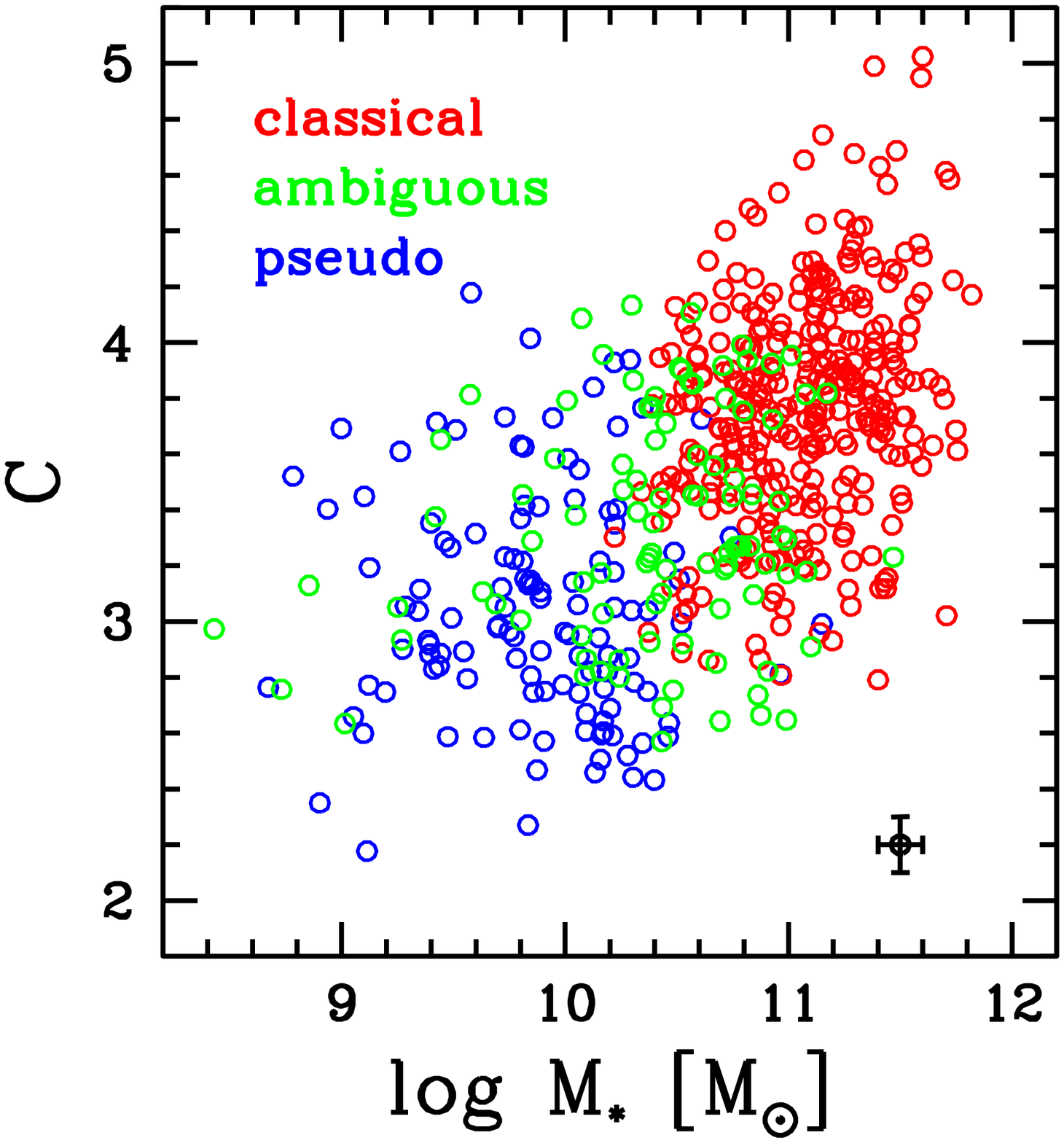}}
\mbox{\includegraphics[trim=0 85 0 0,clip,width=50mm]{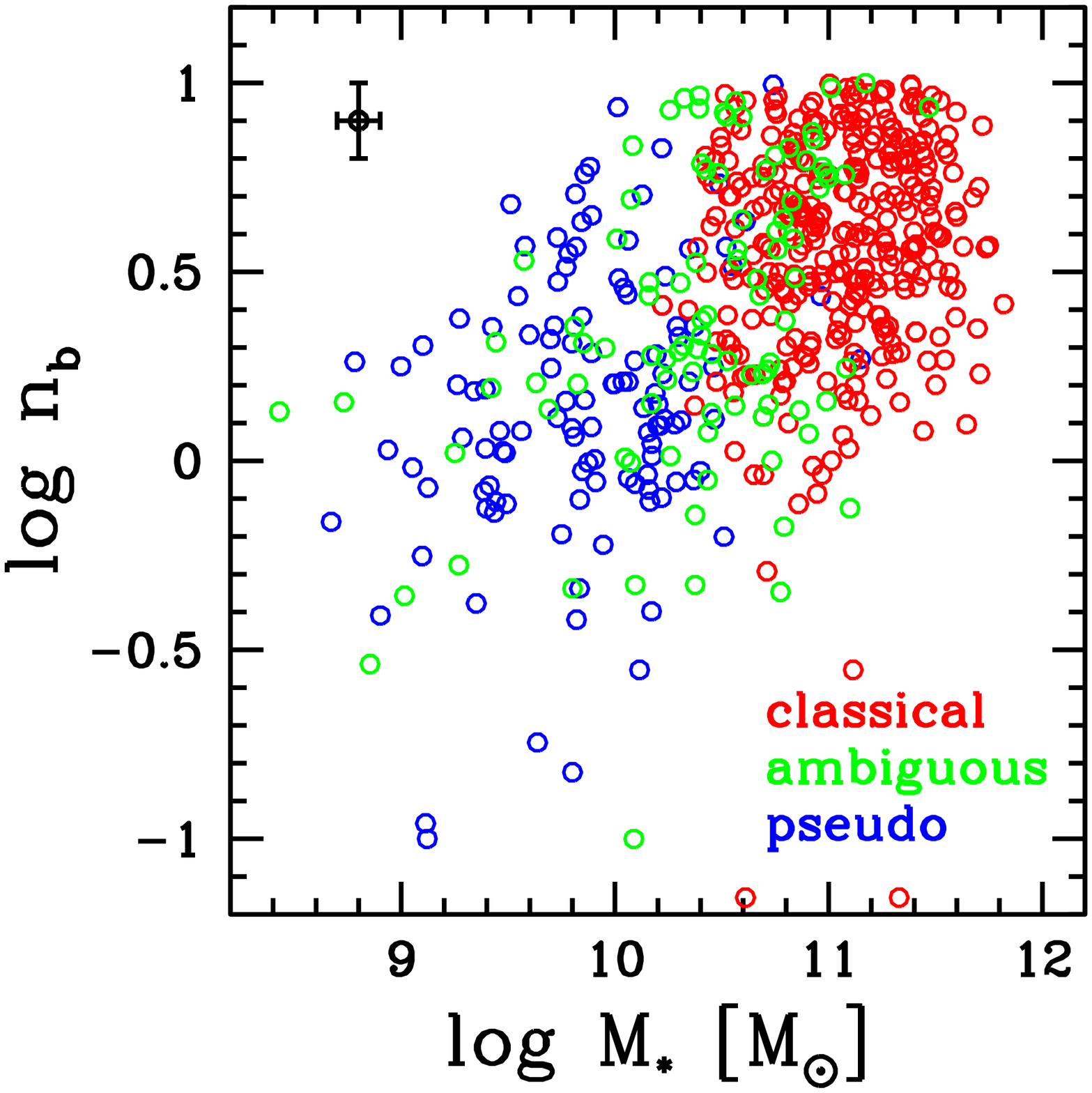}}
\mbox{\includegraphics[trim=0 85 0 0,clip,width=50mm]{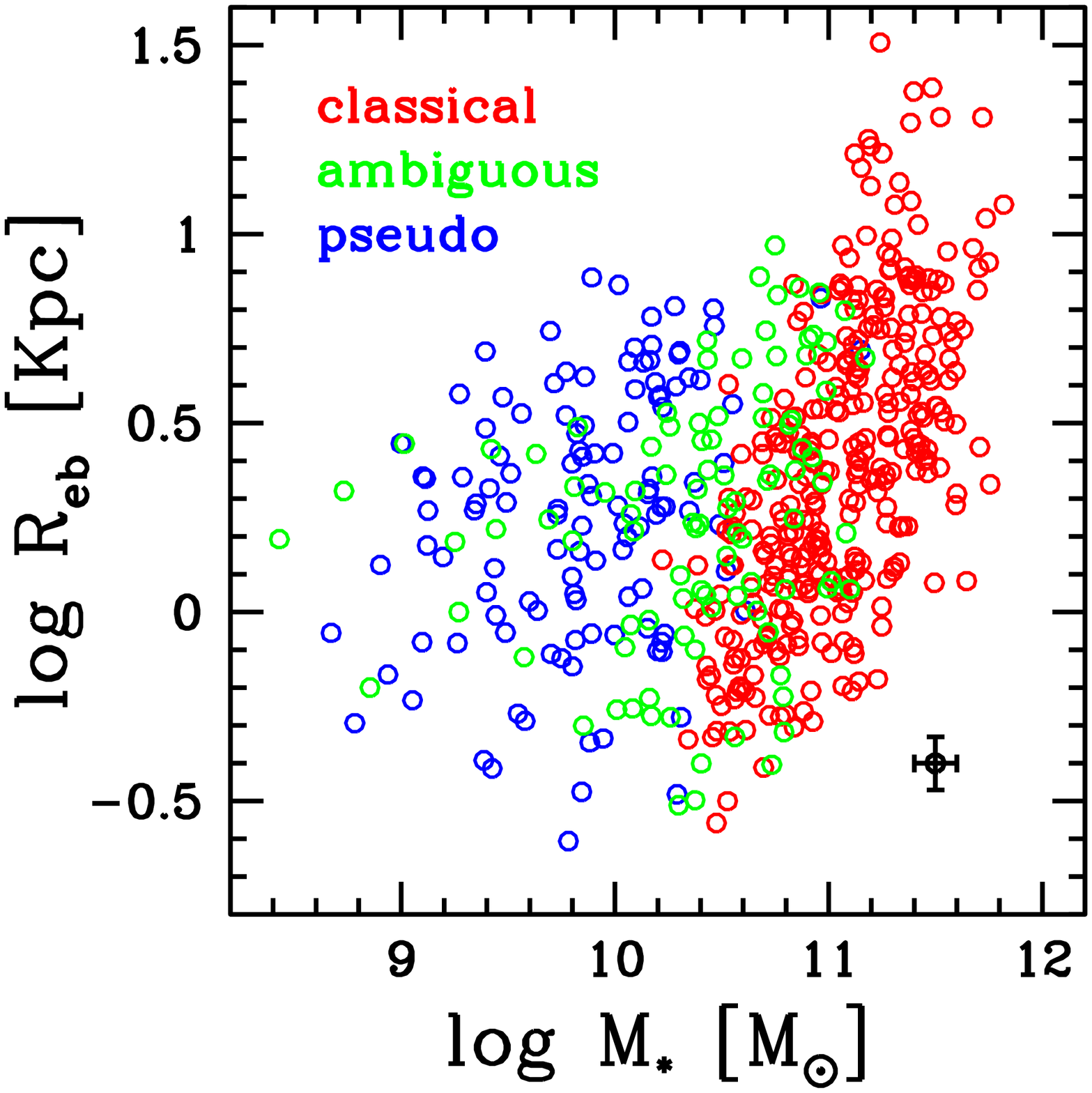}}\\
\mbox{\includegraphics[width=50mm]{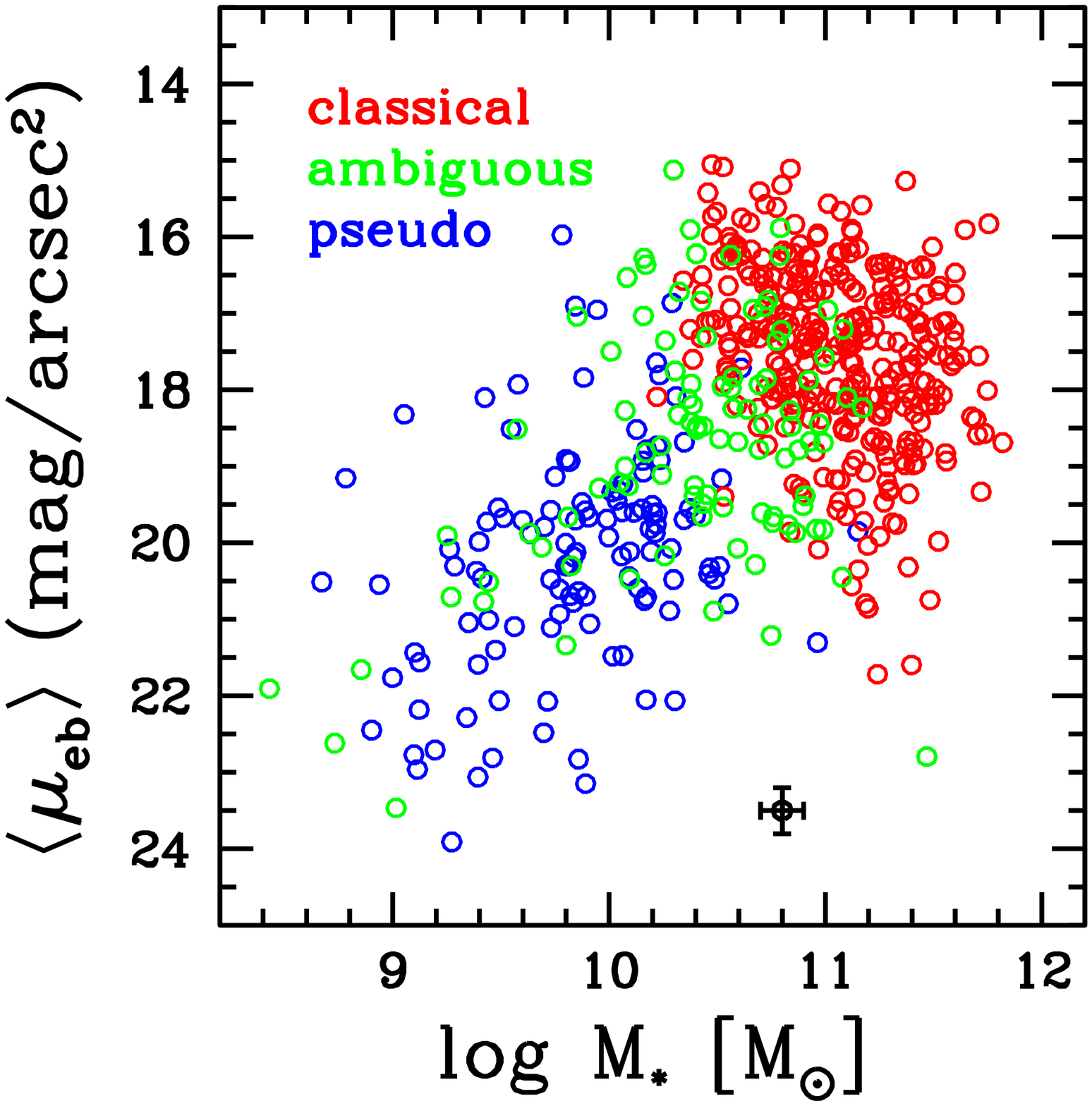}}
\mbox{\includegraphics[width=50mm]{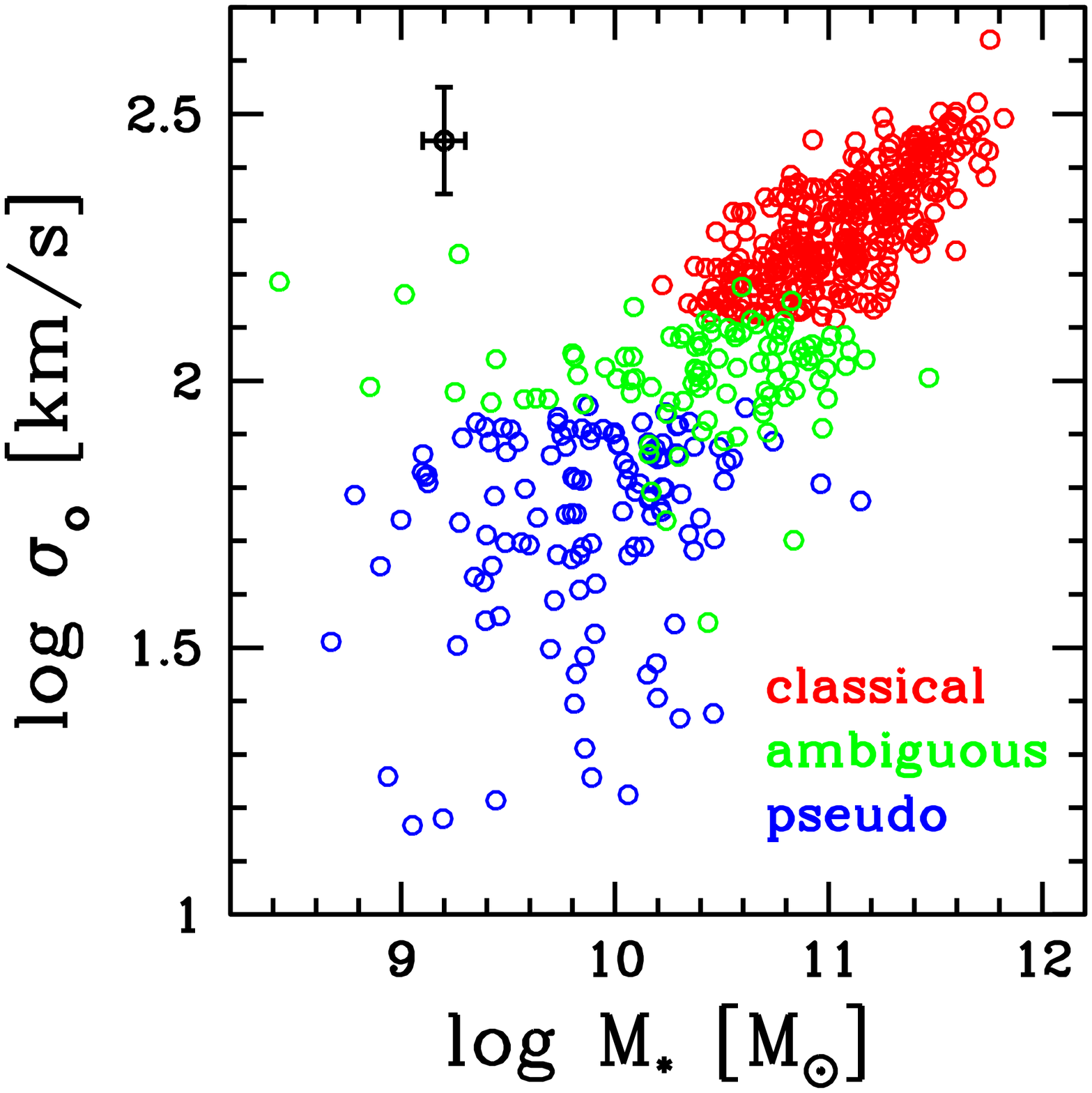}}
\mbox{\includegraphics[width=50mm]{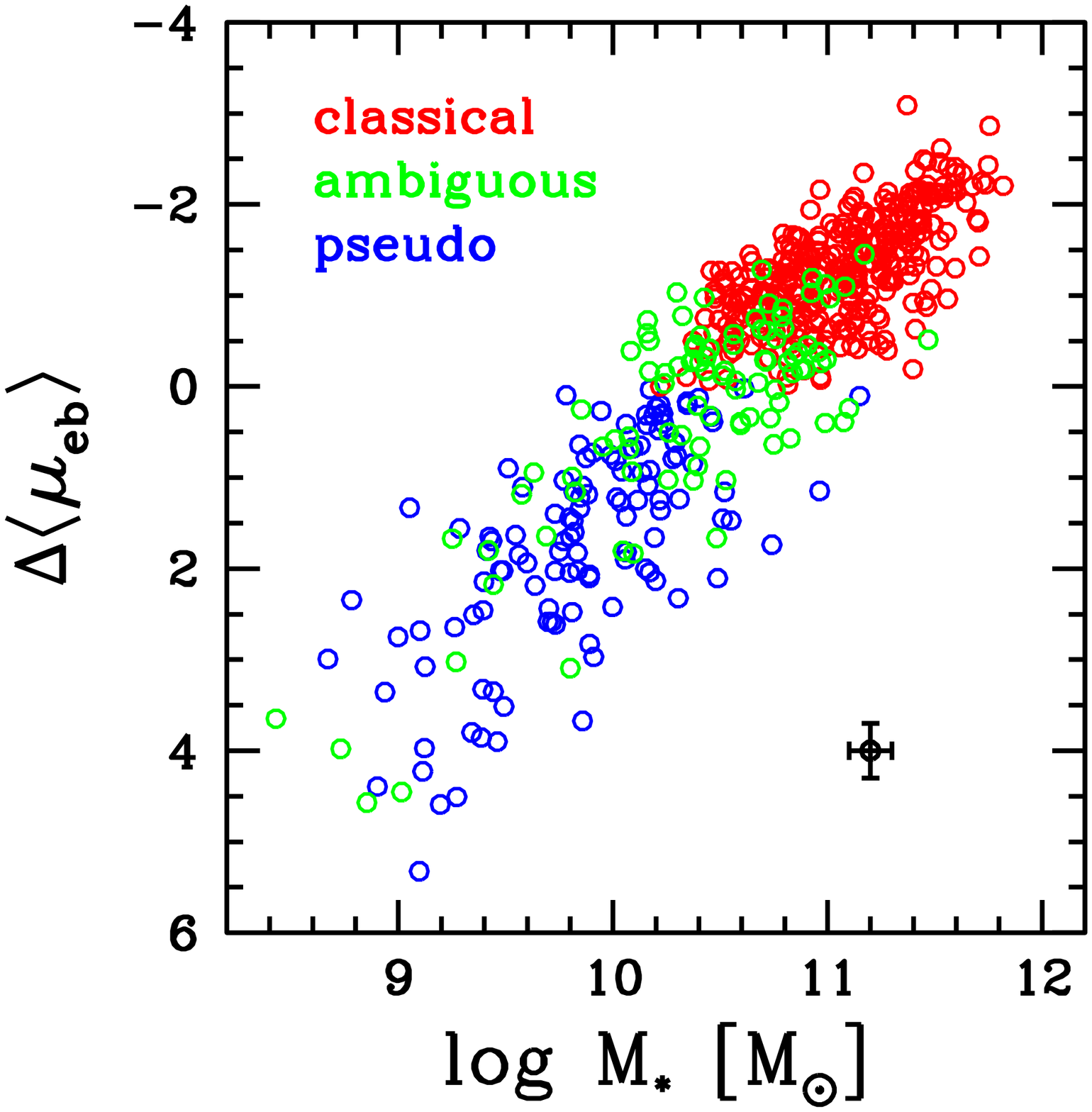}}
\caption{{\bf Complimenting mass:} The correlation of all morphological indicators is examined with respect to the total stellar mass ($M_*$) of the galaxy to select a single morphological indicator which is most adept at separating star forming and quiescent populations. In the first row, the correlation is explored with concentration ($C$), bulge S\'ersic index ($n_b$) and bulge effective radius ($R_{eb}$). In the second row, the correlation is explored with average surface mass density within bulge effective radius ($\langle\mu_{eb}\rangle$), central velocity dispersion ($\sigma_o$) and relative average surface mass density within bulge effective radius ($\Delta$$\langle\mu_{eb}\rangle$). Average error-bars are marked in each plot.}
\label{masscorrelations}
\end{figure*}

We will examine if our indicator $\Delta$$\langle\mu_{eb}\rangle$, based on KR, is successful in solving this issue. Fig.~\ref{masscorrelations} indeed reveals that total stellar mass ($M_*$) of the galaxy has tightest and most well defined correlation with $\Delta$$\langle\mu_{eb}\rangle$ compared to other indicators including $C$, $n_b$, $R_{eb}$, $\langle\mu_{eb}\rangle$ and $\sigma_o$. It depicts that more is the total stellar mass of the galaxy, more is the stellar mass density within the effective radius of the central bulge (Fig.~\ref{masscorrelations}).

\begin{figure}
\centering
\mbox{\includegraphics[trim=0 85 0 0,clip,width=55mm]{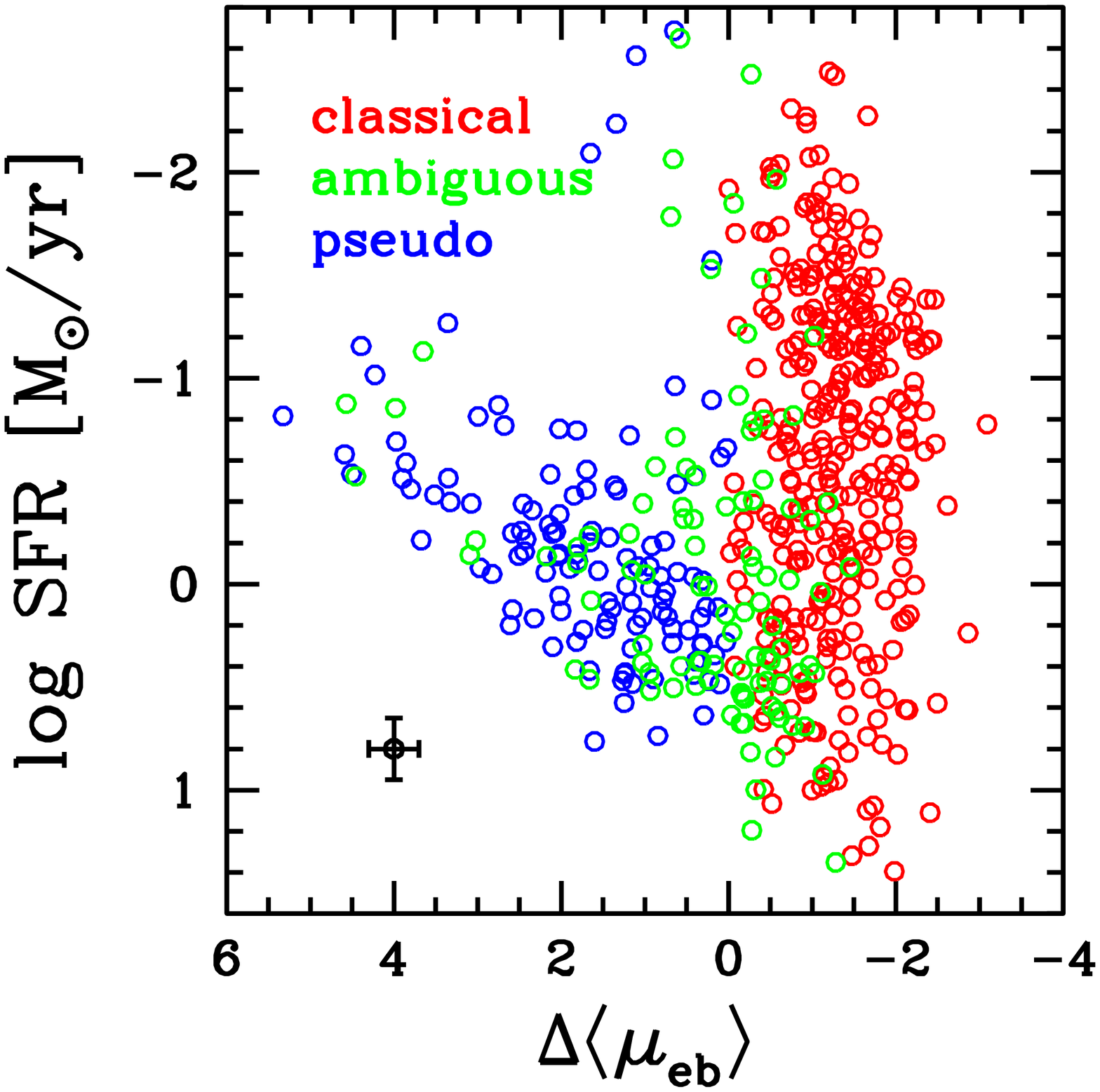}}
\mbox{\includegraphics[trim=0 85 0 0,clip,width=55mm]{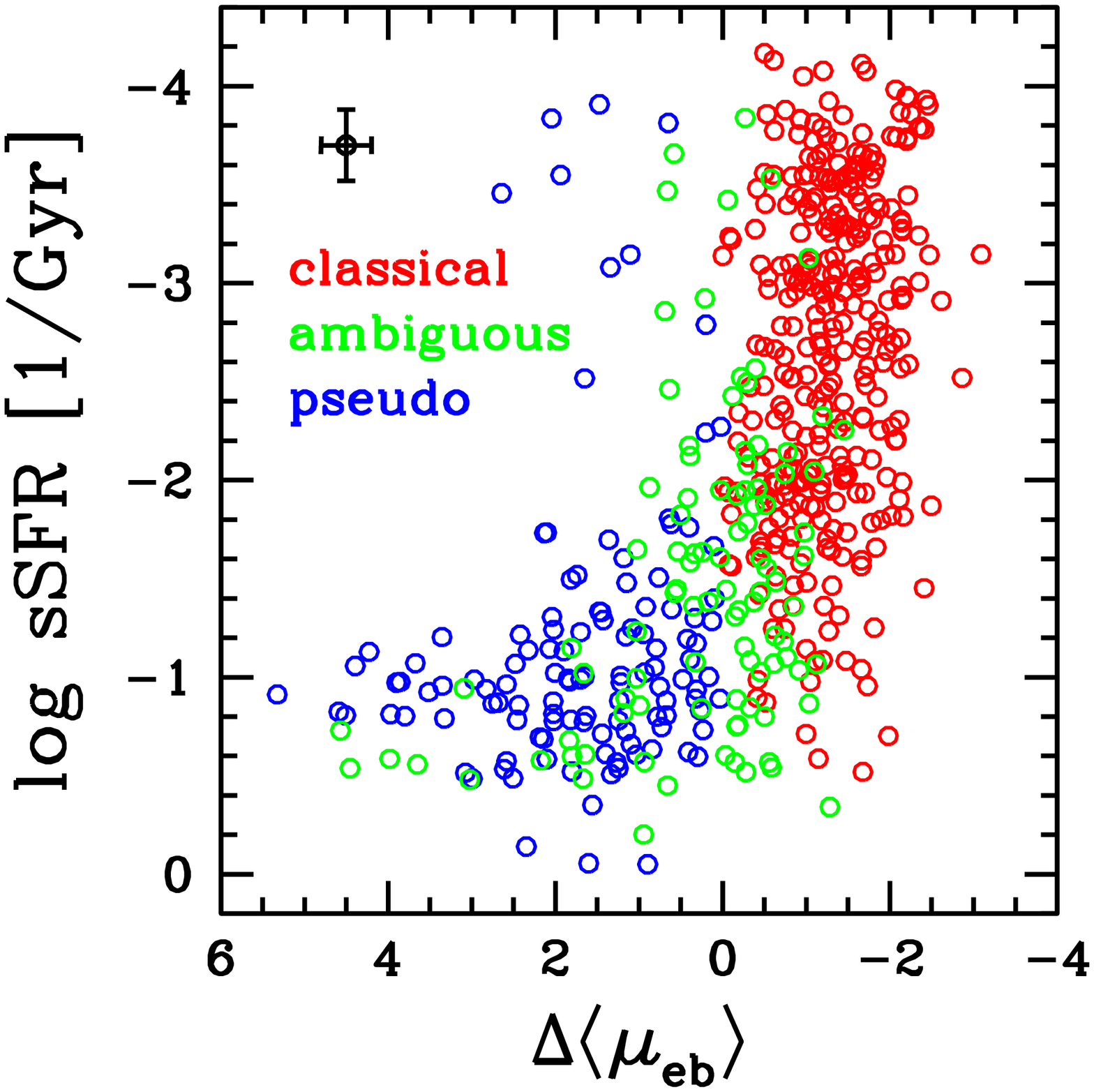}}
\mbox{\includegraphics[width=55mm]{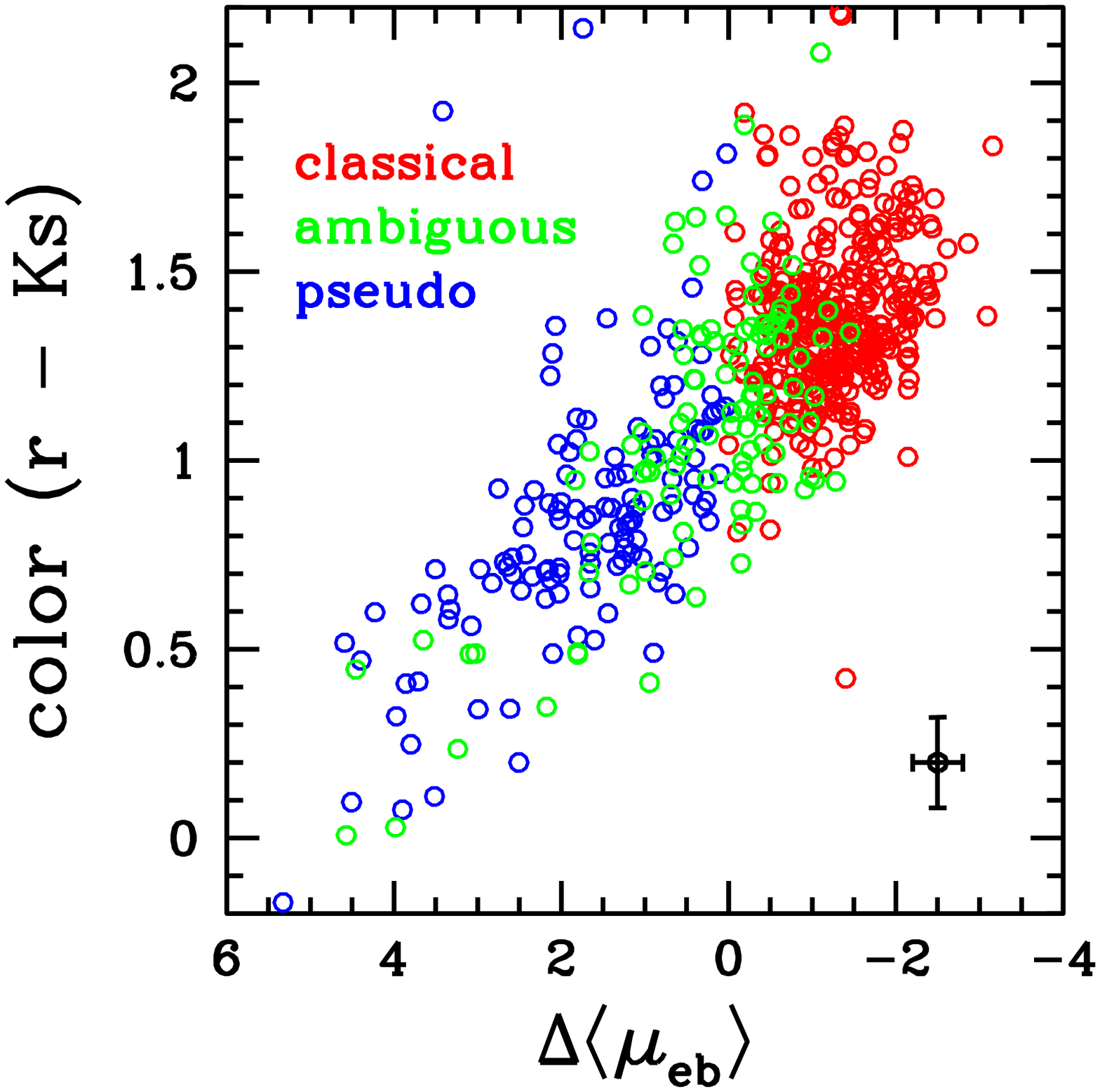}}
\caption{{\bf SFR and colour:} The correlation of the morphology indicator based on the Kormendy relation ($\Delta$$\langle\mu_{eb}\rangle$) is explored with stellar parameters which include star formation rate (SFR), specific SFR (sSFR) and global colour ($r-K_s$) of the galaxy. While in the case of SFR and sSFR, elbow-like pattern appears, there is a straight correlation in the case of colour. The galaxies are marked according to the bulge they host, i.e., pseudo (blue), classical (red) and ambiguous (green). Average error-bars are marked in each plot.}
\label{singleSFR}
\end{figure}

$\Delta$$\langle\mu_{eb}\rangle$, by definition, is a clear demarker of bulge-type of galaxies. We now examine its performance as a predictor of stellar activity in galaxies. In Fig.~\ref{singleSFR} we find that the indicator exhibits an ``elbow-like" correlation with both star formation rate (SFR) and specific SFR (sSFR) of galaxies. Here, the middle region of the elbow is dominated by galaxies with ambiguous bulges (Fig.~\ref{singleSFR}). Thus, as found in the previous section, discs with ambiguous bulges are emerging to be mainly placed in the green valley. The correlation plot of the indicator with global colour ($r-K_s$) of the galaxy provides further evidence in support of this argument. Interestingly, while the correlations with SFR and sSFR are elbow shaped, the indicator exhibits a straight correlation with colour. It shows that more is the central stellar mass density of a galaxy, redder is its colour. {\it Although ambiguous bulge galaxies were depicting an equivalent or higher SFR than pseudo bulge galaxies, they are redder in colour}, suggesting that either they are dust ridden or have composite populations. This could have resulted in the difficulty in resolving these bulges into either pseudo or classical.

\section{Discussion}
\label{sec:discussion} 

In this work, we have explored the connection between structural and stellar properties of local ($z<0.3$) galaxies by comparing the distribution (and correlation) of the parameters describing these properties for discs hosting pseudo and classical bulges. To achieve that, we have performed 2D bulge-disc decomposition of 1263 galaxies in rest-frame $K_s$ band. Bulge type classification has been performed based on multiple stringent complimentary criteria. The major findings of the study are discussed below.

\subsection{Classification of bulge types}

Over the past two decades, it has been established that multiple criteria, based on the structural and kinematic properties of the bulge, are required to be applied in combination to ascertain its type. A particular type of bulges, whether pseudo or classical, will not satisfy all the properties or criteria associated with them \citep[reviewed in][]{FisherandDrory2016,Kormendy2016}. Kormendy relation - which probes the similarity of bulge structure with elliptical galaxies - has been found to be the most effective and reliable differentiator of bulge type \citep{Gadotti2009,Neumannetal2017,Sachdevaetal2019,Gaoetal2020}. Other than the structure, it is crucial that the classification reflects the internal kinematics of the system, i.e., classical bulges, unlike pseudo bulges, should have high velocity dispersion \citep{KormendyandKennicutt2004,FisherandDrory2016}.

Interestingly, in our work, we find that the two criteria, i.e., Kormendy relation (KR) and central velocity dispersion ($\sigma_o$), are highly complimentary to each other. More than 60\% of our sample has $\sigma_o>130$ km/s and all of them ($\sim98\%$) are found to be placed within the $\pm1\sigma$ boundary of the Kormendy relation for ellipticals. For the rest 40\%, a large proportion ($\sim60\%$) has $\sigma_o<90$ km/s and nearly all of them ($\sim90$\%) are found to be outliers to the relation. This highlights the effectiveness of KR, in addition to the accuracy of our bulge parameters, being consistent with the inherent kinematics. We use these two complimentary conditions in combination to obtain indubitably classical and pseudo bulges. Instead of forcibly marking the rest ($<20\%$) sample as either pseudo or classical, we mark them as ``ambiguous", so that pseudo and classical bulge properties can be examined without contamination. The placement of our sample on the Fundamental plane and Faber-Jackson plane provides another confirmation of the classification, where classical bulges are found to be the best adherents of the scaling relations set for elliptical galaxies. Exploration of the properties of ambiguous bulges has revealed that it is an important category in itself. Their placement in terms of KR, $\sigma_o$ and the two planes is in-between the two bulge types, which is the reason for ambiguity in their classification. 

\subsection{Bimodality of bulge properties}

After ensuring that we have a well separated set of indubitably pseudo and classical bulges, we have analyzed the distribution of these two bulges types in terms of their structural and stellar parameters. 

In terms of structural parameters, we have found concentration $C$ to be a better differentiator of bulge types than $n_b$ and $B/T$. This is consistent with the study of bulges in the CALIFA sample \citep{Neumannetal2017} that reported $C$ to be the closest follower of the Kormendy relation compared to all other structural indicators. We report that both $n_b$ and $B/T$ are inefficient classifiers of bulge type. While pseudo bulges are mainly concentrated below $n_b\sim2.0$, classical bulges are uniformly distributed over the full range of $n_b$. This is consistent with earlier studies, both based on statistically large samples as well as individual decomposition of smaller samples \citet{Gadotti2009,Gaoetal2020}. Our results for $B/T$ are also consistent with all previous studies \citep{Gadotti2009,FisherandDrory2016,Neumannetal2017,Gaoetal2020} reporting the inefficiency of this indicator due to substantial overlap between the distribution of the two bulge types. The argument proposed is that the classical bulge - being a merger remnant - is independent of the host disc and can easily be a non-dominant component of the galaxy \citep{FisherandDrory2008,Gadotti2009}. We note that all our findings are consistent with the latest work on classifying bulges \citep{Gaoetal2019,Gaoetal2020}, where, 320 galaxies of a local, well-resolved sample have been individually decomposed into multiple components in the optical.

Our major focus is to examine if morphology of the bulges is correlated with their stellar activity. Bimodality, with regard to stellar parameters, has not been observed by earlier studies, i.e., either a large fraction of pseudo bulges are found in the territory of classical bulges or vice-versa \citep{Gadotti2009,FisherandDrory2016,Luoetal2020}. Crucially, which side contributes to the digression is also not agreed upon. In our study, we find that pseudo and classical bulges are well separated in terms of all stellar parameters. All pseudo bulge galaxies have total stellar mass ($M_*$) less than $10^{10.5}$$M_{\odot}$ and all classical bulge galaxies have $M_*$ more than that, with negligible ($<$5\%) exceptions. Similarly, the specific SFR (sSFR) distribution of the galaxies with pseudo and classical bulges is non-overlapping, where $10^{-1.4}$$Gyr^{-1}$ marks the transition point, again with insignificant ($<$5\%) exceptions. In addition to that, the colour ($r-K_s$) of pseudo and classical bulge host galaxies differ by $\sim0.6$ mag, where the middle point ($1.1$ mag) reports less than $10\%$ digressions. This indicates that if bulges are indubitably classified, their stellar properties will be markedly distinct, where, all pseudo bulges will be blue and star forming and all classical bulges will be red and quiescent, with 5-10\% exceptions. 

\subsection{Single structural indicator for quenching}

The finding that galaxies with different bulge types differ in their stellar properties, highlights that structure of a galaxy is correlated with its stellar parameters. Thus, there is a search for a single structural indicator which is the neatest separator of star forming and quiescent galaxy populations. \citet{Cheungetal2012} explored all potential structural indicators, including $n_b$, $B/T$, $C$, $M_*/R_{eb}^2$, to report that $\Sigma_1$ (stellar mass surface density within 1 kpc) is the best in this regard. Adding to that, recent work has claimed that $\Delta$$\Sigma_1$ ($\Sigma_1$ with mass trend removed) is not only the best predictor of quenching but also the best indicator of galaxy's bulge type \citep{Luoetal2020}. They find that the reason for the success of $\Delta$$\Sigma_1$ lies in its similarity to $\Delta$$\langle\mu_{eb}\rangle$ which is most suitable. Our results based on $\Delta$$\langle\mu_{eb}\rangle$ in $K_s$ band add clarity to this picture.

As observed in the case of $\Delta$$\Sigma_1$ \citep{Luoetal2020}, $\Delta$$\langle\mu_{eb}\rangle$ also portrays an elbow-like pattern with both SFR and sSFR of the galaxy. Based on the ``elbow-like" structure they argued that while all pseudo bulge galaxies are star forming, classical bulge galaxies span the full range of stellar activity, where, those in the middle region are transiting from pseudo to classical stage. Our work reveals that all pseudo bulge galaxies are star forming, all classical bulge galaxies are quiescent and galaxies with ambiguous bulges are mainly in the elbow (or middle) region - possibly the green valley. 

Most importantly, the elbow-like pattern transforms into a straight correlation when SFR (or sSFR) is replaced with the global colour ($r-K_s$) of the galaxy. Thus, although ambiguous bulge galaxies have similar SFR as pseudo bulge galaxies, they are redder in colour suggesting that either they are dust ridden or have composite populations. It is possible that the presence of dust has been the cause for the ambiguity in bulge type determination. Other than that, the placement of ambiguous bulge disc galaxies in our work suggests that they are dominantly in the green valley, constantly found between pseudo bulge hosting (blue) and classical bulge hosting (red) disc galaxies. Green valley galaxies, in addition to being dusty, have been observed to be a complex mixture of morphology and stellar populations \citep{Schawinskietal2014,Kelvinetal2018,Phillippsetal2019,Angthopoetal2020}. A more exhaustive analysis of discs with ambiguous bulges, including the fitting of multiple components and accounting for the dust and gas fraction, has the potential to provide insight regarding the green valley. 

This work is supported by the National Key R\&D Program of China (2016YFA0400702) and the National Science Foundation of China (11721303, 11991052). SS acknowledges support from China Postdoctoral Science Foundation fellowship (2019M660299). We are thankful to the anonymous referee for careful reading and useful comments. SS is thankful to James Geach and Yen-Ting Lin for providing access to the data and detailing image specifications. SS is also thankful to Sandra M. Faber, Yingjie Peng and Hassen M. Yesuf for useful discussions. FS acknowledges partial support from a Leverhulme trust Research Fellowship.

%%%%%%%%%%%%%%%%%%%%%%%%%%%%%%%%%%%%%%%%%%%%%%%%%%%%%%%%%%%%%%%%%%%%%%%%%%%%%%%%%%%%%%%%%%%%%%%%%%%%%%

\bibliographystyle{apj}

\begin{thebibliography}{}
\expandafter\ifx\csname natexlab\endcsname\relax\def\natexlab#1{#1}\fi

\bibitem[{{Aguado} {et~al.}(2019){Aguado}, {Ahumada}, {Almeida}, {Anderson},
  {Andrews}, {Anguiano}, {Aquino Ort{\'\i}z}, {Arag{\'o}n-Salamanca},
  {Argudo-Fern{\'a}ndez}, {Aubert}, {Avila-Reese}, {Badenes}, {Barboza
  Rembold}, {Barger}, {Barrera-Ballesteros}, {Bates}, {Bautista}, {Beaton},
  {Beers}, {Belfiore}, {Bernardi}, {Bershady}, {Beutler}, {Bird}, {Bizyaev},
  {Blanc}, {Blanton}, {Blomqvist}, {Bolton}, {Boquien}, {Borissova}, {Bovy}, \&
  {Zou}}]{Aguadoetal2019}
{Aguado}, D.~S., {Ahumada}, R., {Almeida}, A., {et~al.} 2019, ApJS, 240, 23

\bibitem[{{Akhlaghi} \& {Ichikawa}(2015)}]{AkhlaghiandIchikawa2015}
{Akhlaghi}, M., \& {Ichikawa}, T. 2015, ApJS, 220, 1

\bibitem[{{Angthopo} {et~al.}(2020){Angthopo}, {Ferreras}, \&
  {Silk}}]{Angthopoetal2020}
{Angthopo}, J., {Ferreras}, I., \& {Silk}, J. 2020, MNRAS, 495, 2720

\bibitem[{{Annis} {et~al.}(2014){Annis}, {Soares-Santos}, {Strauss}, {Becker},
  {Dodelson}, {Fan}, {Gunn}, {Hao}, {Ivezi{\'c}}, {Jester}, {Jiang},
  {Johnston}, {Kubo}, {Lampeitl}, {Lin}, {Lupton}, {Miknaitis}, {Seo}, {Simet},
  \& {Yanny}}]{Annisetal2014}
{Annis}, J., {Soares-Santos}, M., {Strauss}, M.~A., {et~al.} 2014, ApJ, 794,
  120

\bibitem[{{Aquino-Ort{\'\i}z} {et~al.}(2018){Aquino-Ort{\'\i}z}, {Valenzuela},
  {S{\'a}nchez}, {Hern{\'a}ndez-Toledo}, {{\'A}vila-Reese}, {van de Ven},
  {Rodr{\'\i}guez-Puebla}, {Zhu}, {Mancillas}, {Cano-D{\'\i}az}, \&
  {Garc{\'\i}a-Benito}}]{Aquino-Ortizetal2018}
{Aquino-Ort{\'\i}z}, E., {Valenzuela}, O., {S{\'a}nchez}, S.~F., {et~al.} 2018,
  MNRAS, 479, 2133

\bibitem[{{Baldry} {et~al.}(2006){Baldry}, {Balogh}, {Bower}, {Glazebrook},
  {Nichol}, {Bamford}, \& {Budavari}}]{Baldryetal2006}
{Baldry}, I.~K., {Balogh}, M.~L., {Bower}, R.~G., {et~al.} 2006, MNRAS, 373,
  469

\bibitem[{{Barden} {et~al.}(2005){Barden}, {Rix}, {Somerville}, {Bell},
  {H{\"a}u{\ss}ler}, {Peng}, {Borch}, {Beckwith}, {Caldwell}, {Heymans},
  {Jahnke}, {Jogee}, {McIntosh}, {Meisenheimer}, {S{\'a}nchez}, {Wisotzki}, \&
  {Wolf}}]{Bardenetal2005}
{Barden}, M., {Rix}, H.-W., {Somerville}, R.~S., {et~al.} 2005, ApJ, 635, 959

\bibitem[{{Bell} {et~al.}(2012){Bell}, {van der Wel}, {Papovich}, {Kocevski},
  {Lotz}, {McIntosh}, {Kartaltepe}, {Faber}, {Ferguson}, {Koekemoer}, {Grogin},
  {Wuyts}, {Cheung}, {Conselice}, {Dekel}, {Dunlop}, {Giavalisco},
  {Herrington}, {Koo}, {McGrath}, {de Mello}, {Rix}, {Robaina}, \&
  {Williams}}]{Belletal2012}
{Bell}, E.~F., {van der Wel}, A., {Papovich}, C., {et~al.} 2012, ApJ, 753, 167

\bibitem[{{Bershady} {et~al.}(2000){Bershady}, {Jangren}, \&
  {Conselice}}]{Bershadyetal2000}
{Bershady}, M.~A., {Jangren}, A., \& {Conselice}, C.~J. 2000, AJ, 119, 2645

\bibitem[{{Bertin} \& {Arnouts}(1996)}]{BertinandArnouts1996}
{Bertin}, E., \& {Arnouts}, S. 1996, A\&AS, 117, 393

\bibitem[{{Blakeslee} {et~al.}(2006){Blakeslee}, {Holden}, {Franx}, {Rosati},
  {Bouwens}, {Demarco}, {Ford}, {Homeier}, {Illingworth}, {Jee}, {Mei},
  {Menanteau}, {Meurer}, {Postman}, \& {Tran}}]{Blakesleeetal2006}
{Blakeslee}, J.~P., {Holden}, B.~P., {Franx}, M., {et~al.} 2006, ApJ, 644, 30

\bibitem[{{Bluck} {et~al.}(2014){Bluck}, {Mendel}, {Ellison}, {Moreno},
  {Simard}, {Patton}, \& {Starkenburg}}]{Blucketal2014}
{Bluck}, A. F.~L., {Mendel}, J.~T., {Ellison}, S.~L., {et~al.} 2014, MNRAS,
  441, 599

\bibitem[{{Bluck} {et~al.}(2016){Bluck}, {Mendel}, {Ellison}, {Patton},
  {Simard}, {Henriques}, {Torrey}, {Teimoorinia}, {Moreno}, \&
  {Starkenburg}}]{Blucketal2016}
---. 2016, MNRAS, 462, 2559

\bibitem[{{Bluck} {et~al.}(2019){Bluck}, {Bottrell}, {Teimoorinia},
  {Henriques}, {Mendel}, {Ellison}, {Thanjavur}, {Simard}, {Patton},
  {Conselice}, {Moreno}, \& {Woo}}]{Blucketal2019}
{Bluck}, A. F.~L., {Bottrell}, C., {Teimoorinia}, H., {et~al.} 2019, MNRAS,
  485, 666

\bibitem[{{Bolton} {et~al.}(2012){Bolton}, {Schlegel}, {Aubourg}, {Bailey},
  {Bhardwaj}, {Brownstein}, {Burles}, {Chen}, {Dawson}, {Eisenstein}, {Gunn},
  {Knapp}, {Loomis}, {Lupton}, {Maraston}, {Muna}, {Myers}, {Olmstead},
  {Padmanabhan}, {P{\^a}ris}, {Percival}, {Petitjean}, {Rockosi}, {Ross},
  {Schneider}, {Shu}, {Strauss}, {Thomas}, {Tremonti}, {Wake}, {Weaver}, \&
  {Wood-Vasey}}]{Boltonetal2012}
{Bolton}, A.~S., {Schlegel}, D.~J., {Aubourg}, {\'E}., {et~al.} 2012, AJ, 144,
  144

\bibitem[{{Bottrell} {et~al.}(2019){Bottrell}, {Simard}, {Mendel}, \&
  {Ellison}}]{Bottrelletal2019}
{Bottrell}, C., {Simard}, L., {Mendel}, J.~T., \& {Ellison}, S.~L. 2019, MNRAS,
  486, 390

\bibitem[{{Brinchmann} {et~al.}(2004){Brinchmann}, {Charlot}, {White},
  {Tremonti}, {Kauffmann}, {Heckman}, \& {Brinkmann}}]{Brinchmannetal2004}
{Brinchmann}, J., {Charlot}, S., {White}, S.~D.~M., {et~al.} 2004, MNRAS, 351,
  1151

\bibitem[{{Bundy} {et~al.}(2006){Bundy}, {Ellis}, {Conselice}, {Taylor},
  {Cooper}, {Willmer}, {Weiner}, {Coil}, {Noeske}, \&
  {Eisenhardt}}]{Bundyetal2006}
{Bundy}, K., {Ellis}, R.~S., {Conselice}, C.~J., {et~al.} 2006, ApJ, 651, 120

\bibitem[{{Cameron} {et~al.}(2009){Cameron}, {Driver}, {Graham}, \&
  {Liske}}]{Cameronetal2009}
{Cameron}, E., {Driver}, S.~P., {Graham}, A.~W., \& {Liske}, J. 2009, ApJ, 699,
  105

\bibitem[{{Cheung} {et~al.}(2012){Cheung}, {Faber}, {Koo}, {Dutton}, {Simard},
  {McGrath}, {Huang}, {Bell}, {Dekel}, {Fang}, {Salim}, {Barro}, {Bundy},
  {Coil}, {Cooper}, {Conselice}, {Davis}, {Dom{\'\i}nguez}, {Kassin},
  {Kocevski}, {Koekemoer}, {Lin}, {Lotz}, {Newman}, {Phillips}, {Rosario},
  {Weiner}, \& {Willmer}}]{Cheungetal2012}
{Cheung}, E., {Faber}, S.~M., {Koo}, D.~C., {et~al.} 2012, ApJ, 760, 131

\bibitem[{{Conselice}(2003)}]{Conselice2003}
{Conselice}, C.~J. 2003, ApJS, 147, 1

\bibitem[{{Conselice}(2014)}]{Conselice2014}
---. 2014, ARA\&A, 52, 291

\bibitem[{{Cortese} {et~al.}(2014){Cortese}, {Fogarty}, {Ho}, {Bekki},
  {Bland-Hawthorn}, {Colless}, {Couch}, {Croom}, {Glazebrook}, {Mould},
  {Scott}, {Sharp}, {Tonini}, {Allen}, {Bloom}, {Bryant}, {Cluver}, {Davies},
  {Drinkwater}, {Goodwin}, {Green}, {Kewley}, {Kostantopoulos}, {Lawrence},
  {Mahajan}, {Medling}, {Owers}, {Richards}, {Sweet}, \&
  {Wong}}]{Corteseetal2014}
{Cortese}, L., {Fogarty}, L.~M.~R., {Ho}, I.~T., {et~al.} 2014, ApJL, 795, L37

\bibitem[{{Cowie} {et~al.}(1996){Cowie}, {Songaila}, {Hu}, \&
  {Cohen}}]{Cowieetal1996}
{Cowie}, L.~L., {Songaila}, A., {Hu}, E.~M., \& {Cohen}, J.~G. 1996, AJ, 112,
  839

\bibitem[{{Dawson} {et~al.}(2013){Dawson}, {Schlegel}, {Ahn}, {Anderson},
  {Aubourg}, {Bailey}, {Barkhouser}, {Bautista}, {Beifiori}, {Berlind},
  {Bhardwaj}, {Bizyaev}, {Blake}, {Blanton}, {Blomqvist}, {Bolton}, {Borde},
  {Bovy}, {Brandt}, {Brewington}, {Brinkmann}, {Brown}, {Brownstein}, {Bundy},
  {Busca}, {Carithers}, {Carnero}, {Carr}, {Zehavi}, {Zhao}, \&
  {Zheng}}]{Dawsonetal2013}
{Dawson}, K.~S., {Schlegel}, D.~J., {Ahn}, C.~P., {et~al.} 2013, AJ, 145, 10

\bibitem[{{Driver} {et~al.}(2006){Driver}, {Allen}, {Graham}, {Cameron},
  {Liske}, {Ellis}, {Cross}, {De Propris}, {Phillipps}, \&
  {Couch}}]{Driveretal2006}
{Driver}, S.~P., {Allen}, P.~D., {Graham}, A.~W., {et~al.} 2006, MNRAS, 368,
  414

\bibitem[{{Drory} \& {Fisher}(2007)}]{DroryandFisher2007}
{Drory}, N., \& {Fisher}, D.~B. 2007, ApJ, 664, 640

\bibitem[{{Eisenstein} {et~al.}(2011){Eisenstein}, {Weinberg}, {Agol},
  {Aihara}, {Allende Prieto}, {Anderson}, {Arns}, {Aubourg}, {Bailey},
  {Balbinot}, {Barkhouser}, {Beers}, {Berlind}, {Bickerton}, {Bizyaev},
  {Blanton}, {Bochanski}, {Bolton}, {Bosman}, {Bovy}, {Brandt}, {Breslauer},
  {Brewington}, {Brinkmann}, {Brown}, {Brownstein}, {Burger}, {Busca},
  {Campbell}, {Cargile}, \& {Zhao}}]{Eisensteinetal2011}
{Eisenstein}, D.~J., {Weinberg}, D.~H., {Agol}, E., {et~al.} 2011, AJ, 142, 72

\bibitem[{{Faber} \& {Jackson}(1976)}]{FaberandJackson1976}
{Faber}, S.~M., \& {Jackson}, R.~E. 1976, ApJ, 204, 668

\bibitem[{{Fang} {et~al.}(2013){Fang}, {Faber}, {Koo}, \&
  {Dekel}}]{Fangetal2013}
{Fang}, J.~J., {Faber}, S.~M., {Koo}, D.~C., \& {Dekel}, A. 2013, ApJ, 776, 63

\bibitem[{{Fisher} \& {Drory}(2008)}]{FisherandDrory2008}
{Fisher}, D.~B., \& {Drory}, N. 2008, \aj, 136, 773

\bibitem[{{Fisher} \& {Drory}(2016)}]{FisherandDrory2016}
---. 2016, ASSL, Vol. 418, {An Observational Guide to Identifying Pseudobulges
  and Classical Bulges in Disc Galaxies}, ed. E.~{Laurikainen}, R.~{Peletier},
  \& D.~{Gadotti}, 41

\bibitem[{{Franx} {et~al.}(2008){Franx}, {van Dokkum}, {F{\"o}rster Schreiber},
  {Wuyts}, {Labb{\'e}}, \& {Toft}}]{Franxetal2008}
{Franx}, M., {van Dokkum}, P.~G., {F{\"o}rster Schreiber}, N.~M., {et~al.}
  2008, ApJ, 688, 770

\bibitem[{{Gadotti}(2009)}]{Gadotti2009}
{Gadotti}, D.~A. 2009, MNRAS, 393, 1531

\bibitem[{{Gallazzi} {et~al.}(2006){Gallazzi}, {Charlot}, {Brinchmann}, \&
  {White}}]{Gallazzietal2006}
{Gallazzi}, A., {Charlot}, S., {Brinchmann}, J., \& {White}, S. D.~M. 2006,
  MNRAS, 370, 1106

\bibitem[{{Gao} {et~al.}(2019){Gao}, {Ho}, {Barth}, \& {Li}}]{Gaoetal2019}
{Gao}, H., {Ho}, L.~C., {Barth}, A.~J., \& {Li}, Z.-Y. 2019, ApJS, 244, 34

\bibitem[{{Gao} {et~al.}(2020){Gao}, {Ho}, {Barth}, \& {Li}}]{Gaoetal2020}
---. 2020, ApJS, 247, 20

\bibitem[{{Geach} {et~al.}(2017){Geach}, {Lin}, {Makler}, {Kneib}, {Ross},
  {Wang}, {Hsieh}, {Leauthaud}, {Bundy}, {McCracken}, {Comparat}, {Caminha},
  {Hudelot}, {Lin}, {Van Waerbeke}, {Pereira}, \& {Mast}}]{Geachetal2017}
{Geach}, J.~E., {Lin}, Y.~T., {Makler}, M., {et~al.} 2017, ApJS, 231, 7

\bibitem[{{Graham} \& {Driver}(2005)}]{GrahamandDriver2005}
{Graham}, A.~W., \& {Driver}, S.~P. 2005, PASA, 22, 118

\bibitem[{{Graham} {et~al.}(2005){Graham}, {Driver}, {Petrosian}, {Conselice},
  {Bershady}, {Crawford}, \& {Goto}}]{Grahametal2005}
{Graham}, A.~W., {Driver}, S.~P., {Petrosian}, V., {et~al.} 2005, AJ, 130, 1535

\bibitem[{{Graham} {et~al.}(2018){Graham}, {Cappellari}, {Li}, {Mao},
  {Bershady}, {Bizyaev}, {Brinkmann}, {Brownstein}, {Bundy}, {Drory}, {Law},
  {Pan}, {Thomas}, {Wake}, {Weijmans}, {Westfall}, \& {Yan}}]{Grahametal2018}
{Graham}, M.~T., {Cappellari}, M., {Li}, H., {et~al.} 2018, MNRAS, 477, 4711

\bibitem[{{Hill} {et~al.}(2008){Hill}, {Gebhardt}, {Komatsu}, {Drory},
  {MacQueen}, {Adams}, {Blanc}, {Koehler}, {Rafal}, {Roth}, {Kelz}, {Gronwall},
  {Ciardullo}, \& {Schneider}}]{Hilletal2008}
{Hill}, G.~J., {Gebhardt}, K., {Komatsu}, E., {et~al.} 2008, ASPC, Vol. 399,
  {The Hobby-Eberly Telescope Dark Energy Experiment (HETDEX): Description and
  Early Pilot Survey Results}, ed. T.~{Kodama}, T.~{Yamada}, \& K.~{Aoki}, 115

\bibitem[{{Hodge} {et~al.}(2011){Hodge}, {Becker}, {White}, {Richards}, \&
  {Zeimann}}]{Hodgeetal2011}
{Hodge}, J.~A., {Becker}, R.~H., {White}, R.~L., {Richards}, G.~T., \&
  {Zeimann}, G.~R. 2011, AJ, 142, 3

\bibitem[{{Huang} {et~al.}(2013){Huang}, {Ho}, {Peng}, {Li}, \&
  {Barth}}]{Huangetal2013}
{Huang}, S., {Ho}, L.~C., {Peng}, C.~Y., {Li}, Z.-Y., \& {Barth}, A.~J. 2013,
  ApJ, 766, 47

\bibitem[{{Jedrzejewski}(1987)}]{Jedrzejewski1987}
{Jedrzejewski}, R.~I. 1987, MNRAS, 226, 747

\bibitem[{{Jorgensen} {et~al.}(1995){Jorgensen}, {Franx}, \&
  {Kjaergaard}}]{Jorgensenetal1995}
{Jorgensen}, I., {Franx}, M., \& {Kjaergaard}, P. 1995, MNRAS, 276, 1341

\bibitem[{{Kauffmann} {et~al.}(2006){Kauffmann}, {Heckman}, {De Lucia},
  {Brinchmann}, {Charlot}, {Tremonti}, {White}, \&
  {Brinkmann}}]{Kauffmannetal2006}
{Kauffmann}, G., {Heckman}, T.~M., {De Lucia}, G., {et~al.} 2006, MNRAS, 367,
  1394

\bibitem[{{Kauffmann} {et~al.}(2003){Kauffmann}, {Heckman}, {White}, {Charlot},
  {Tremonti}, {Peng}, {Seibert}, {Brinkmann}, {Nichol}, {SubbaRao}, \&
  {York}}]{Kauffmannetal2003}
{Kauffmann}, G., {Heckman}, T.~M., {White}, S. D.~M., {et~al.} 2003, MNRAS,
  341, 54

\bibitem[{{Kelvin} {et~al.}(2018){Kelvin}, {Bremer}, {Phillipps}, {James},
  {Davies}, {De Propris}, {Moffett}, {Percival}, {Baldry}, {Collins},
  {Alpaslan}, {Bland-Hawthorn}, {Brough}, {Cluver}, {Driver}, {Hashemizadeh},
  {Holwerda}, {Laine}, {Lara-Lopez}, {Liske}, {Maciejewski}, {Napolitano},
  {Penny}, {Popescu}, {Sansom}, {Sutherland}, {Taylor}, {van Kampen}, \&
  {Wang}}]{Kelvinetal2018}
{Kelvin}, L.~S., {Bremer}, M.~N., {Phillipps}, S., {et~al.} 2018, MNRAS, 477,
  4116

\bibitem[{{Kormendy}(1977)}]{Kormendy1977}
{Kormendy}, J. 1977, ApJ, 218, 333

\bibitem[{{Kormendy}(2016)}]{Kormendy2016}
---. 2016, ASSL, Vol. 418, {Elliptical Galaxies and Bulges of Disc Galaxies:
  Summary of Progress and Outstanding Issues}, ed. E.~{Laurikainen},
  R.~{Peletier}, \& D.~{Gadotti}, 431

\bibitem[{{Kormendy} \& {Kennicutt}(2004)}]{KormendyandKennicutt2004}
{Kormendy}, J., \& {Kennicutt}, Robert~C., J. 2004, ARA\&A, 42, 603

\bibitem[{{Lang} {et~al.}(2016){Lang}, {Hogg}, \& {Schlegel}}]{Langetal2016}
{Lang}, D., {Hogg}, D.~W., \& {Schlegel}, D.~J. 2016, AJ, 151, 36

\bibitem[{{Lang} {et~al.}(2014){Lang}, {Wuyts}, {Somerville}, {F{\"o}rster
  Schreiber}, {Genzel}, {Bell}, {Brammer}, {Dekel}, {Faber}, {Ferguson},
  {Grogin}, {Kocevski}, {Koekemoer}, {Lutz}, {McGrath}, {Momcheva}, {Nelson},
  {Primack}, {Rosario}, {Skelton}, {Tacconi}, {van Dokkum}, \&
  {Whitaker}}]{Langetal2014}
{Lang}, P., {Wuyts}, S., {Somerville}, R.~S., {et~al.} 2014, ApJ, 788, 11

\bibitem[{{Lotz} {et~al.}(2004){Lotz}, {Primack}, \& {Madau}}]{Lotzetal2004}
{Lotz}, J.~M., {Primack}, J., \& {Madau}, P. 2004, AJ, 128, 163

\bibitem[{{Luo} {et~al.}(2020){Luo}, {Faber}, {Rodr{\'\i}guez-Puebla}, {Woo},
  {Guo}, {Koo}, {Primack}, {Chen}, {Yesuf}, {Lin}, {Barro}, {Fang}, {Pand ya},
  {Huertas-Company}, \& {Mao}}]{Luoetal2020}
{Luo}, Y., {Faber}, S.~M., {Rodr{\'\i}guez-Puebla}, A., {et~al.} 2020, MNRAS,
  493, 1686

\bibitem[{{Mendel} {et~al.}(2013){Mendel}, {Simard}, {Ellison}, \&
  {Patton}}]{Mendeletal2013}
{Mendel}, J.~T., {Simard}, L., {Ellison}, S.~L., \& {Patton}, D.~R. 2013,
  MNRAS, 429, 2212

\bibitem[{{Miyazaki} {et~al.}(2012){Miyazaki}, {Komiyama}, {Nakaya}, {Kamata},
  {Doi}, {Hamana}, {Karoji}, {Furusawa}, {Kawanomoto}, {Morokuma}, {Ishizuka},
  {Nariai}, {Tanaka}, {Uraguchi}, {Utsumi}, {Obuchi}, {Okura}, {Oguri},
  {Takata}, {Tomono}, {Kurakami}, {Namikawa}, {Usuda}, {Yamanoi}, {Terai},
  {Uekiyo}, \& {Yamada}}]{Miyazakietal2012}
{Miyazaki}, S., {Komiyama}, Y., {Nakaya}, H., {et~al.} 2012, SPIE, Vol. 8446,
  {Hyper Suprime-Cam}, 84460Z

\bibitem[{{Neumann} {et~al.}(2017){Neumann}, {Wisotzki}, {Choudhury},
  {Gadotti}, {Walcher}, {Bland-Hawthorn}, {Garc{\'\i}a-Benito}, {Gonz{\'a}lez
  Delgado}, {Husemann}, {Marino}, {M{\'a}rquez}, {S{\'a}nchez}, {Ziegler}, \&
  {Califa Collaboration}}]{Neumannetal2017}
{Neumann}, J., {Wisotzki}, L., {Choudhury}, O.~S., {et~al.} 2017, A\&A, 604,
  A30

\bibitem[{{Papovich} {et~al.}(2012){Papovich}, {Gebhardt}, {Behroozi},
  {Bender}, {Blanc}, {Ciardullo}, {DePoy}, {de Jong}, {Drory}, {Evans},
  {Fabricius}, {Finkelstein}, {Gawiser}, {Greene}, {Gronwall}, {Hill}, {Hopp},
  {Jogee}, {Lacy}, {Landriau}, {Marshall}, {Tuttle}, {Somerville}, {Steinmetz},
  {Suntzeff}, {Tran}, {Wechsler}, \& {Wisotzki}}]{Papovichetal2012}
{Papovich}, C.~J., {Gebhardt}, K., {Behroozi}, P., {et~al.} 2012, in AAS, Vol.
  219, American Astronomical Society Meeting Abstracts \#219, 424.09

\bibitem[{{Peng} {et~al.}(2002){Peng}, {Ho}, {Impey}, \& {Rix}}]{Pengetal2002}
{Peng}, C.~Y., {Ho}, L.~C., {Impey}, C.~D., \& {Rix}, H.-W. 2002, AJ, 124, 266

\bibitem[{{Peng} {et~al.}(2010){Peng}, {Ho}, {Impey}, \& {Rix}}]{Pengetal2010}
---. 2010, AJ, 139, 2097

\bibitem[{{Phillipps} {et~al.}(2019){Phillipps}, {Bremer}, {Hopkins}, {De
  Propris}, {Taylor}, {James}, {Davies}, {Cluver}, {Driver}, {Eales},
  {Holwerda}, {Kelvin}, \& {Sansom}}]{Phillippsetal2019}
{Phillipps}, S., {Bremer}, M.~N., {Hopkins}, A.~M., {et~al.} 2019, MNRAS, 485,
  5559

\bibitem[{{Prugniel} \& {Soubiran}(2001)}]{PrugnielandSoubiran2001}
{Prugniel}, P., \& {Soubiran}, C. 2001, A\&A, 369, 1048

\bibitem[{{Ravindranath} {et~al.}(2004){Ravindranath}, {Ferguson}, {Conselice},
  {Giavalisco}, {Dickinson}, {Chatzichristou}, {de Mello}, {Fall}, {Gardner},
  {Grogin}, {Hornschemeier}, {Jogee}, {Koekemoer}, {Kretchmer}, {Livio},
  {Mobasher}, \& {Somerville}}]{Ravindranathetal2004}
{Ravindranath}, S., {Ferguson}, H.~C., {Conselice}, C., {et~al.} 2004, ApJL,
  604, L9

\bibitem[{{Rettura} {et~al.}(2006){Rettura}, {Rosati}, {Strazzullo},
  {Dickinson}, {Fosbury}, {Rocca-Volmerange}, {Cimatti}, {di Serego Alighieri},
  {Kuntschner}, {Lanzoni}, {Nonino}, {Popesso}, {Stern}, {Eisenhardt},
  {Lidman}, \& {Stanford}}]{Retturaetal2006}
{Rettura}, A., {Rosati}, P., {Strazzullo}, V., {et~al.} 2006, A\&A, 458, 717

\bibitem[{{Sachdeva}(2013)}]{Sachdeva2013}
{Sachdeva}, S. 2013, MNRAS, 435, 1186

\bibitem[{{Sachdeva} {et~al.}(2015){Sachdeva}, {Gadotti}, {Saha}, \&
  {Singh}}]{Sachdevaetal2015}
{Sachdeva}, S., {Gadotti}, D.~A., {Saha}, K., \& {Singh}, H.~P. 2015, MNRAS,
  451, 2

\bibitem[{{Sachdeva} {et~al.}(2019){Sachdeva}, {Gogoi}, {Saha}, {Kembhavi}, \&
  {Raychaudhury}}]{Sachdevaetal2019}
{Sachdeva}, S., {Gogoi}, R., {Saha}, K., {Kembhavi}, A., \& {Raychaudhury}, S.
  2019, MNRAS, 487, 1795

\bibitem[{{Sachdeva} \& {Saha}(2016)}]{SachdevaandSaha2016}
{Sachdeva}, S., \& {Saha}, K. 2016, ApJL, 820, L4

\bibitem[{{Sachdeva} {et~al.}(2017){Sachdeva}, {Saha}, \&
  {Singh}}]{Sachdevaetal2017}
{Sachdeva}, S., {Saha}, K., \& {Singh}, H.~P. 2017, ApJ, 840, 79

\bibitem[{{Salim} {et~al.}(2018){Salim}, {Boquien}, \& {Lee}}]{Salimetal2018}
{Salim}, S., {Boquien}, M., \& {Lee}, J.~C. 2018, ApJ, 859, 11

\bibitem[{{Salim} {et~al.}(2005){Salim}, {Charlot}, {Rich}, {Kauffmann},
  {Heckman}, {Barlow}, {Bianchi}, {Byun}, {Donas}, {Forster}, {Friedman},
  {Jelinsky}, {Lee}, {Madore}, {Malina}, {Martin}, {Milliard}, {Morrissey},
  {Neff}, {Schiminovich}, {Seibert}, {Siegmund}, {Small}, {Szalay}, {Welsh}, \&
  {Wyder}}]{Salimetal2005}
{Salim}, S., {Charlot}, S., {Rich}, R.~M., {et~al.} 2005, ApJL, 619, L39

\bibitem[{{Salim} {et~al.}(2007){Salim}, {Rich}, {Charlot}, {Brinchmann},
  {Johnson}, {Schiminovich}, {Seibert}, {Mallery}, {Heckman}, {Forster},
  {Friedman}, {Martin}, {Morrissey}, {Neff}, {Small}, {Wyder}, {Bianchi},
  {Donas}, {Lee}, {Madore}, {Milliard}, {Szalay}, {Welsh}, \&
  {Yi}}]{Salimetal2007}
{Salim}, S., {Rich}, R.~M., {Charlot}, S., {et~al.} 2007, ApJS, 173, 267

\bibitem[{{Salim} {et~al.}(2016){Salim}, {Lee}, {Janowiecki}, {da Cunha},
  {Dickinson}, {Boquien}, {Burgarella}, {Salzer}, \& {Charlot}}]{Salimetal2016}
{Salim}, S., {Lee}, J.~C., {Janowiecki}, S., {et~al.} 2016, ApJS, 227, 2

\bibitem[{{Schawinski} {et~al.}(2014){Schawinski}, {Urry}, {Simmons},
  {Fortson}, {Kaviraj}, {Keel}, {Lintott}, {Masters}, {Nichol}, {Sarzi},
  {Skibba}, {Treister}, {Willett}, {Wong}, \& {Yi}}]{Schawinskietal2014}
{Schawinski}, K., {Urry}, C.~M., {Simmons}, B.~D., {et~al.} 2014, MNRAS, 440,
  889

\bibitem[{{Shen} {et~al.}(2003){Shen}, {Mo}, {White}, {Blanton}, {Kauffmann},
  {Voges}, {Brinkmann}, \& {Csabai}}]{Shenetal2003}
{Shen}, S., {Mo}, H.~J., {White}, S. D.~M., {et~al.} 2003, MNRAS, 343, 978

\bibitem[{{Simard} {et~al.}(2011){Simard}, {Mendel}, {Patton}, {Ellison}, \&
  {McConnachie}}]{Simardetal2011}
{Simard}, L., {Mendel}, J.~T., {Patton}, D.~R., {Ellison}, S.~L., \&
  {McConnachie}, A.~W. 2011, ApJS, 196, 11

\bibitem[{{Strateva} {et~al.}(2001){Strateva}, {Ivezi{\'c}}, {Knapp},
  {Narayanan}, {Strauss}, {Gunn}, {Lupton}, {Schlegel}, {Bahcall}, {Brinkmann},
  {Brunner}, {Budav{\'a}ri}, {Csabai}, {Castander}, {Doi}, {Fukugita},
  {Gy{\H{o}}ry}, {Hamabe}, {Hennessy}, {Ichikawa}, {Kunszt}, {Lamb}, {McKay},
  {Okamura}, {Racusin}, {Sekiguchi}, {Schneider}, {Shimasaku}, \&
  {York}}]{Stratevaetal2001}
{Strateva}, I., {Ivezi{\'c}}, {\v{Z}}., {Knapp}, G.~R., {et~al.} 2001, AJ, 122,
  1861

\bibitem[{{Teimoorinia} {et~al.}(2016){Teimoorinia}, {Bluck}, \&
  {Ellison}}]{Teimooriniaetal2016}
{Teimoorinia}, H., {Bluck}, A. F.~L., \& {Ellison}, S.~L. 2016, MNRAS, 457,
  2086

\bibitem[{{Thomas} {et~al.}(2013){Thomas}, {Steele}, {Maraston}, {Johansson},
  {Beifiori}, {Pforr}, {Str{\"o}mb{\"a}ck}, {Tremonti}, {Wake}, {Bizyaev},
  {Bolton}, {Brewington}, {Brownstein}, {Comparat}, {Kneib}, {Malanushenko},
  {Malanushenko}, {Oravetz}, {Pan}, {Parejko}, {Schneider}, {Shelden},
  {Simmons}, {Snedden}, {Tanaka}, {Weaver}, \& {Yan}}]{Thomasetal2013}
{Thomas}, D., {Steele}, O., {Maraston}, C., {et~al.} 2013, MNRAS, 431, 1383

\bibitem[{{Timlin} {et~al.}(2016){Timlin}, {Ross}, {Richards}, {Lacy}, {Ryan},
  {Stone}, {Bauer}, {Brandt}, {Fan}, {Glikman}, {Haggard}, {Jiang}, {LaMassa},
  {Lin}, {Makler}, {McGehee}, {Myers}, {Schneider}, {Urry}, {Wollack}, \&
  {Zakamska}}]{Timlinetal2016}
{Timlin}, J.~D., {Ross}, N.~P., {Richards}, G.~T., {et~al.} 2016, ApJS, 225, 1

\bibitem[{{Valiante} {et~al.}(2016){Valiante}, {Smith}, {Eales}, {Maddox},
  {Ibar}, {Hopwood}, {Dunne}, {Cigan}, {Dye}, {Pascale}, {Rigby}, {Bourne},
  {Furlanetto}, \& {Ivison}}]{Valianteetal2016}
{Valiante}, E., {Smith}, M.~W.~L., {Eales}, S., {et~al.} 2016, MNRAS, 462, 3146

\bibitem[{{van de Sande} {et~al.}(2018){van de Sande}, {Scott},
  {Bland-Hawthorn}, {Brough}, {Bryant}, {Colless}, {Cortese}, {Croom},
  {d'Eugenio}, {Foster}, {Goodwin}, {Konstantopoulos}, {Lawrence}, {McDermid},
  {Medling}, {Owers}, {Richards}, \& {Sharp}}]{vandeSandeetal2018}
{van de Sande}, J., {Scott}, N., {Bland-Hawthorn}, J., {et~al.} 2018, NatAs, 2,
  483

\bibitem[{{van den Bosch}(2016)}]{vandenBosch2016}
{van den Bosch}, R. C.~E. 2016, ApJ, 831, 134

\bibitem[{{van der Wel}(2008)}]{vanderWel2008}
{van der Wel}, A. 2008, ApJL, 675, L13

\bibitem[{{Viero} {et~al.}(2014){Viero}, {Asboth}, {Roseboom}, {Moncelsi},
  {Marsden}, {Mentuch Cooper}, {Zemcov}, {Addison}, {Baker}, {Beelen}, {Bock},
  {Bridge}, {Conley}, {Devlin}, {Dor{\'e}}, {Farrah}, {Finkelstein},
  {Font-Ribera}, {Geach}, {Gebhardt}, {Gill}, {Glenn}, {Hajian}, {Halpern},
  {Jogee}, {Kurczynski}, {Lapi}, {Negrello}, {Oliver}, {Papovich}, {Quadri},
  {Ross}, {Scott}, {Schulz}, {Somerville}, {Spergel}, {Vieira}, {Wang}, \&
  {Wechsler}}]{Vieroetal2014}
{Viero}, M.~P., {Asboth}, V., {Roseboom}, I.~G., {et~al.} 2014, ApJS, 210, 22

\bibitem[{{Wake} {et~al.}(2012){Wake}, {van Dokkum}, \& {Franx}}]{Wakeetal2012}
{Wake}, D.~A., {van Dokkum}, P.~G., \& {Franx}, M. 2012, ApJL, 751, L44

\bibitem[{{Wuyts} {et~al.}(2011){Wuyts}, {F{\"o}rster Schreiber}, {van der
  Wel}, {Magnelli}, {Guo}, {Genzel}, {Lutz}, {Aussel}, {Barro}, {Berta},
  {Cava}, {Graci{\'a}-Carpio}, {Hathi}, {Huang}, {Kocevski}, {Koekemoer},
  {Lee}, {Le Floc'h}, {McGrath}, {Nordon}, {Popesso}, {Pozzi}, {Riguccini},
  {Rodighiero}, {Saintonge}, \& {Tacconi}}]{Wuytsetal2011}
{Wuyts}, S., {F{\"o}rster Schreiber}, N.~M., {van der Wel}, A., {et~al.} 2011,
  ApJ, 742, 96

\end{thebibliography}

\end{document}